\documentclass{pasa}%

\usepackage{graphicx}

\title[$m$-mode Imaging the Southern Sky at 159\,MHz with the EDA2]{Imaging the Southern Sky at 159\,MHz using Spherical Harmonics with the Engineering Development Array 2}

\author[M. A. Kriele et al.]{%
Michael A. Kriele$^{1,2,3}$\thanks{E-mail: mike.kriele@curtin.edu.au}, %
Randall B. Wayth$^{1,2}$, %
Mark J. Bentum$^{3,4}$, %
Budi Juswardy$^{1}$ and Cathryn M. Trott$^{1,2}$
\affil{$^{1}$International Centre for Radio Astronomy Research (ICRAR), Curtin University, Bentley, WA, Australia, 6102}%
\affil{$^{2}$ARC Centre of Excellence for All-Sky Astrophysics in 3 Dimensions (ASTRO 3D)}%
\affil{$^{3}$Eindhoven University of Technology, 5612 AZ Eindhoven, Netherlands}%
\affil{$^{4}$ASTRON, the Netherlands Institute for Radio Astronomy, 7991 PD Dwingeloo, Netherlands}%
}%

\jid{PASA}
\doi{10.1017/pas.\the\year.xxx}
\jyear{\the\year}

\usepackage{aas_macros}
\usepackage[unicode]{hyperref} 
\hypersetup{colorlinks,citecolor=blue,linkcolor=blue,urlcolor=blue}

\usepackage{graphicx}	
\usepackage[nolist,nohyperlinks]{acronym}    
\usepackage{amsmath}	
\usepackage{amssymb}	
\usepackage[authormarkup=none]{changes}    
\usepackage{enumitem}   
\usepackage{subcaption}

\definechangesauthor[name=Michael Kriele, color=red]{MK}

\newcommand\Beta{\mathrm{\mathcal{B}}}

\begin{document}

\begin{acronym}
    \acro{CD}{cosmic dawn}
    \acro{CMB}{cosmic microwave background}
    \acro{DA}{dark ages}
    \acro{EDA}{Engineering Development Array}
    \acro{EDA2}{Engineering Development Array 2}
    \acro{EoR}{epoch of reionisation}
    \acro{FoV}{field of view}
    \acro{SKA}{Square Kilometre Array}
    \acro{LWA}{Long Wavelength Array}
    \acro{MWA}{Murchison Widefield Array}
    \acro{AAVS}{Aperture Array Verification System}
    \acro{LST}{local sidereal time}
    \acro{PSF}{point-spread function}
    \acro{MRO}{Murchison Radio-astronomy Observatory}
    \acro{ICRAR}{Internation Centre of Radio Astronomy Research}
    \acro{RA}{right ascension}
    \acro{DEC}{declination}
    \acro{HEALPix}{Hierarchical Equal Area isoLatitude Pixelation of a sphere}
    \acro{FWHM}{Full-Width Half-Maximum}
    \acro{UTC}{Coordinated Universal Time}
    \acro{ASTRO3D}{Australian Research Council Centre of Excellence for All Sky Astrophysics in Three Dimensions}
    \acro{LAMDA}{Legacy Archive for Microwave Background Data Analysis}
    \acro{HEASARC}{High Energy Astrophysics Science Archive Center}
    \acro{IGM}{intergalactic medium}
    \acro{AGN}{active galactic nuclei}
    \acro{SHBC}{spherical harmonic beam coverage}
    \acro{SH-space}{spherical-harmonic space}
    \acro{LLS}{linear least squares}
    \acro{SI}{spectral indices}
    \acro{FWHM}{full-width half-maximum}
    \acro{GSM}{global sky model}
    \acro{LFSS}{low-frequency sky survey}
    \acro{RFI}{radio frequency interference}
    \acro{OVRO-LWA}{Owens Valley Radio Observatory Long Wavelength Array}
    \acro{RMS}{root mean square}
\end{acronym}

\begin{frontmatter}
\maketitle

\begin{abstract}
One of the major priorities of international radio astronomy is to study the early universe through the detection of the 21 cm HI line from the \acf{EoR}. Due to the weak nature of the 21 cm signal, an important part in the detection of the \acs{EoR} is removing contaminating foregrounds from our observations as they are multiple orders of magnitude brighter. In order to achieve this, sky maps spanning a wide range of frequencies and angular scales are required for calibration and foreground subtraction. Complementing the existing low-frequency sky maps, we have constructed a Southern Sky map through spherical harmonic transit interferometry utilising the \acf{EDA2}, a \acf{SKA} low-frequency array prototype system.  We use the $m$-mode formalism to create an all-sky map at 159\,MHz with an angular resolution of 3 degrees, with data from the \acs{EDA2} providing information over +60 degrees to -90 degrees in declination. We also introduce a new method for visualising and quantifying how the baseline distribution of an interferometer maps to the spherical harmonics, and discuss how prior information can be used to constrain spherical harmonic components that the interferometer is not sensitive to.
\end{abstract}

\end{frontmatter}

\section{Introduction}
One of the primary focuses in radio astronomy science is understanding the early Universe formation history. Although the redshift boundary where ionised gas recombined into neutral hydrogen is well measureable through the \ac{CMB} \citep{Komatsu2009}, the redshift region of the birth of the first stars, during the \ac{CD}, and the absolute redshift boundaries of the \acf{EoR} remain uncertain. Observing and better constraining the redshift regions of these epochs is therefore key to gain a better understating of the formation history of our Universe; as well as early stages of cosmic structure formation, where presently almost nothing is known. \citep{Furlanetto2006}.

Much effort has been put in to attempt to constrain the boundaries of the \ac{EoR} through directly probing the \ac{IGM}, \textit{e.g.} determining the anisotropies in the \ac{CMB} due to Thompson Scattering and calculating the optical depth limit \citep{Holder2003, PlanckCollaborationXIII2016}, measuring the scattering effects caused by Lyman-Alpha emitters \citep{Dijkstra2016}, measuring Gunn-Peterson absorption at high-redshift quasar spectra caused by quasar damping wings \citep{Fan2006}, and probing the 21\,cm spectral line of neutral hydrogen \citep{Furlanetto2004}. Results from recent studies lead us to believe the bounds of the \ac{EoR} are somewhere between a redshift of $z\sim\{5.4-10\}$ \citep{Kulkarni2019, Nasir2020, PlanckCollaborationXIII2016}.

However, detecting the 21\,cm line does not come without challenges. One of the biggest challenges to detect this signal is that it is hidden within the foregrounds. Foregrounds consist of relatively compact emission from extra-galactic sources (\ac{AGN} and star-forming galaxies) and diffuse polarised emission due to galactic synchrotron radiation, which have angular scales from 10's of arcminutes to the entire sky. These foregrounds are generally three to five orders of magnitude brighter than the signal we desire to detect~\citep{Morales2010, Vedantham2012, Mertens2018, Trott2020, Ghosh2020}. In order to solve for this, foreground mitigation methods have to be employed. Ideally, one would characterise these foregrounds on a wide range of angular scales and frequencies, through the generation of high resolution all-sky maps~\citep{Eastwood2017} and create a model for foreground subtraction.

Generating a comprehensive \ac{EoR} foreground model requires a combination of both compact components and a diffuse component. Compact source information can be obtained from one of the many interferometric surveys that cover large areas of the sky, \textit{e.g.} GLEAM~\citep{Wayth2015}, TGSS~\citep{Intema2017}, and LoTSS~\citep{Williams2019}, and the source catalogues that result from them. However, compact source catalogues do not describe diffuse emission and the galactic plane is often excluded due to the complexity of imaging those regions with interferometers.

Mapping the diffuse emission, simultaneously in the galactic plane and outside it, is a challenging process. The well-known Haslam sky map~\citep{Haslam1981,Haslam1982} has been the most prominent in use over the past few decades as a low frequency sky model. However, the need for more diffuse sky maps over a wider frequency range sparked the increase in these maps through a multitude of sky surveys. The most prominent diffuse sky maps generated since the Haslam map are the GSM~\citep{OLIVEIRA-COSTA2008}, S-PASS 2.3 GHz Polarisation Survey~\citep{Carretti2019}, CHIPASS~\citep{calabretta2014},  LWA-LFSS~\citep{Dowell2017}, recalibrated versions of the 150\,MHz diffuse sky map by~\citet{Landecker1970}~\citep{Patra2015, Monsalve2021}, and the 45\,MHz diffuse map by \citet{Guzman2010}. Although a significant improvement, more maps have to be generated at more areas on the sky and at more frequencies to provide an accurate diffuse foreground model; especially the lower frequency regime and the southern hemisphere. 

\citet{Eastwood2017} started to address these issues by generating low-frequency high resolution (around $\sim15$\,arcminutes) northern sky maps using the \ac{OVRO-LWA}~\citep{Kassim2005} interferometer array using a whole different imaging method altogether. \citet{Eastwood2017} employed a method known as the Tikhonov-regularised $m$-mode formalism; an adaption to the spherical harmonic transit interferometric imaging method suggested by \citet{Shaw2014,Shaw2015}. Opposite to traditional radio interferometry, the $m$-mode formalism no longer utilises snapshot visibilities to image the sky, but uses components of timescale variations within these visibilities instead. The formalism uses these components to rebuild the sky using spherical harmonic basis functions. Since these basis functions operate over the full celestial sphere, tracking the time variance components across a full sidereal day allows one to reconstruct the full sky in a single imaging step whilst maintaining exact widefield accuracy.

In this paper, we aim to complement the existing diffuse low-frequency sky maps by generating a low-frequency southern sky map at 159\,MHz using the \acf{EDA2}~\citep{Wayth2021}, which is a prototype station of the future \ac{SKA}-Low~\citep{Braun2015}.
For the generation of this sky map we also employ the $m$-mode formalism. This allows us to take full advantage of the \ac{EDA2}'s wide \ac{FoV}, allowing us to hyper-resolve the spatial scales on the sky with full widefield accuracy. We also introduce the concept of spherical-harmonic beam coverage, an analogy to measure for completeness on the sky similar to the $u,v$-coverage in traditional interferometry. Additionally, an image-based spherical harmonic CLEANing algorithm is presented, significantly reducing the number of unique point-sources that need to be generated, whilst maintaining the ability to accurately deconvolve \acp{PSF}.

\section{All-Sky Interferometry}
\label{sec:interferometry}
In radio astronomy--and interferometry imaging algorithms in general--the goal is to determine the sky brightness temperature ${\rm T}_{{\rm b},\nu}\left(\hat{\mathbf n}\right)$ or sky intensity ${\rm I}_{\nu}\left(\hat{\mathbf n}\right)$ for a specific pointing $\hat{\mathbf n}$ on the celestial sphere, with quasi-monochromatic frequency~$\nu$. However, an interferometer cannot measure this directly; instead it sees the correlated voltage response between two antenna elements ${\big\langle{\textsf{U}}_{\nu}^{\rm i}(t){\textsf{U}}_{\nu}^{\rm j*}(t)\big\rangle}$. Ignoring the polarisation responses, we can define the relation between the measured response and the brightness temperature of the sky as \citep{Thompson2017, Shaw2014}

\begin{align}
    \label{eq:measurementeq}
    V_{\nu}^{\rm ij} &= \int B_{\nu}^{\rm ij}\left(\hat{\mathbf n}\right){\rm T}_{\nu}\left(\hat{\mathbf n}\right)d^2\hat{\mathbf n}\,,\,\text{and}\\[10pt]
    \label{eq:beam_func}
    B_{\nu}^{\rm ij}\left(\hat{\mathbf n}\right) &= \frac{1}{\sqrt{\Omega_{\rm i}\Omega_{\rm j}}}A_{\nu}^{\rm i}\left(\hat{\mathbf n}\right)A_{\nu}^{\rm j*}\left(\hat{\mathbf n}\right)e^{2\pi i\hat{\mathbf n}\cdot{\mathbf u_{\rm ij}}}\,.
\end{align}

\noindent In \autoref{eq:measurementeq} $B_{\nu}^{\rm ij}$ is the transfer function that encapsulates the system response on the sky, $A_{\nu}^{\rm i}$ is the primary beam voltage response of antenna element $\rm i$, ${\mathbf u}_{\rm ij}$ is the separation between two antenna elements--in the $u,v$-plane--normalised by the wavelength $\lambda$, and $\Omega_{\rm i}$ is the beam solid angle of antenna element $\rm i$.

Although \autoref{eq:measurementeq} is a perfectly valid description of the measured voltage response over the full celestial sphere, \citet{Eastwood2017} has shown that directly solving for the three-dimensional equation, due to computational cost and complexity, is not a tangible solution. Alternatively, one could assume a flat sky reducing the measurement equation to a two-dimensional Fourier transform. However, this concept starts breaking down for wide \acp{FoV} as the curvature of the sky starts playing a role, which the 2D transform does not account for \citep{Carozzi2015,Presley2015,Singh2015,Thyagarajan2015a,Thyagarajan2015b,Thompson2017}. Additionally, the flat-sky approach restricts the capability to image the full-sky in a single imaging sweep, as one cannot distinguish between points on different hemispheres. As a result, one has to resort to making individual snapshots of the sky and \textit{e.g.} mosaic them together; restricting one to information only available in each individual snapshot. 

\subsection{Spherical Harmonic Transit Interferometry}
To address the issues and complexities of all-sky interferometry, \citet{Shaw2014} proposed spherical harmonic transit interferometry as an alternate solution. Instead of phase-tracking sources with a narrow \ac{FoV} and mosaicing/stacking the resulting images together to create an image of the full sky, the issues of solving \autoref{eq:measurementeq} are avoided all-together. 

With transit interferometry, the element's wide \acp{FoV} is pointed at zenith to observe the sky in transit over a sidereal day. This allows one to then utilise spherical harmonics to describe the interaction of radio waves emitted by celestial bodies on the observed celestial sphere.

\subsection{Spherical Harmonics}
\begin{figure*}
    \centering
    \begin{minipage}{.2\columnwidth}\hfill\end{minipage}%
    \begin{minipage}{1.6\columnwidth}
    \centering
        \includegraphics[width=0.9\linewidth]{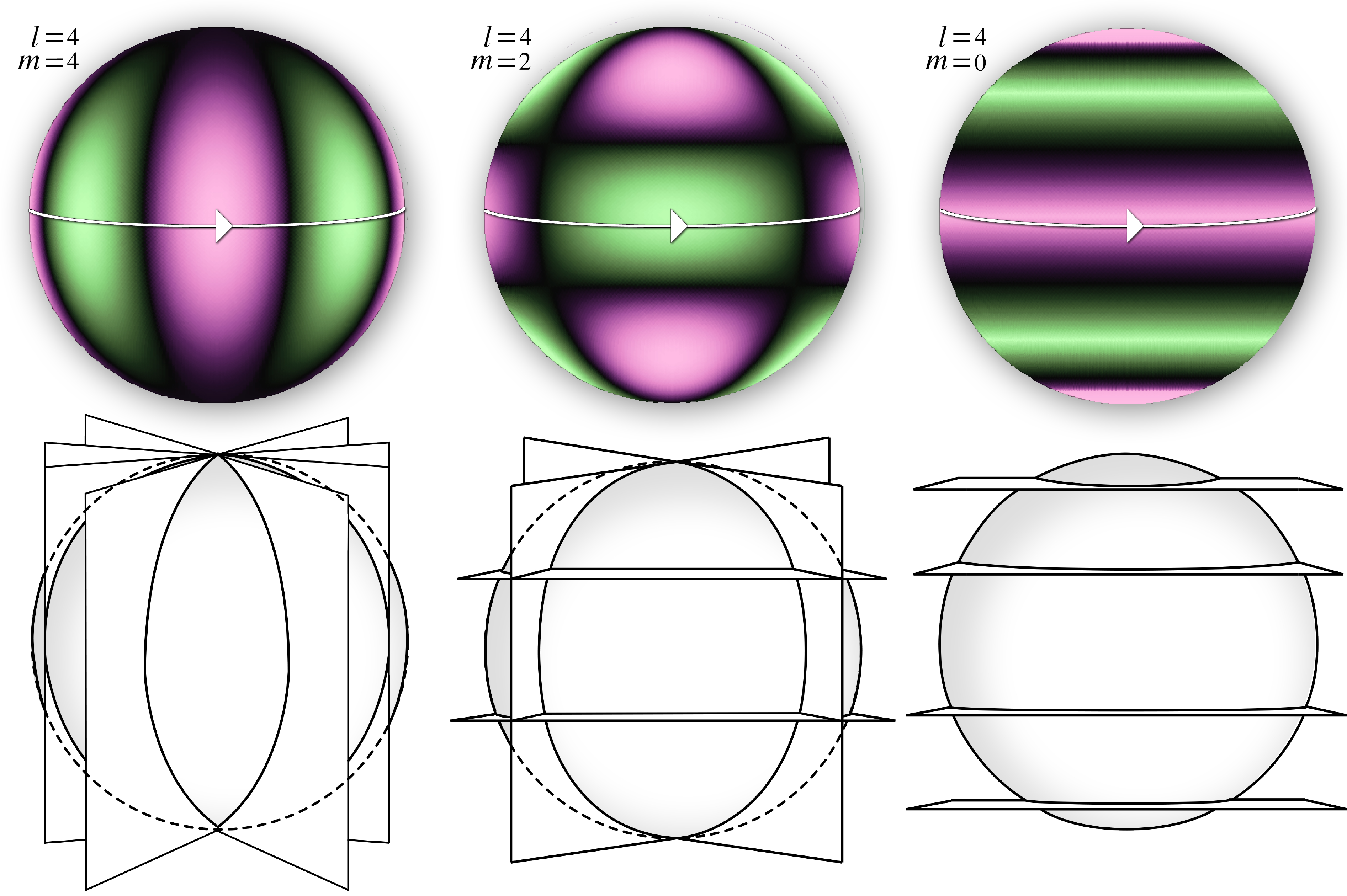}
        \caption{Types of spherical harmonics. Left: sectoral spherical harmonic (a function of $e^{im\varphi}$) with $m=4$, Middle: tesseral spherical harmonic (a function of $P_{l}^{|m|}\left(\cos\theta\right)e^{im\varphi}$) with $l=4$ and $m=2$, Right: zonal spherical harmonic (a function of $P_{l}^{|m|}\left(\cos\theta\right)$) with $l=4$ and $m=0$. Angular velocity of the basis functions, with respect to \acf{RA}, are a function of $e^{im\phi}$.}
        \label{fig:spherical_harmonics}
    \end{minipage}%
    \begin{minipage}{.2\columnwidth}\hfill\end{minipage}%
\end{figure*}

Much similar to how the Fourier series describes how periodic functions interact on a circle, spherical harmonics describe the rate of change (angular frequency) of functions on a sphere. Especially the Laplacian spherical harmonics--derived by expanding the Laplacian in three dimensions--are of interest, as they form an orthonormal basis. In other words, any function that acts on a spherical surface can be expanded into a sum of these Laplacian spherical harmonics; akin to how varying functions restricted to a circle can be expanded into a series of circular functions--\textit{i.e.} sines and cosines--using a Fourier transform. An example of the Laplacian spherical harmonics can be seen in \autoref{fig:spherical_harmonics}.

Spherical Harmonics can be divided into three different categories with regards to how wave forms interact on the sphere. The zonal spherical harmonics (\autoref{fig:spherical_harmonics}, right image) only have variation in the latitudinal direction ($\theta$), the sectoral spherical harmonics (\autoref{fig:spherical_harmonics}, left image) only have variation in the longitudinal direction ($\varphi$), and the tesseral spherical harmonics (\autoref{fig:spherical_harmonics}, middle image) have variation in both the latitudinal and longitudinal direction. This is convenient, as this allows us to express any continuous waveform into subsets of basis functions that describes the waveform's angular dependencies on both coordinates axes~$(\varphi,\theta)$

\begin{align}
    \label{eq:spharm}
    Y_{l}^{m}\left(\varphi,\theta\right) &= \sqrt{\frac{2l+1}{4\pi}\frac{\left(l-|m|\right)!}{\left(l+|m|\right)!}}P_{l}^{|m|}\left(\cos\theta\right)e^{im\varphi}\,.
\end{align}

\noindent Here, $Y_{l}^{m}\left(\varphi,\theta\right)$ are the complex Laplace spherical harmonics, $P_{l}^{|m|}\left(\cos\theta\right)$ are the associated Legendre polynomials\footnote{It should be noted that throughout this paper the Condon-Shortley phase $\left(-1\right)^{|m|}$ is included in the Legendre polynomial notation and is not to be confused with the quantum physical notation $P_{l|m|}\left(\cos\theta\right)$; for which the Condon-Shortley phase still has to be included.}, and $e^{im\varphi}$ is Euler's formula with dependence on $m$. The components $l$ and $m$ describe the degree and rank of the function respectively, where $l\,\in\,\mathbb{Z}^{0+}$ and $-l\,\leq\,m\,\leq\,l$. Simply said, the degree $l$ describes the number of cross-sections on the sphere, of which $|m|$ are longitudinal functions and $l-|m|$ are latitudinal functions.

\subsection{The \texorpdfstring{$m$}{\textit{m}}-mode formalism}
\label{subsec:m-modes}
In order to leverage spherical harmonics to describe functions across the sky, we align the spherical coordinate system with the celestial sphere. Doing so, $\varphi$ describes the celestial plane's \acf{RA} and $\theta$ describes the \ac{DEC}. Given that the Earth's rotation is periodic over a sidereal day (LST) and is explicitly dependent on the azimuthal axis ($\phi$), \citet{Shaw2014} concluded that one can take advantage of this Earth rotational symmetry in plane with the sectoral spherical harmonics; using wide \ac{FoV} interferometers through transit interferometry.

The topic of spherical harmonic transit interferometry has been extensively covered by \citet{Shaw2014,Shaw2015} and \citet{Eastwood2017}. As of such, we will only briefly review the key points concerning the $m$-mode formalism and solving for the sky. However, we highly recommend the interested reader to refer to the aforementioned papers for a more complete overview.

Using zenith pointed observations and measure over a full sidereal day would therefore make the three-dimensional measurement equation a function of $\phi$ \citep{Shaw2014}:

\begin{align}
    \label{eq:measurementeq_tf_rot}
    V_{\nu}^{\rm ij}(\phi) &= \int B_{\nu}^{\rm ij}\left(\hat{\mathbf n}; \phi\right){\rm T}_{\nu}\left(\hat{\mathbf n}\right)d^2\hat{\mathbf n}\,.
\end{align}

\noindent Expanding the beam-transfer function (beam $\times$ fringe) into its spherical harmonic coefficients provides a description of the angular and spatial coverage of the interferometer system for a specific baseline and frequency. The beam-transfer function can be expressed as

\begin{align}
\label{eq:beam_decomp}
    B_{\nu}^{\rm ij}\left(\hat{\mathbf n}; \phi\right) &= \sum\limits_{l=0}^{l_{\rm max}}\sum\limits_{m=-l_{\rm max}}^{l_{\rm max}} b_{lm, \nu}^{\rm ij}(\phi)Y_{l}^{m*}(\hat{\mathbf n})\,,
\end{align}

\noindent where $b_{lm, \nu}^{\rm ij}(\phi)$ are the spherical harmonic coefficients of the decomposed beam-transfer function. Similarly, we can decompose the sky brightness temperature into a set of spherical harmonic coefficients describing the full spatial coverage across the celestial sphere

\begin{align}
\label{eq:sky_decomp}
    {\rm T}_{\nu}\left(\hat{\mathbf n}\right) &= \sum\limits_{l=0}^{l_{\rm max}}\sum\limits_{m=-l_{\rm max}}^{l_{\rm max}} a_{lm, \nu}Y_{l}^{m}(\hat{\mathbf n})\,,
\end{align}

\noindent where $a_{lm, \nu}$ are the spherical harmonic coefficients of the decomposed sky. It should be noted that in the $m$-mode definitions by \citet{Shaw2014} the spherical harmonic basis functions in the beam transfer function and sky decomposition are conjugates of each other to ``simplify later notation''. As a result, due to the fact that spherical harmonics are orthonormal functions, the basis functions drop out of the $m$-mode equation given that

\begin{align}
    \label{eq:sh-orthonormality}
    \int Y_{l}^{m}(\hat{\mathbf n})Y_{l}^{m*}(\hat{\mathbf n})d^2\hat{\mathbf n} &= \delta_{ll'}\delta_{mm'},
\end{align}

\noindent with $\delta_{ll'}$ and $\delta_{mm'}$ being Kronecker delta functions. 

Closely observing \autoref{eq:spharm}, it is clear that $m$ only affects the $\varphi$ coordinate via the phase term $e^{im\varphi}$, which is a rotation in the plane of \ac{RA}. Any \ac{RA} displacement in spherical harmonic space is therefore only a function of $e^{im\varphi}$. As a result, the rotation ($\phi$) of the sky for a transit telescope only affects components of the spherical harmonics that change with $m$. \citet{Shaw2014} concluded that we can thus track components that vary across different timescales over a sidereal day on a per-$m$-basis. In order to relate what we measure on the sky to the spherical harmonics we can Fourier transform our visibilities with respect to $m$, such that

\begin{align}
\label{eq:m-mode_transform}
    v_{m,\nu}^{{\rm ij}} &= \frac{1}{2\pi}\int V_{\nu}^{\rm ij}\left(\phi\right)e^{-im\phi}d\phi\,.
\end{align}

\noindent This provides us with a formalism where we no longer use information of individual snapshots across the sky, but rather the Fourier components that describe the varying timescales of what of we observe across the sky over a full sidereal day. These components are also known as $m$-modes. The complete spherical harmonic relation can therefore be defined as

\begin{align}
\label{eq:m-modes}
    v_{m,\nu}^{{\rm ij}} &= \sum\limits_{l=0}^{l_{\rm max}} b_{lm,\nu}^{\rm ij}a_{lm,\nu}.
\end{align}

\noindent which describes the encoded observed spatial frequency of the whole sky observed through the physical system on an $m$-by-$m$ basis, also known as the $m$-mode formalism. The sky coefficients $a_{lm,\nu}$ are then used to reconstruct the sky, acting as weightings for the spherical harmonic basis functions. This allows for us to image the full sky in a single imaging step, maintaining exact widefield accuracy without the introduction of regridding artifacts, since we no longer use individual snapshots. \autoref{eq:m-modes} also reduces the measurement equation into a simple linear relation, describing how the sky maps to data obtained through the physical system \citep{Shaw2014}

\begin{align}
    \label{eq:m-mode-mat}
    \mathbf{v} &= \mathbf{B}\mathbf{a}\,,
\end{align}

\noindent where $\mathbf{v}$ is the column-vector describing the $m$-mode response for each baseline, $\mathbf{B}$ is a block-diagonal matrix describing the physical system, and $\mathbf{a}$ is the column-vector describing the spherical harmonic sky coefficients.

\citet{Eastwood2017} described the block diagonal structure of $\mathbf{B}$. Here we note some properties of $\mathbf{B}$ that were not explicitly made in \citep{Shaw2014,Eastwood2017}.
As per \citep{Shaw2014}, the $m$-mode equation described by \autoref{eq:m-mode-mat} is sorted per $|m|$. Following \citep{Shaw2014,Eastwood2017} we can therefore determine the shapes of the vectors and matrix as

\begin{itemize}
    \item[]{shape $\mathbf{v}$: $\left[|m|\times1\right]\rightarrow\left[\left(m\times2N\right)\times1\right]$}
    \item[]{shape $\mathbf{B}$: $\left[|m|\times|m|\right]\rightarrow\left[\left(m\times2N\right)\times\left(m\times l\right)\right]$}
    \item[]{shape $\mathbf{a}$: $\left[\left(m\times l\right)\times 1\right]$}
\end{itemize}

\noindent where N is the total number of baselines.
For every value for $|m|$ the $m$-modes $\mathbf{v_{|m|}}$ have $2N$ components,
with $N$ coming from $+m$, and $N$ coming from $-m$.
Similarly, for every value of $|m|$ in a block on the block-diagonal, there are $2N\times l$ spherical harmonic coefficients of the beam transfer function; an example of one such block is shown below

\begin{align}
    \label{eq:b_block}
    \mathbf{B}_{|m|} &= \begin{bmatrix}{\rm b}_{l=0,+m}^{0,1} & \dots & {\rm b}_{l=l_{\rm max}, +m}^{0,1}\\
    \vdots & \vdots & \vdots\\
    {\rm b}_{l=0,+m}^{N-1,N} & \dots & {\rm b}_{l=l_{\rm max}, +m}^{N-1,N}\\
    \color[rgb]{0.6,0.6,0.6}\text{-- -- -- -- --}&\color[rgb]{0.6,0.6,0.6}\text{-- -- -- -- --}&\color[rgb]{0.6,0.6,0.6}\text{-- -- -- -- --}\\
    {\rm b}_{l=0,-m}^{0,1*} & \dots & {\rm b}_{l=l_{\rm max}, -m}^{0,1*}\\
    \vdots & \vdots & \vdots\\
    {\rm b}_{l=0,-m}^{N-1,N*} & \dots & {\rm b}_{l=l_{\rm max}, -m}^{N-1,N*}
    \end{bmatrix}\,.
\end{align}

\noindent In \autoref{eq:b_block} ${\rm b}_{l,\pm m}^{i,j}$ describes beam transfer coefficients for spherical harmonic order $l$ increasing on a per-column-basis up to $l_{\rm max}$, $(i,j)$ describes the baseline indices increasing on a per-row-basis until the maximum baseline configuration $(N-1,N)$ has been reached. The matrix is split in half vertically with the upper half populating $+m$ and the bottom half populating $-m$, where in the $-m$ section the coefficients have been conjugated to conform with definitions defined by \citet{Shaw2014}. Lastly, the sky coefficients $\mathbf{a}$ (\autoref{eq:m-mode-mat}) are only a function of $+m$, as due to the real-sky relation it satisfies that $a_{l,m}=(-1)^{m}a_{l,-m}^{*}$.

Ideally, one would solve for $\mathbf{a}$ through inverting \autoref{eq:m-mode-mat} by estimating the sky brightness temperatures using the visibilities, minimising $||\mathbf{v}-\mathbf{B}\mathbf{a}||^{2}$ \citep{Shaw2014, Eastwood2017}

\begin{align}
    \label{eq:sky-inverse}
    \hat{\mathbf{a}} &= \left(\mathbf{B}^{\dagger}\mathbf{B}\right)^{-1}\mathbf{B}^{\dagger}\mathbf{v}\,,
\end{align}

\noindent where $\hat{\mathbf{a}}$ is the estimate of $\mathbf{a}$; also known as the linear least-squares solution, where $\dagger$ denotes the conjugate transpose. 

\subsection{Spatial Coverage}
\citet{Eastwood2017} has shown that the maximum number of spherical harmonic orders $l_{\rm max}$ that an interferometer is sensitive to is proportional to the  array's diameter. The maximum spherical harmonic order is given by

\begin{align}
    \label{eq:lmax}
    l_{\rm max} &= \frac{2\pi{\rm r}_{{\rm ij},{\rm max}}}{\lambda}\,,
\end{align}

\noindent where ${\rm r}_{{\rm ij},{\rm max}}$ is the maximum baseline separation.
We note that this is different to the original definition of \citet{Eastwood2017} by  a factor of 2, as omitting this term causes the equation to no longer satisfy the Nyquist sampling rate required to represent the smallest variations across the sky measured by the longest baselines.

In standard radio interferometry the $u,v$-plane coverage defines the spatial frequencies that have been measured by the array. Since in spherical harmonic transit interferometry the $u,v$-plane is omitted altogether, and a completely different coordinate system is used, one cannot sample the $u,v$-plane to test for completeness. Consequently, an alternative formalism has to be employed to quantitatively describe the interferometer's completeness of measurements in spherical harmonic $l$ and $m$ space. 

As explained in \autoref{subsec:m-modes}, the spherical harmonic coefficients of the beam-transfer functions can be used to visualise the spatial coverage information of the sky on a per-baseline basis (examples using \ac{EDA2} beam models and baselines are shown in \hyperref[apdx:beam-coverage]{Appendix A}).
Similar to $u,v$-coverage, we combine the beam-transfer function coefficients $b_{lm}$ from all baselines together:

\begin{align}
    \Beta_{lm} &= \sum\limits_{n=1}^{N}|b_{lm}^{n}|,
\end{align}

\noindent where $\Beta_{lm}$ is the total \ac{SHBC} contribution in \ac{SH-space} from all baselines for each $l,m$.
This can be translated to percentage relative mode sensitivity in \ac{SH-space} as:

\begin{align}
    \Beta_{lm,\%} &= 100\%\times\frac{\Beta_{lm}}{\mathrm{max}(\Beta_{lm})}.
\end{align}

\noindent An example of how this SH-space looks like is shown in \autoref{fig:SH-space}. 

Since these coefficients are weighting functions of the absolute spherical harmonic basis functions that operate across the full-sky, an  equal (uniform) contribution of each \ac{SHBC} mode in the \ac{SH-space} coefficient plane would therefore result in a system that is `equisensitive' to all spatial frequencies and phase components across the whole sky; i.e. the interferometer would measure the true sky. 
Such a uniform coverage would be equivalent to a complete $u,v$-coverage for conventional interferometry.
The \ac{SHBC} can therefore be considered a powerful tool to visualise the interferometer's coverage spherical harmonic coefficient space.

\subsection{Tikhonov-regularised \texorpdfstring{$m$}{\textit{m}}-modes}
In the general case telescopes cannot see some parts of the sky. As a result the square matrix $\mathbf{B}^\dagger\mathbf{B}$ is not full-rank and \autoref{eq:sky-inverse} cannot be solved as $\mathbf{B}^\dagger\mathbf{B}$ cannot be inverted.

One way resolve this is to use either the Moore-Penrose pseudo-inverse or \ac{LLS} instead \citep{Shaw2014,Eastwood2017}. However, if the measurement data ($\mathbf{v}$) is noise dominated or has too many unknowns, due to missing information on the sky, to fit the data properly, these methods might put emphasis on modes that vary on shorter timescales across the sky; suppressing the modes that vary on longer timescales instead \citep{Eastwood2017}.  This is because the conditions on which the pseudo-inverse and \ac{LLS} satisfies the minimisation of the error and the estimated solution is only based on known information.

Conversely, \citet{Eastwood2017} proposed one could bias the known data by slightly increasing the initial error, but ultimately reducing the variance on what is being fit. This is also known as Tikhonov regularisation

\begin{align}
    \label{eq:tikhonov-inverse}
    \hat{\mathbf{a}} &= \left(\mathbf{B}^{\dagger}\mathbf{B}+\varepsilon\mathbf{I}\right)^{-1}\mathbf{B}^{\dagger}\mathbf{v}\,.
\end{align}

\noindent where $\varepsilon\ge0$ is the ridge-parameter and $\mathbf{I}$ is the identity matrix; forcing $\varepsilon$ to $0$ would again yield \autoref{eq:sky-inverse}. \autoref{eq:tikhonov-inverse} is also known as Tikhonov-regularised $m$-mode analysis \citep{Eastwood2017}.

\subsection{Optimal ridge-parameter (\texorpdfstring{$\varepsilon$}{ε}) selection}
\begin{figure}
    \centering
    \includegraphics[width=\linewidth]{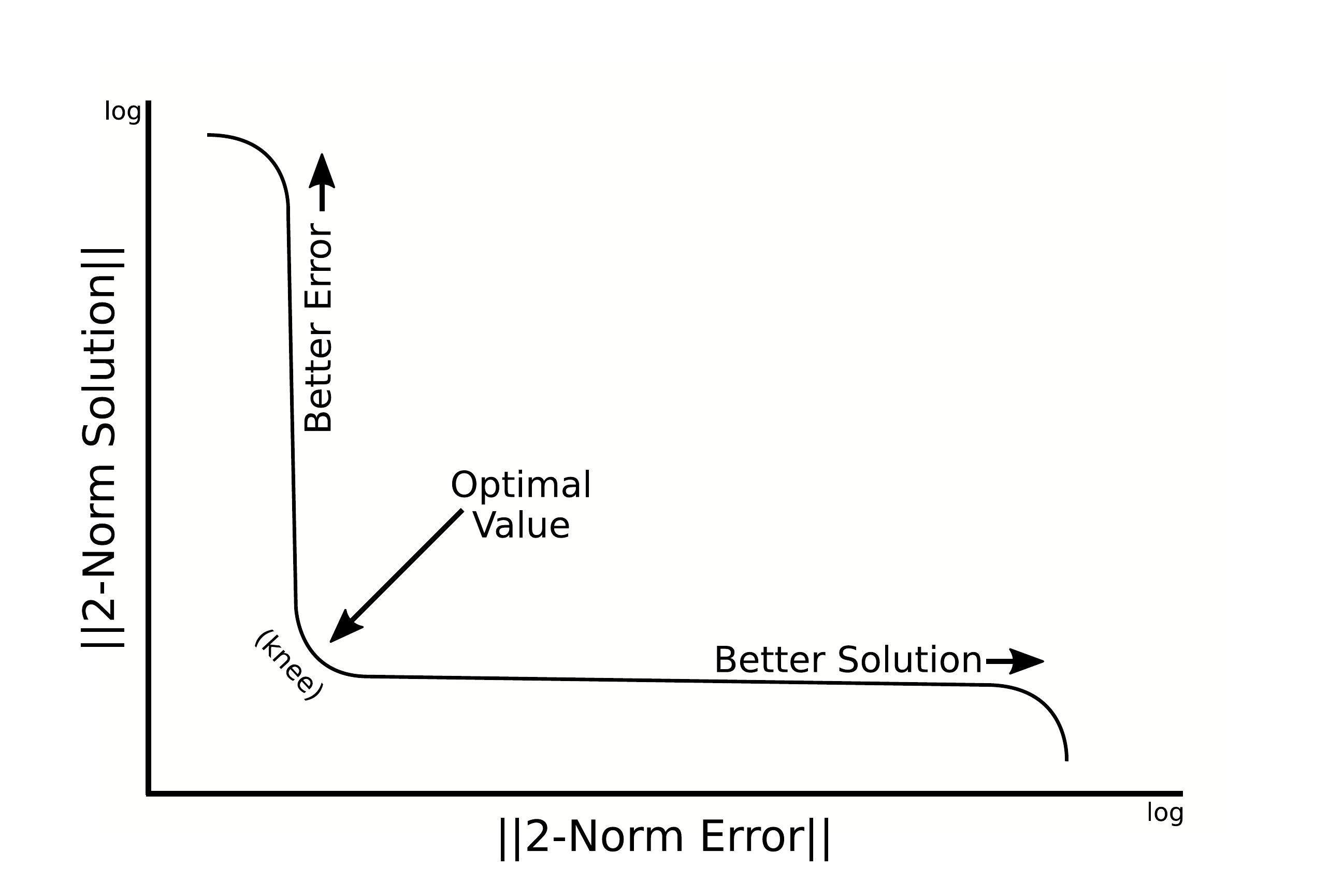}
    \caption{Example of an L-curve measurement plot in log-log space, the ``knee'' indicates the optimal ridge regression value.}
    \label{fig:L-curve}
\end{figure}

Tikhonov regularisation regularises the data by forcing the matrix to be full rank by positively biasing $\mathbf{B}$ on the diagonals enforcing preference for a solution where both $||\mathbf{v}-\mathbf{B}\mathbf{a}||^{2}$ and $||\hat{\mathbf{a}}||^{2}$ are jointly at their lowest possible norm \citep{Eastwood2017}. In order to find a solution where both the 2-norm of the solution and the error are both at their minimum, an optimal value for $\varepsilon$ has to be selected. To achieve this, \citet{Eastwood2017} suggested the use of L-curves as a graphical tool for determining the optimal solution for $\varepsilon$. In this case a graph plotting the norm of the solution against the norm of the residual error provides a characteristic L-shape with the minimum of both norms at the `elbow' or `knee' (\autoref{fig:L-curve}). Contrary to \citep{Eastwood2017}, however, we decided to follow the definition by \citep{Hansen2001} instead. Here the L-curve is defined as a log-log graph with the axes inverted relative to \citep{Eastwood2017}. This allows for steeper transitions outside the optimum region and a better distinction of the `knee'. 

\subsection{Prior knowledge to constrain the ridge parameter}
\label{subsubsec:priorfit}
In case one has prior knowledge of the sky, \citet{Eastwood2017} has shown that coefficients of a prior map can be used to better minimise for the estimated sky coefficients $\hat{\mathbf{a}}$, such that $||\hat{\mathbf
a}-\mathbf{a}_{\rm prior}||$ is used instead and the solution is estimated by

\begin{align}
    \label{eq:tik_inv_prior}
    \hat{\mathbf{a}} &= \left(\mathbf{B}^{\dagger}\mathbf{B}+\varepsilon\mathbf{I}\right)^{-1}\mathbf{B}^{\dagger}\left(\mathbf{v}-\mathbf{B}\mathbf{a}_{\rm prior}\right) + \mathbf{a}_{\rm prior}\,.
\end{align}

\noindent However, this assumption is only valid if the prior coefficients have similar magnitudes to the coefficients of the measured sky. In our experience having a too large difference between the measured and model monopole component $a_{00}$ resulted in instability of the system and in improper constraints on the $a_{00}$ mode. Because of this we've opted to constrain our measured modes with a prior where ${\rm a}_{00}$ is subtracted. The global component can then be re-inserted after the regression step, such that

\begin{align}
    \label{eq:tik_inv_prior_noglobal}
    \hat{\mathbf{a}} &= \left(\mathbf{B}^{\dagger}\mathbf{B}+\varepsilon\mathbf{I}\right)^{-1}\mathbf{B}^{\dagger}\left(\mathbf{v}-\mathbf{B}\left[\mathbf{a}_{\rm prior}-a_{00}\right]\right) + \mathbf{a}_{\rm prior}\,.
\end{align}

\noindent To determine the system's insensitivity  to specific modes in \ac{SH-space}, the \ac{SHBC} should be inspected. We note that this issue is a weakness of this form of regularisation as it disproportionately affects the cost function underlying the regularisation process for coefficients with large relative magnitude, \textit{e.g.} the monopole and dipole components for a low-frequency all-sky map.

\section{Methods \& Data}
\label{sec:observations}
In order to observe the Southern sky at low frequencies with the $m$-mode formalism, the \ac{EDA2} has been used.

\subsection{The Engineering Development Array 2}
The \ac{EDA2}, located at the \ac{MRO} in Western Australia and a successor to the \ac{EDA}~\citep{Wayth2017}, is a 256-element dipole interferometer array~\citep{Wayth2021}. The antenna elements are identical to the bow-tie dipole elements used in the \ac{MWA}~\citep{Tingay2013}, but distributed in a pseudo-random configuration spanning 35 meters in diameter; much similar to what a station of the future \ac{SKA} will look like. Each antenna is a dual-polarised antenna, allowing the \ac{EDA2} to measure a total of 512 signal paths. The \ac{EDA2} has a frequency range in accordance with the \ac{SKA}-low specification of 50\,MHz--350\,MHz. Different to both the \ac{MWA} and \ac{EDA}, the \ac{EDA2} is not beamformed in analog domain, but instead digitised on a per-antenna basis. This allows the \ac{EDA2} to be run as a station beamformer, or as a small 256-element interferometer, which is the mode used for this paper.

\begin{figure}
    \centering
    \includegraphics[width=\linewidth]{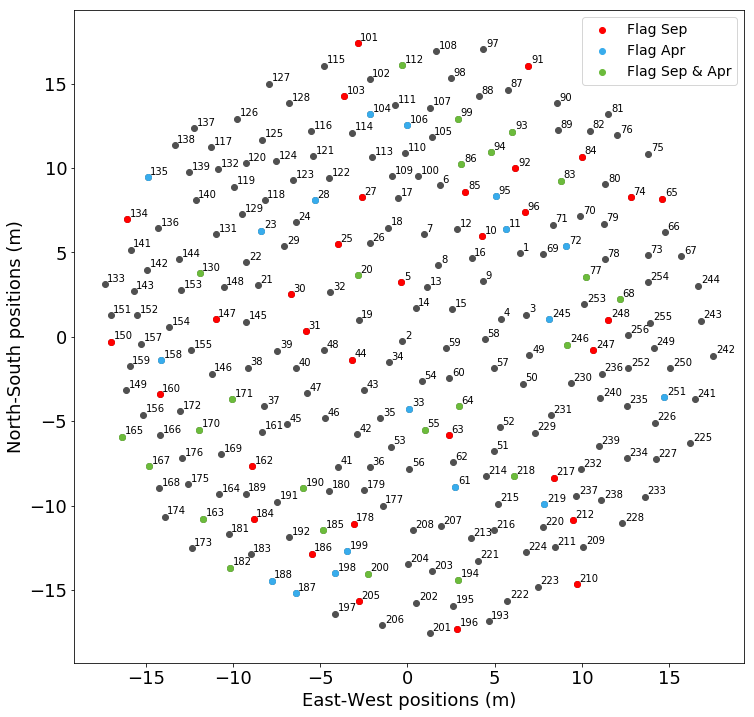}
    \caption{EDA2 array 256 element layout in local (North-South, East-West) coordinates.}
    \label{fig:EDA-layout}
\end{figure}

\begin{figure*}
    \centering
    \begin{minipage}{0.95\columnwidth}
    \centering
        \includegraphics[width=0.9\linewidth]{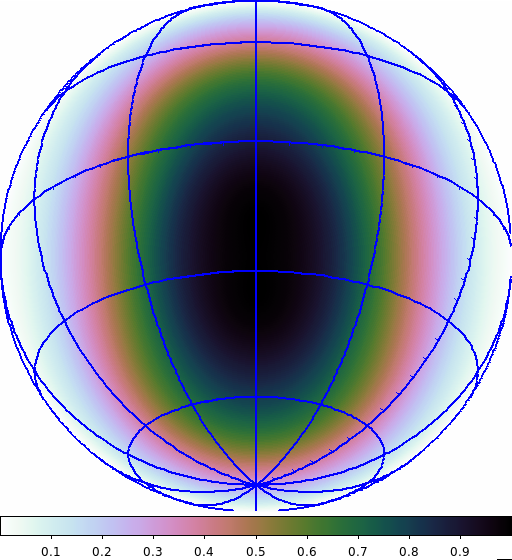}
        \caption{Normalised x-polarisation FEKO-simulated single-element beam pattern of the \acs{EDA2}; orthographic projection on a hemisphere.}
        \label{fig:xpol_beam}
    \end{minipage}%
    \begin{minipage}{0.1\columnwidth}\hfill\end{minipage}%
    \begin{minipage}{0.95\columnwidth}
    \centering
        \includegraphics[width=0.9\linewidth]{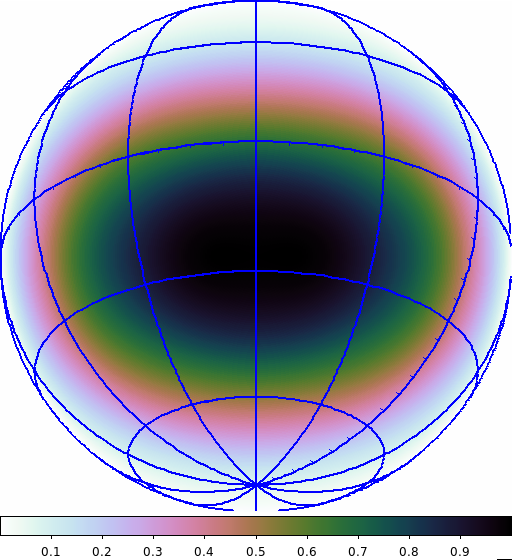}
        \caption{Normalised y-polarisation FEKO-simulated single-element beam pattern of the \acs{EDA2}; orthographic projection on a hemisphere.}
        \label{fig:ypol_beam}
    \end{minipage}%
\end{figure*}

\subsection{Data \& Calibration}
For this paper, we use two data sets spanning each between 25-28 hours continuous zenith-pointed drift-scan observations. These two observations were separated by 7 months to have enough spatial separation of the sun between both observations for easier removal. The first observation was performed from 2019/09/16-05:35:19 until 2019/09/17-09:37:10 in \ac{UTC}, the second observation was performed from 2020/04/10-11:30:10 until 2020/04/11-12:17:36 in \ac{UTC}. 

The integration time for these observations was 1.98181 seconds. The observations were then averaged four times into a time-resolution of 7.92724 seconds and stored in sets of 12, containing 95.1269 seconds per file. To fulfil the sidereal day periodicity discussed in \hyperref[subsec:m-modes]{Section 2.1.2}, 906 consecutive data files were selected to encompass a full sidereal day per observation. For this paper a single 0.926\,MHz band at 159.375\,MHz (center-band) was selected to generate a Southern sky image, since this band has very little interference. 

The EDA2 is phase stable for days, so a single calibration is applied for each dataset. We use the Sun as a strong compact radio source with known flux density. The quiet sun has well measured radio flux density over a large frequency range~\citep{Benz2009}, and we use this to define the flux density of the quiet sun to be 56,195\,Jy for these observations at 159.375\,MHz. The data were taken during solar minimum, hence the quiet sun model is appropriate. The calibration method used in this work has already been demonstrated by \citep{Sokolowski2021, Wayth2021} for science and system verification purposes.

Phase and amplitude calibration was performed during a 10-minute interval centred on solar transit for each dataset, using an arbitrarily chosen antenna in the array as a reference, and discarding baselines shorter than $5\lambda$ to avoid bias from Galactic emission. The X and Y polarisations were independently calibrated. The solutions were transferred to the entire dataset. 

The flux scale was corrected for the Sun's location in the FEKO\footnote{\hyperlink{https://www.altair.com/feko}{www.altair.com/feko}}-generated dipole radiation power pattern~\citep{Ung2019}\footnote{Beam patterns have been derived from accurate method-of-moments modelling of the antennas over an infinite ground plane.}, which is different for X (\autoref{fig:xpol_beam}) and Y (\autoref{fig:ypol_beam}) and is different for the two datasets because the Sun was at different declinations. The apparent flux density of the sun is reduced away from the peak directivity of the antenna (zenith in this case), hence we scale the data by factors of 0.856 and 0.624 for the X and Y dipoles respectively in September, and 0.780 and 0.481 for the X and Y dipoles respectively in April. Although the $m$-mode method can in principle embed different beam patterns for the elements in the array, the \ac{EDA2} patterns are very similar. \citet{Jones2021} have shown that a single element beam model for the whole array is sufficient for better than $1\%$ accuracy. As such, all elements are assumed to have the same beam patterns for their corresponding polarisations and have been normalised to be unity at their maxima.

During commissioning work for EDA2, an additional temperature-dependent gain amplitude variation was found as described in~\citep{Wayth2021}. This effect modulates the gain amplitude by approximately +/- 10\% over a solar day. We use an identical procedure to model and correct for this variation as described in~\citep{Wayth2021}.

To ensure we do not include bad data in our imaging phase, we flag timestamps where anomalies occur. In order to ascertain we do not void the periodicity assumption in \autoref{subsec:m-modes}, we flag the full 24-hour observation for a baseline if flagging results in gaps in the observations spanning more than our angular resolution on the sky (in our case 12 minutes). If antennas consistently misbehave we remove all observations using this antenna from our imaging process. During the observations, 42 and 56 antennas were offline or flagged in the September and April data respectively, and are depicted in \autoref{fig:EDA-layout}. 

\subsection{Spherical Harmonic Beam Coverage}
\label{subsec:SHBC}
\begin{figure*}
    \centering
    \includegraphics[width=\textwidth]{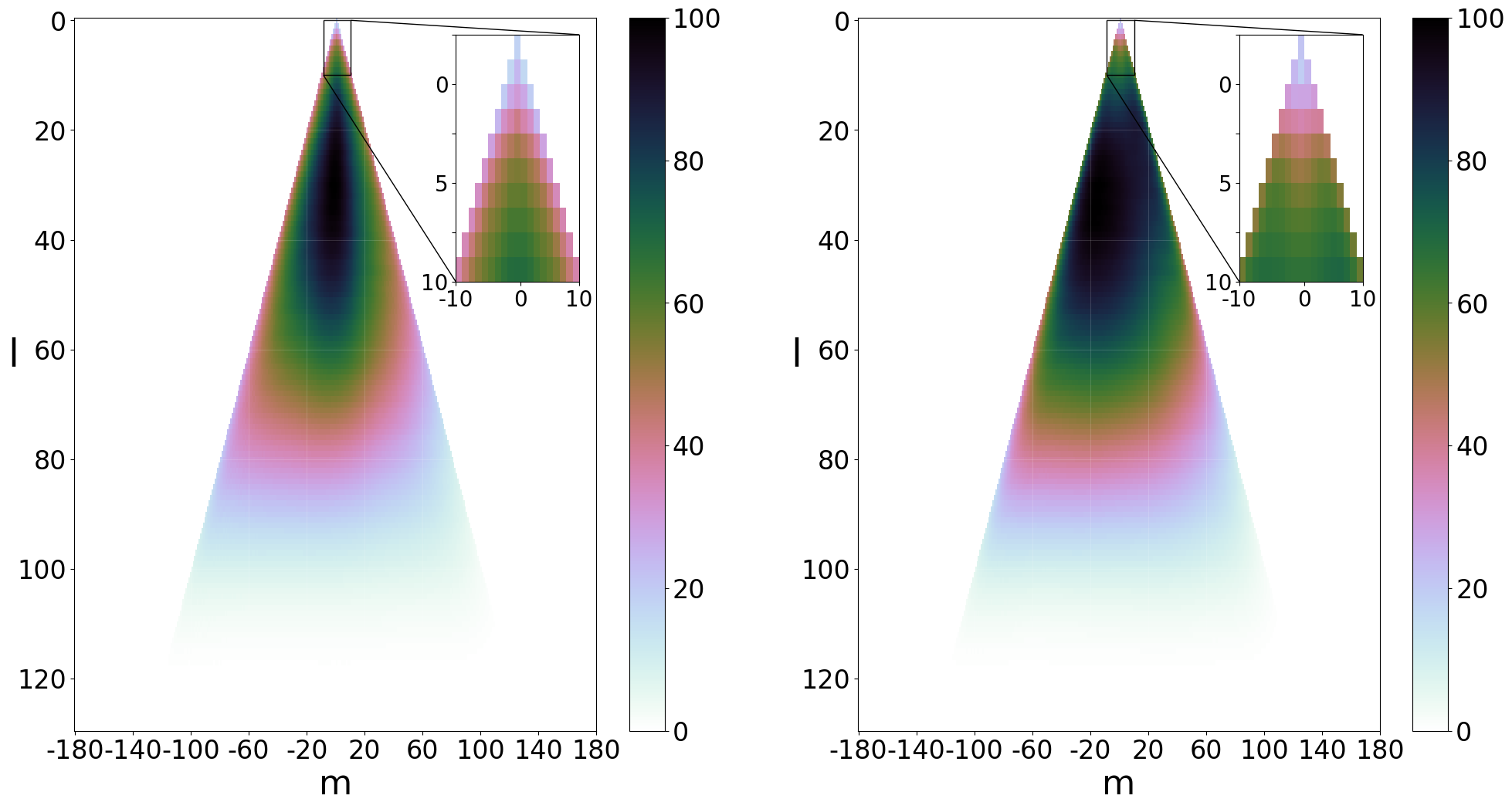}
    \caption{The spherical harmonic beam coverage, with contribution per mode in percentage relative mode sensitivity, in SH-space. A homogeneous array was assumed. The overall contribution is asymmetric in $m$ due to the fact the beam transfer function is a complex waveform. This is a similar phenomenon one sees when plotting the $u,v$-coverage in standard radio interferometry when the conjugate is not included. Left: x-polarisation, Right: y-polarisation. Zoomed areas show the first ten spherical harmonic beam coverage coefficient contributions.}
    \label{fig:SH-space}
\end{figure*}

The \ac{SHBC} in \ac{SH-space} has been generated for both the X-polarisation and Y-polarisation and are depicted in \autoref{fig:SH-space}. From the beam coverage plots three observations can be made.

Firstly, due to our array diameter being limited to 35 meters, we lack sensitivity in the higher $l$-modes. As can be seen, sensitivity it limited to a maximum of $l\approx 117$ and the drop off from our higher modes to zero is quite steep.

Secondly, when looking at the coefficient contribution across the first 10 spherical harmonic orders $l$, it can be seen that we have partial sensitivity at the lower orders, with 19\% coverage on the $l=0$ mode. This is significant as this shows the $m$-mode formalism allows recovery on scales we would not be sensitive to with traditional snapshot imaging methods. Besides avoiding regridding artifacts in our imaging step, this is one of the primary advantages of spherical harmonic transit interferometry over the traditional method when it comes to diffuse all-sky imaging.

In contrast, with traditional interferometric snapshot imaging we expect to be limited to scales corresponding 1$\lambda$, or approximately $60^{\circ}$ in angular resolution, which is equivalent to a lowest mode sensitivity of $l=5$ in spherical harmonic imaging. We verified this by calculating $l_{\rm min}$ for a single snapshot of the sky, which is achieved by substituting ${\rm r}_{{\rm ij},{\rm max}}$ with ${\rm r}_{{\rm ij},{\rm min}}$ (in our case approximately 1.5\,m) in \autoref{eq:lmax}.

Thirdly, the overall contribution is asymmetric in $m$. This is due to the fact the beam transfer function is a complex waveform that interacts with the conjugated complex spherical harmonic basis functions. During decomposition (depending on baseline vector orientation) the coefficients will shift either to positive or negative rank $m$ in baselines that have East-West contributions. The same asymmetry is also also encoded into the $m$-modes and will resolve itself when generating the coefficients of the real-valued sky, which are a product of $|m|$. This is a similar phenomenon one sees when plotting the $u,v$-coverage in standard radio interferometry when the conjugate is not included.

\subsection{Ionosphere}
In the operating bandwidth of the \ac{EDA2} (50\,MHz--350\,MHz) the spatial offsets caused by ionospheric effects is measured to be in the order of arcminutes \citep{Jordan_2017}. Since the angular resolution of the \ac{EDA2} at 159\,MHz is approximately $3^{\circ}$, ionospheric offsets are negligible.

\subsection[Scintillation and Refraction]{Scintillation and Refraction}
\citet{Eastwood2017} has shown that scintillation and refraction offsets decrease when frequency increases, where these effects reduced to <5\% at 73.152\,MHz. Since we observe the sky at 159\,MHz we assume these effects negligible and therefore do not consider them in our sky map imaging process. In addition, no scintillation was observed in snapshot images made from our data.

\subsection{Coordinate System and Pixel Grid}
For our coordinate system, we have opted to format all sky maps in the \ac{HEALPix}\footnote{\hyperlink{https://healpix.jpl.nasa.gov}{healpix.jpl.nasa.gov}} \citep{Gorski2005}, as this is an equal-area representation and naturally works with spherical harmonics.

Although \ac{HEALPix} is convenient for correctly depicting discretely sampled functions on a sphere, it should be taken into account that the resolution of a \ac{HEALPix} grid is dictated by its N$_{side}$ (\textit{i.e.} the number of times a base pixel is divided along its sides) which operates in a power of 2 fashion. Ideally, based on our physical system dimensions our N$_{side}$ would be defined as

\begin{align}
    \mathrm{N}_{side} &= \frac{l_{\rm max}+1}{3},
\end{align}

\noindent which in our case with a maximum diameter of 35\,m and a frequency of 159\,MHz would result in an N$_{side}$ of 39. However, since this is not a power of 2, our real N$_{side}$ needs to be increased to the nearest power of 2; which is N$_{side}=64$; resulting in a pixel area of $\sim 0.84$ square degrees.

\subsection{Sky Coefficients}
In order to retrieve the spherical harmonic sky coefficients $(\mathbf{a})$ to generate images of the Southern sky, we have to solve for \autoref{eq:tikhonov-inverse}. The beam function $\mathbf{B}$ is generated by multiplying the individual x-pol and y-pol beam models with the respective fringes on the sky for each baseline. The fringes are calculated by solving for the exponent in \autoref{eq:measurementeq} on the whole sky. 

Instead of inverting $\mathbf{B}$ as block-diagonal matrix, $\mathbf{B}$ is split in blocks of independent $m$'s (${\rm B}_{m}$), where each block is defined as in \autoref{eq:b_block}; to reduce size of matrices required in computer memory. Since the $m$-modes are grouped on an $m$-by-$m$ basis, we can invert each $m$-mode and ${\rm B}_{m}$ matrix independently to solve for each individual $\hat{a}_{m}$. However, given that our shortest baseline separation (1.5\,m) at 159\,MHz is not much shorter than our wavelength ($\lambda=1.89\,$m), we are no longer fully sensitive to the lowest spherical harmonic orders and global signal component ($\hat{a}_{0,0}$); as shown in \autoref{fig:SH-space}. Therefore, we need a prior model to better fit for the solutions of $\hat{\mathbf{a}}$.

\subsubsection{Prior Model}
\label{subsubsec:prior}
In order to properly regularise for diffuse emission in the sky map, \autoref{eq:tik_inv_prior} will be used instead. The model used as prior information is the desourced, destriped Haslam map from \citet{Remazeilles2015}. Since the Haslam map itself is at 408\,MHz, the map has been rescaled to match the correct brightness temperature using pyGDSM \citep{Price2016}, a python interface for diffuse global sky models \ac{SI} for downscaling to 159\, MHz are embedded in the package and, for the Haslam map, extracted from \citep{Mozdzen2016}. We've selected Haslam as our prior as it is still the most prevalent in use; furthermore, in most sky map papers it's a common approach to compute \ac{SI} between a single sky map frequency and the 408\,MHz Haslam map. This should provide others a reference frame to compare their maps to our prior constrained map and map without prior.

\begin{figure*}
    \centering
    \includegraphics[width=\linewidth]{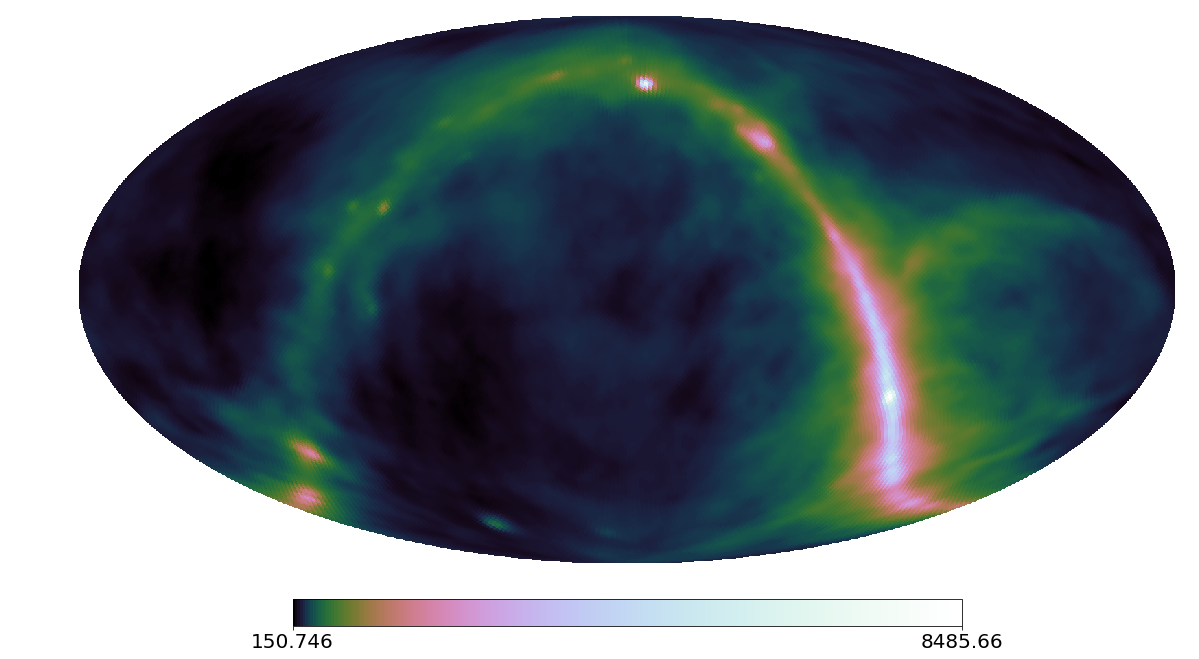}
    \caption{159\,MHz diffuse model map, 98\% log-scaling (N$_{side}=64$). Generated from the 2014 desourced and destriped reprocessed 408\,MHz Haslam map \citep{Remazeilles2015}.}
    \label{fig:model_map}
\end{figure*}

The map is also downscaled to the correct HEALPix N$_{side}$ in order to match our measurements and angular resolution. Additionally, we employed a Gaussian smoothing kernel at the \ac{FWHM} of our synthesised beam. This is achieved by applying \ac{HEALPix}'s \texttt{ud\_grade()} and \texttt{smoothing()} functions respectively; in the case of changing pixel scale, \ac{HEALPix} sets the rescaled superpixel to be the mean of the children pixels. The resulting model map is shown in \autoref{fig:model_map}. We also need to account for the brightness of the sun in both the September and April observations, the model map is used to create two prior maps for \autoref{eq:tik_inv_prior_noglobal} with a model of the sun at the correct location in the sky for both observations. We opted only to remove the global component as the discrepancy between $a_{00}$ on the prior and the measured sky was too large. This mode has later been reinserted to assure a global sky component is present in the maps. These model map coefficients are calculated in accordance to \autoref{eq:sky_decomp}. 

\begin{figure}
    \centering
    \includegraphics[width=\linewidth]{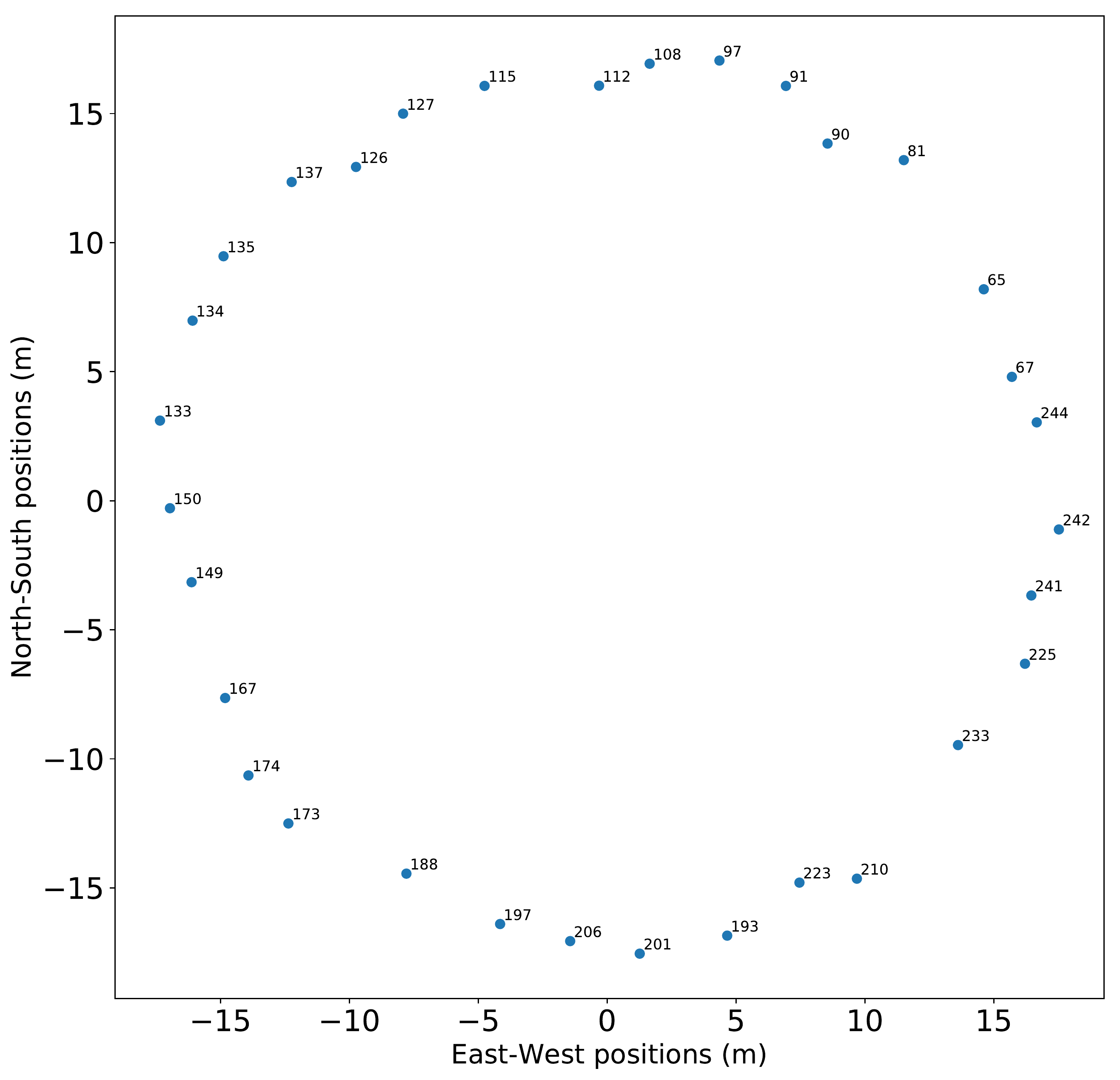}
    \caption{EDA2  array  32  element  outer ring layout  in  local  (North-South, East-West) coordinates.}
    \label{fig:EDA-ringlayout}
\end{figure}

\subsubsection{ridge parameter selection}
In order to properly constrain the inversion of the $m$-modes to solve for the sky $\hat{\mathbf{a}}$, we need to determine the proper value for $\varepsilon$ given the measured data. Since we have two 24h observations, each with two polarisations, four ridge parameters have to be selected. Since solving for $\varepsilon$ requires us to solve for the norms of the error and the solution multiple times to determine the optimal value.

To select the $\varepsilon$ to best fit our data, ideally one would like to sample for as many values for $\varepsilon$ as possible; preferably infinite. However, due to the computational complexity of solving for \autoref{eq:tik_inv_prior_noglobal}, choosing for an increasingly large sample of $\varepsilon$ quickly becomes a non-feasible endeavour; a clear trade-off exists between time spent to constrain $\varepsilon$ and the accuracy that we gain from it. To reduce the time spent, but still maintain accuracy, we propose an alternate method to constrain $\varepsilon$ instead; a two-stage constraining scenario has been created to ease computational strain and time-constraints. Finely sampling for a reduced subset of a full array, yet with still sufficient \ac{SHBC} in \ac{SH-space}, provides us a rough estimation where the optimum must lie for the full array. Coarsely recalculating $\varepsilon$ across this sample-space for the full array will give us a close estimate for the best overall fit. Tweaking the recalculated $\varepsilon$ around this close estimate should therefore yield us our optimal solution.

\begin{figure*}
    \centering
    \begin{subfigure}[b]{0.9\columnwidth}
        \includegraphics[width=0.9\linewidth]{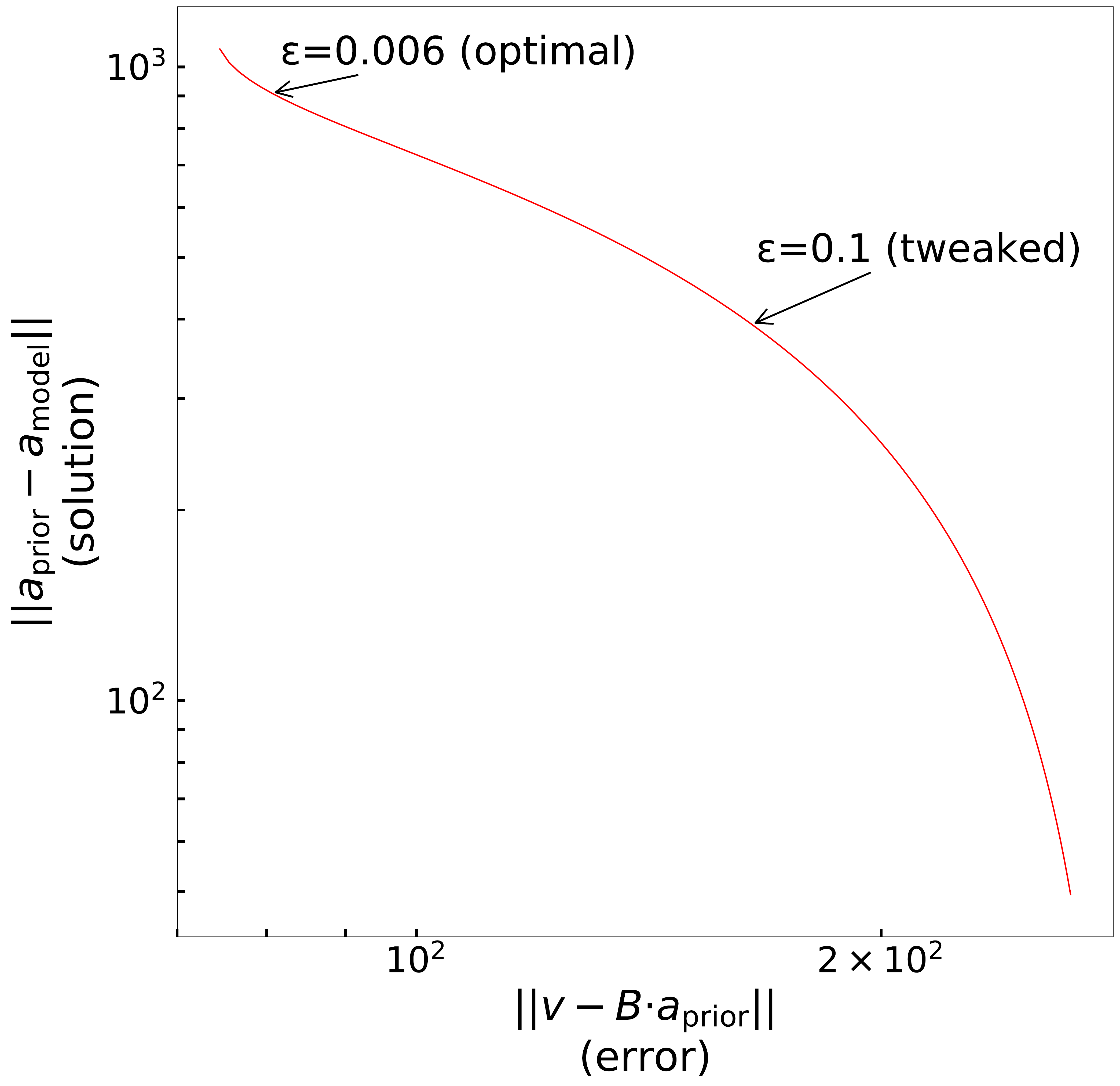}
        \caption{X-pol September Data L-curve}
        \label{subfig:xpol_sep_curve}
    \end{subfigure}
    \hfill
    \begin{subfigure}[b]{0.9\columnwidth}
        \includegraphics[width=0.9\linewidth]{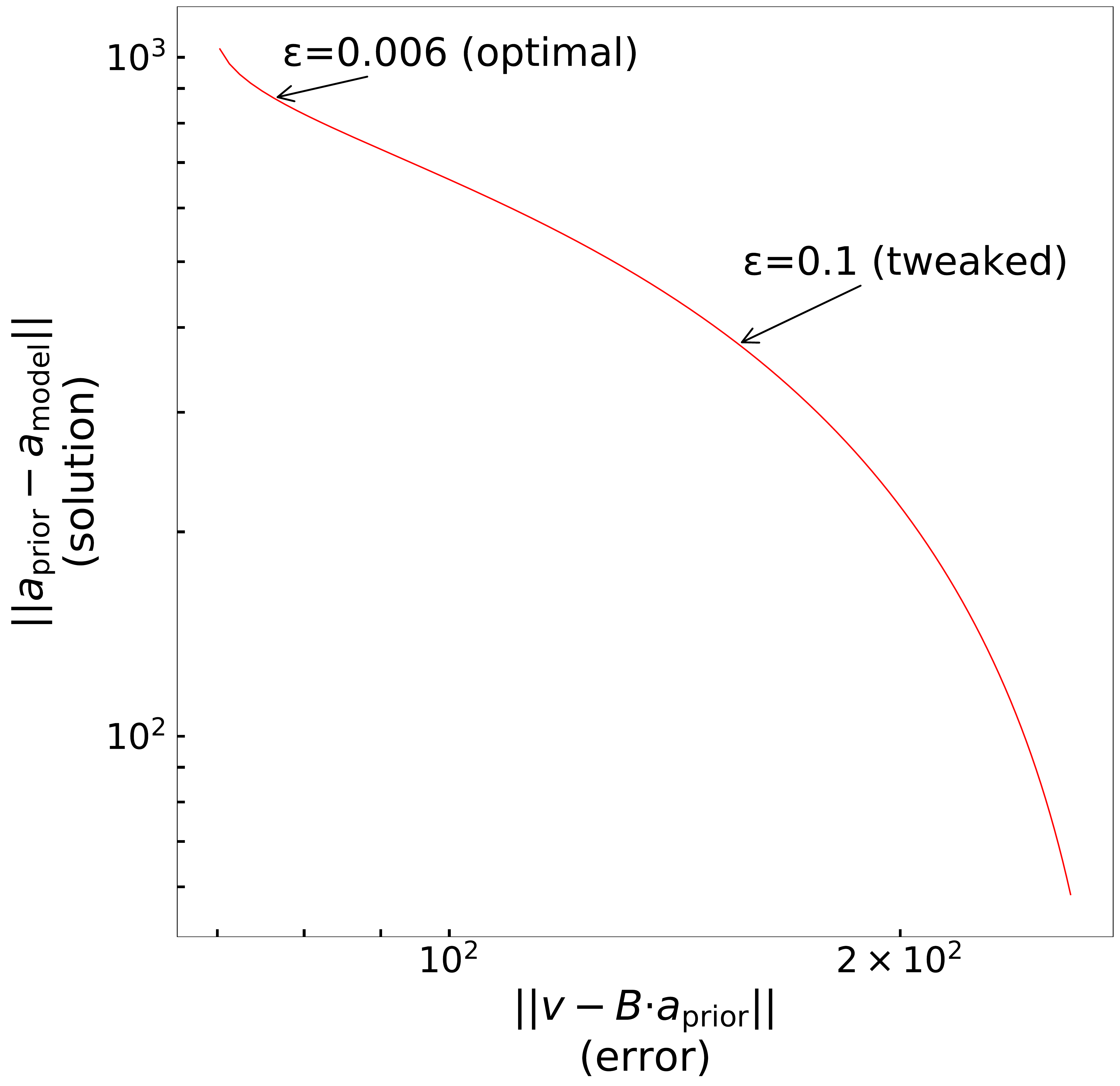}
        \caption{Y-pol September Data L-curve}
        \label{subfig:ypol_sep_curve}
    \end{subfigure}
    \\
    \begin{subfigure}[b]{0.9\columnwidth}
        \includegraphics[width=0.9\linewidth]{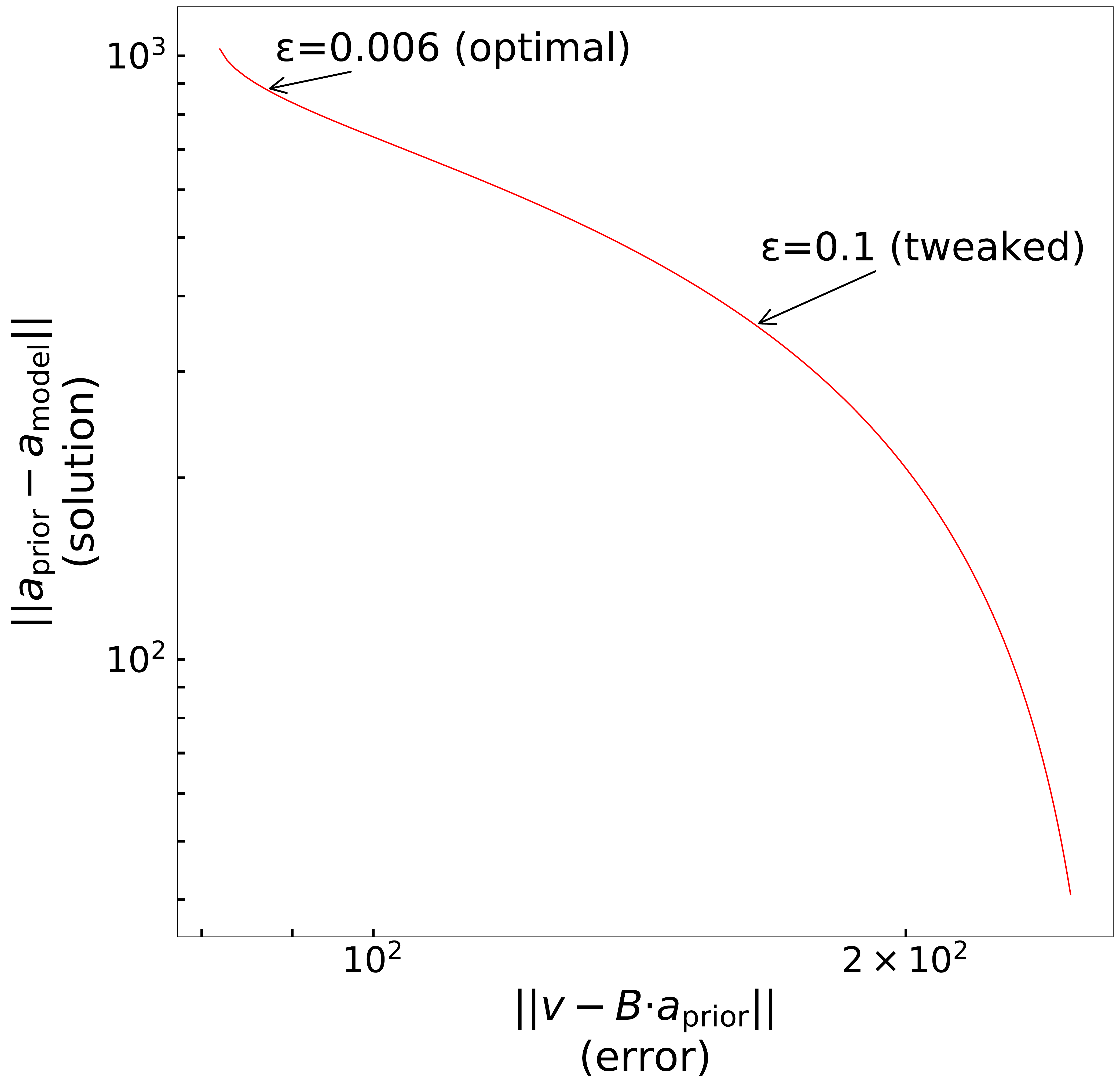}
        \caption{X-pol April Data L-curve}
        \label{subfig:xpol_apr_curve}
    \end{subfigure}
    \hfill
    \begin{subfigure}[b]{0.9\columnwidth}
        \includegraphics[width=0.9\linewidth]{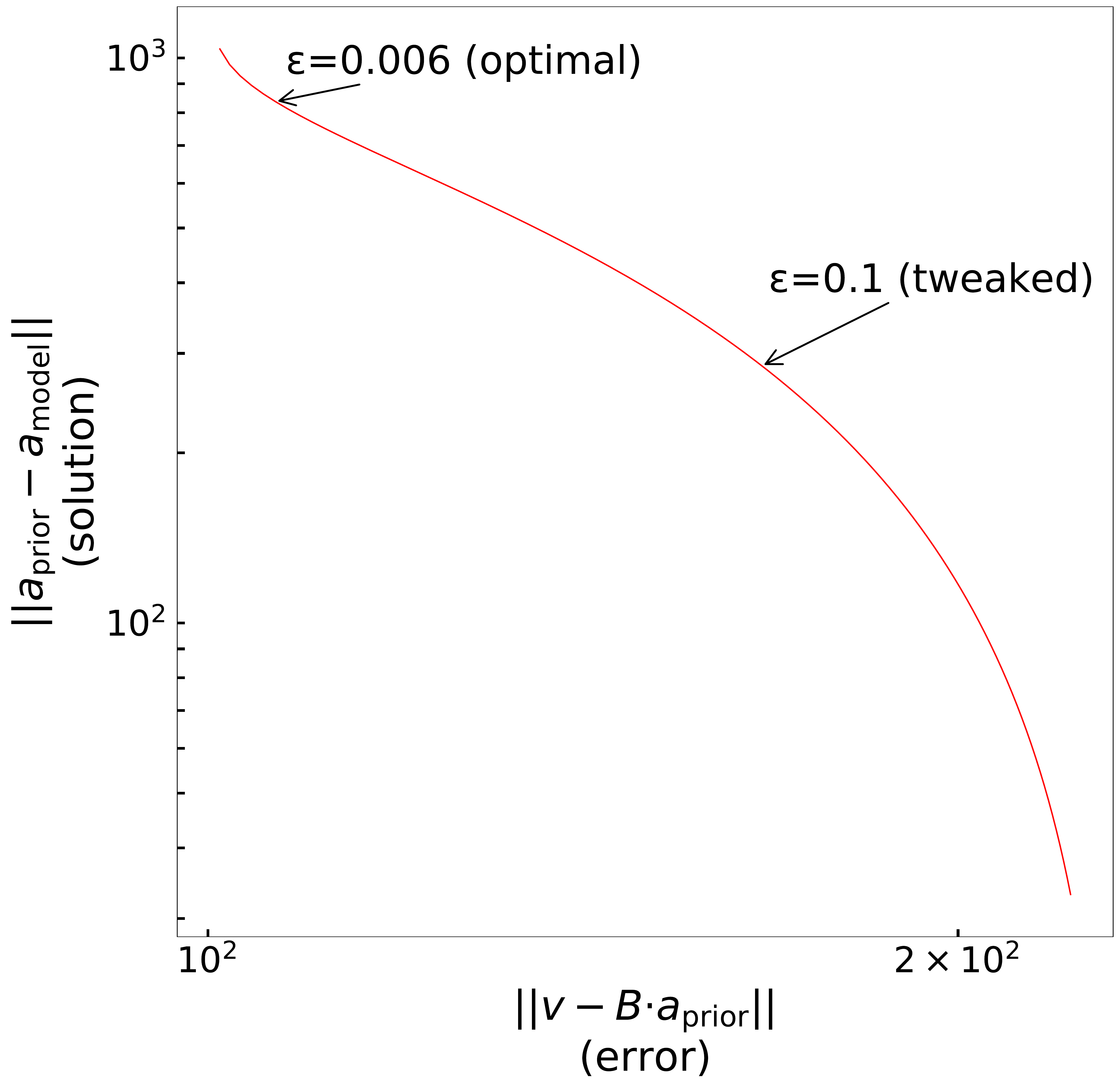}
        \caption{Y-pol April Data L-curve}
        \label{subfig:ypol_apr_curve}
    \end{subfigure}
    \caption{L-curves computed for the \ac{EDA2} data for both September and April observations; and X and Y polarisations. The data was generated by first trialing 2000 samples of $\varepsilon$ for a 32-element subset array ($\varepsilon_{32}$) and then fit for the full array by coarsely re-sampling at 20 evenly spaced points within the linear regime in the 32-element L-curve. The final $\varepsilon$ is then obtained tough tweaking around the best coarse fit for an optimum 256 element solution ($\varepsilon_{256}$). The data is represented following \autoref{fig:L-curve}, where the 2000 sample points generated an L-curve from the `knee' down; \textit{i.e.} the bottom half of \autoref{fig:L-curve} is therefore only shown in this representation.} 
    \label{fig:L-curves}
\end{figure*}

For the \ac{EDA2} array subset, to make sure we still maintain a proper baseline distribution, only the outer rings of elements have been selected for the first stage (\autoref{fig:EDA-ringlayout}). To determine the rough optimal range of $\varepsilon$ for this subset, a fine-scale inversion process of \autoref{eq:tik_inv_prior} is ran 2000 times for each observation and each polarisation, minimising for $||\mathbf{v}-\mathbf{B}\hat{\mathbf{a}}||$ and $||\hat{\mathbf{a}}-\mathbf{a}_{\rm prior}||$ each time. The resulting L-curves are shown in \autoref{fig:L-curves}, which represent the L-curve from the `knee' down. The optimal $\varepsilon$ values resulting from the L-curves are $\varepsilon_{32}=0.006$ for the 32-element data.

In order to then better constrain $\varepsilon$ for the 256-element array, in the second stage the norms are recalculated for 20 $\varepsilon$ data points equidistantly spaced on the linear regime in the L-curves (between approximately 70 and 200 on the horizontal axis). For the lowest-norm outcomes in each polarisation, the $\varepsilon$ are then tweaked to assure the lowest norms possible providing 
the best fit for the solutions of the data. The final $\varepsilon$ values obtained were the same for all polarisations and are $\varepsilon_{256}=0.1$.

\subsection{Deconvolution of Compact Sources}
Since we have a finite array that physically limits the angular resolution and has incomplete sampling of the sky, much similar to holes in the $u,v$-sampling space in traditional interferometry, we expect our sky sources to be convolved with a \ac{PSF}. In order to counteract a \ac{PSF} we can try to minimise their contribution through deconvolution, traditionally done via the CLEAN algorithm in radio astronomy~\citep{Thompson2017}.

Originally, CLEAN~\citep{Hogbom1974} was designed to work with Fourier-synthesis imaging methods. In spherical harmonic transit interferometry, we instead work in spherical harmonic domain and cannot directly apply standard CLEAN algorithms. \citet{Eastwood2017} showed how to directly relate the \acp{PSF} in spherical harmonic coefficient space:

\begin{align}
    \hat{\mathbf{a}}_{\rm PSF}(\theta) &= \left(\mathbf{B}^{\dagger}\mathbf{B}+\varepsilon\mathbf{I}\right)^{-1}\mathbf{B}^{\dagger}\mathbf{B}\mathbf{a}_{\rm ps}(\theta)\,.
\end{align}

\noindent where $\hat{\mathbf{a}}_{\rm PSF}$ are the spherical harmonic coefficients of the \ac{PSF}, and $\mathbf{a}_{\rm ps}$ is the spherical harmonic decomposition of a single point on the sky.

\begin{figure*}
    \centering
    \includegraphics[width=\linewidth]{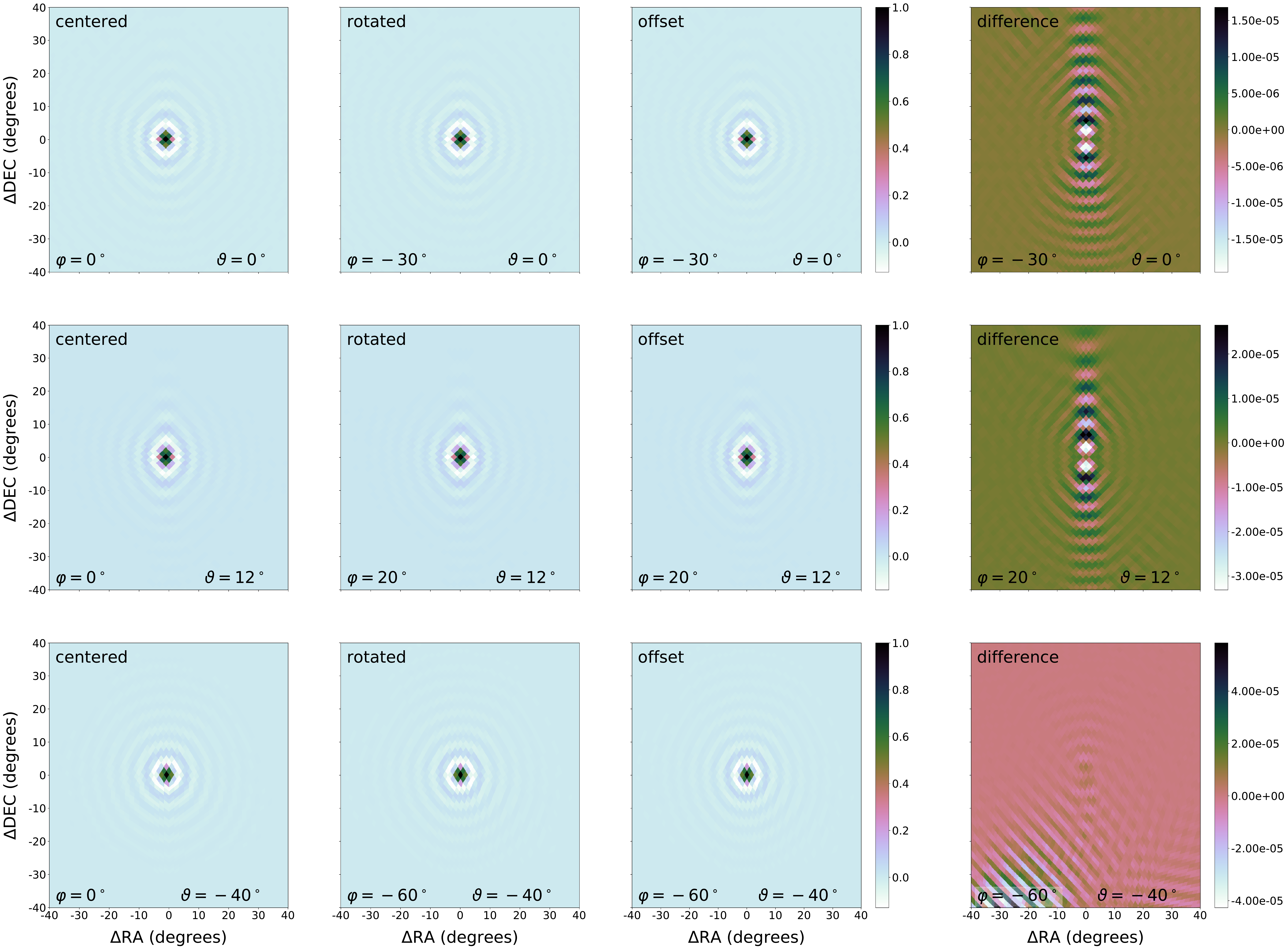}
    \caption{EDA2 array \acfp{PSF}. Left column: \acp{PSF} generated at an \acf{RA} ($\varphi$) of 0 degrees, Middle left column: \acp{PSF} of the left column rotated in coefficient space to a specific \acs{RA}, Middle right column: \acp{PSF} generated at an offset \acs{RA} ($\varphi$), Right column: difference between the rotated and the offset \acp{PSF}. Top row: \acp{PSF} generated at a \acf{DEC} ($\vartheta$) of  0 degrees, Middle row: \acp{PSF} generated at a \acs{DEC} ($\vartheta$) of 12 degrees, Bottom row: \acp{PSF} generated at a \acs{DEC} ($\vartheta$) of -40 degrees.}
    \label{fig:PSFs}
\end{figure*}

As part of the proposed CLEANing algorithm by \citet{Eastwood2017}, it was noted the \acp{PSF} are shift invariant in \ac{RA}, \textit{i.e.} \acp{PSF} only have to be generated on a per-declination basis. Unlike \citet{Eastwood2017}, instead of pre-identifying CLEAN components and precalcuting the \acp{PSF} for each of those components, we simply store a \ac{PSF} image for each possible declination on the sky and do all deconvolution in image space. To deconvolve the ``dirty images'', bright compact sources are selected and windowed in image space to determine the brightest pixel. For each bright pixel a \ac{PSF} is selected based on declination and rotated to the correct longitude. We tested that  each \ac{PSF} generated via rotation was consistent with its equivalent \ac{PSF}, generated via spherical harmonics for a point source, at that location. Residuals and imaging artifacts propagated by imaging the rotated \acp{PSF} are $<0.01\%$ and therefore considered negligible. Examples of these residuals are shown in \autoref{fig:PSFs}.

After deconvolving the \acp{PSF}, the identified pixels used to model the \acp{PSF} are then convolved with a restoring Gaussian (which is an identical Gaussian fit to the \ac{PSF} central region to preserve the flux) and reinserted into the residual image to create the CLEANed sky map. It should be noted that this image-based CLEANing method closely follows the CLEAN steps outlined by \citep{Eastwood2017}, however, deviates at the stages where \acp{PSF} selection and deconvolution occurs. The complete step-by-step selective image-based CLEANing process we employed is outlined below:

\vspace*{1em}
\hrule
\begin{center}
Selective Image-based Spherical\\ Harmonic CLEANing process
\end{center}
\hrule
\begin{enumerate}[leftmargin=*,labelindent=6pt,label=\bfseries \arabic*.]
    \item Identify bright compact sources in image.
    \item Mask these sources and isolate in separate map to determine the brightest pixels to be used as model.
    \item Extract the model pixel's coordinates ($\varphi, \theta$).
    \item For each model pixel:
    \begin{enumerate}[leftmargin=*,label=\bfseries \arabic{enumi}.\arabic*.]
        \item Select \ac{PSF} image at correct declination and convert to normalised \ac{PSF} sky map.
        \item Rotate the \ac{PSF} sky map to the correct longitude~ ($\varphi$) using \texttt{rotate\_alm()\footnote{See \hyperlink{https://healpix.jpl.nasa.gov/html/subroutinesnode90.htm}{HEALPix rotator functions documentation}}}.
        \item For selected model pixel, set subtraction threshold.
        \item Until threshold is met:
        \begin{enumerate}[leftmargin=*,label=\bfseries \arabic{enumi}.\arabic{enumii}.\arabic*.]
            \item Subtract~\ac{PSF}~$\times$ maximum pixel brightness~$\times$~$\gamma$ from the source in the ``dirty image''.
            \item Append subtracted source pixel value to a model sky map.
        \end{enumerate}
    \item Convolve total model pixel in model map with restoring beam
    \end{enumerate}
    \item Add model map with restored beam back to residual sky map.
\end{enumerate}
\hrule
\vspace*{1em}

\begin{figure*}
    \centering
    \begin{minipage}{0.95\columnwidth}
    \centering
        \includegraphics[width=0.9\linewidth]{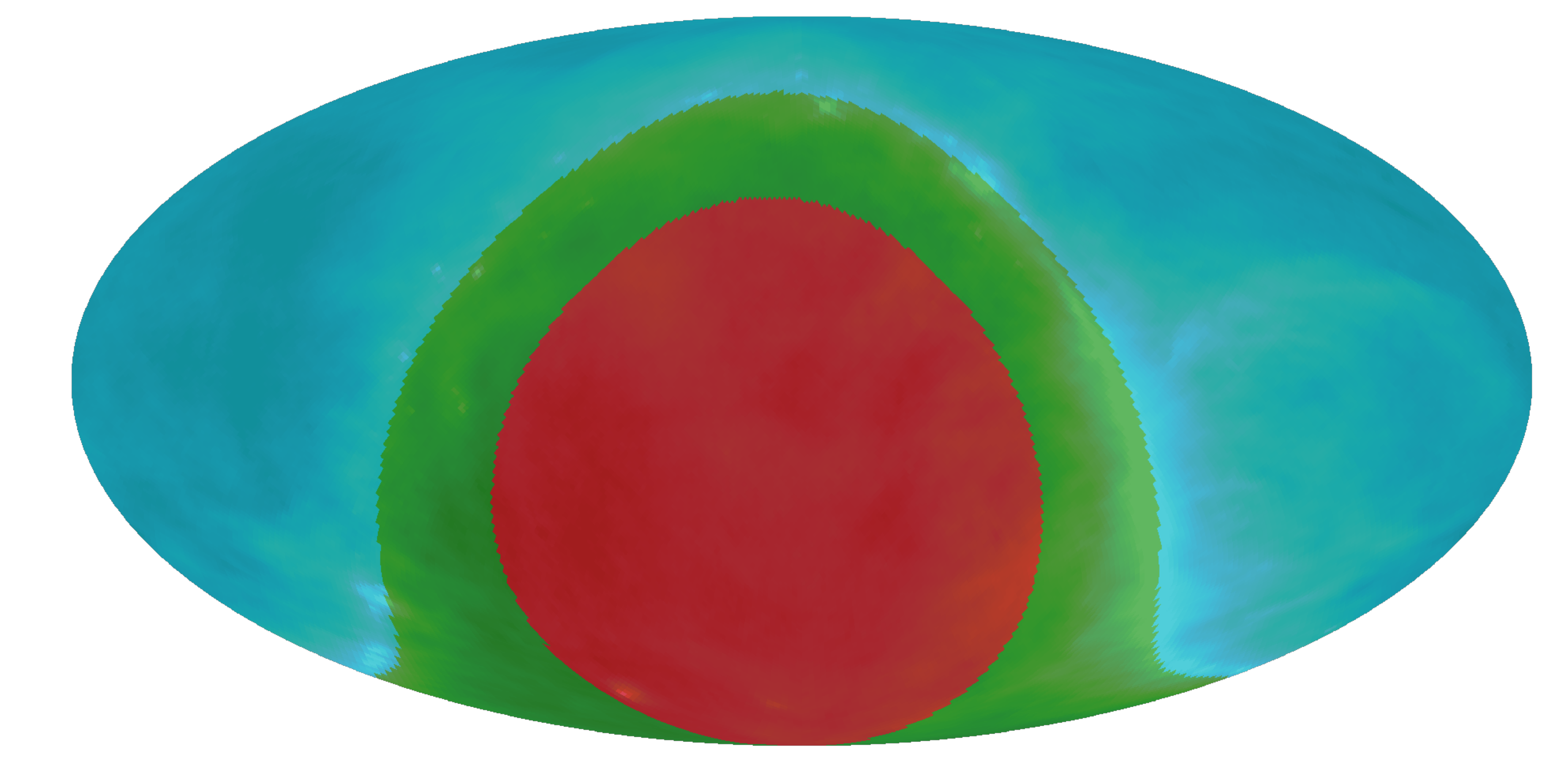}
        \caption{Contour map for combining the April and September data. Cyan: April data, Green: linearly-weighted combination of April and September data, Red: September data. Overlaid with the model map from Figure \ref{fig:model_map} as a reference for overlap of regions.}
        \label{fig:epoch-weight}
    \end{minipage}%
    \begin{minipage}{0.1\columnwidth}\hfill\end{minipage}%
    \begin{minipage}{0.95\columnwidth}
    \centering
        \includegraphics[width=0.9\linewidth]{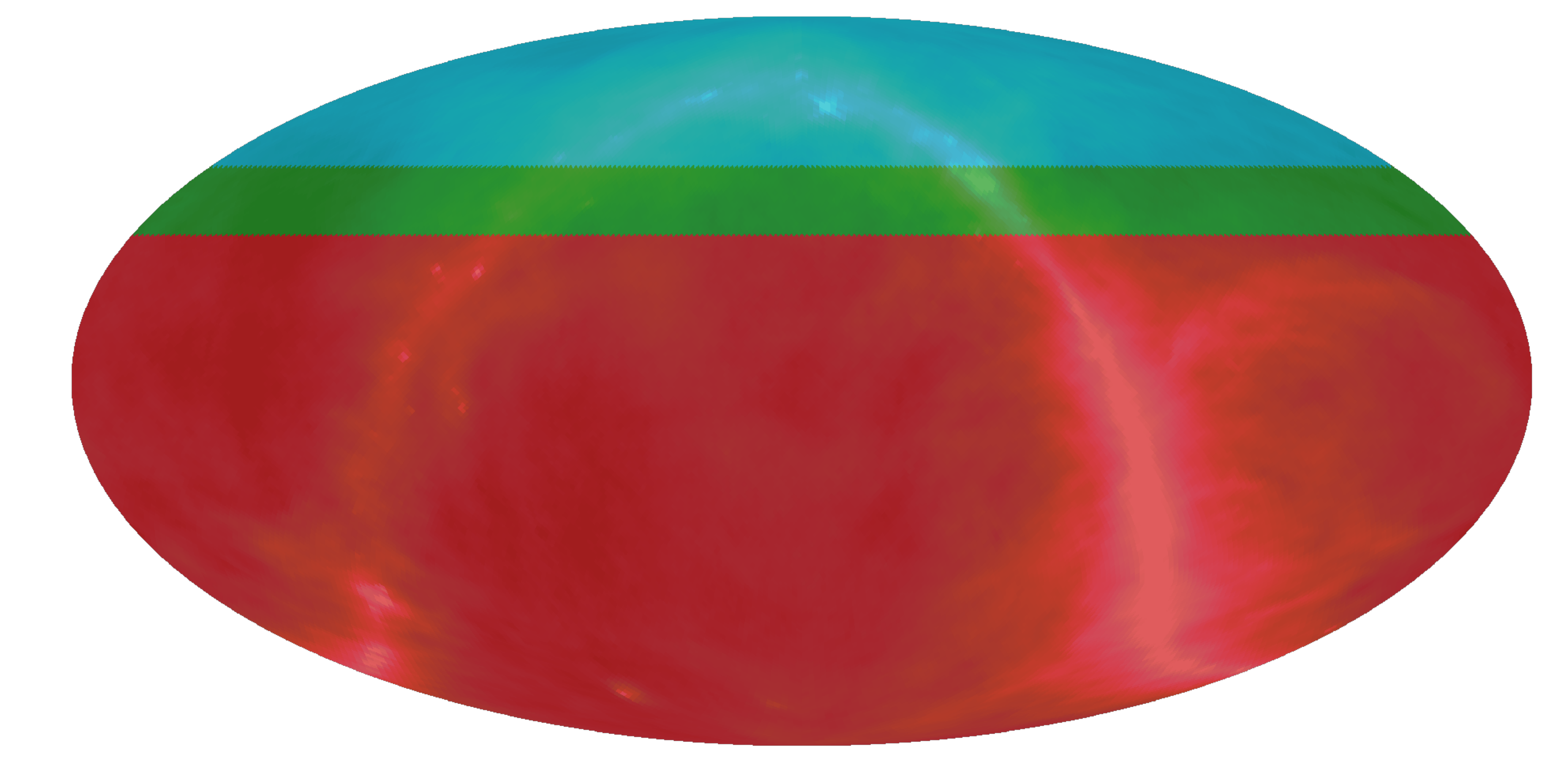}
        \caption{Contour map for combining the intensity map and model data. Cyan: model data, Green: linearly-weighted combination of intensity map and model data, Red: intensity map. Overlaid with the model map from Figure \ref{fig:model_map} as a reference for overlap of regions.}
        \label{fig:polarisation-weight}
    \end{minipage}%
\end{figure*}

\subsection{Source Removal}
\label{subsec:weighting}
\subsubsection{The Sun}
The Sun in our images is very strong and slightly smeared out, meaning we cannot perfectly deconvolve the \ac{PSF} from our images, leaving residual ripples into our sky maps. This has been the main motivation to image the sky during two different months (April and September) where the sun manifests at different locations on the sky. This gives us the opportunity to combine the two epochs such that the sun no longer is present. We achieve this by averaging the two sky maps together, after CLEANing, according to \autoref{fig:epoch-weight}. The blue region contains the September data, the red region contains the April data, and the green region is a linear-gradient weighted average of both maps; the green region closer to the red region contains more contribution from the April observations and vice versa. This action is performed on both x-polarisation and y-polarisation.

\subsubsection{Field of View Edge}
Regions at the sky near the northern horizon are poorly sampled and passed through a low-sensitivity part of the beam. Although these regions of the sky do appear in our output images they are not trustworthy. Instead of truncating the sky at an arbitrary \ac{DEC}, we apply difference weighting between the intensity map we generate using our observations, and the prior model (for the prior fit map) or global ($a_{00}$) emission (for the non-fit map) to provide a smoother transition towards the region of the sky no longer observable to the system. This has a small trade-off of sacrificing a small region of our measurable sky to better constrain the edges of our \ac{FoV}. These regions are depicted in \autoref{fig:polarisation-weight}, in the red region the intensity map is 100\% represented, the green region is again a linear-gradient weighted average of both the intensity map and the model maps (Haslam/diffuse) between $+50^{\circ}$ and $+60^{\circ}$; the green region closer to red denotes a higher contribution of the intensity map, whereas the green region near the cyan region has more weight on the model. The cyan region has solely contribution of the model map depicted in \autoref{fig:model_map} or the $a_{00}$ diffuse mode map.

\section{Results}
\label{sec:results}
\begin{figure*}
    \centering
    \includegraphics[width=\linewidth]{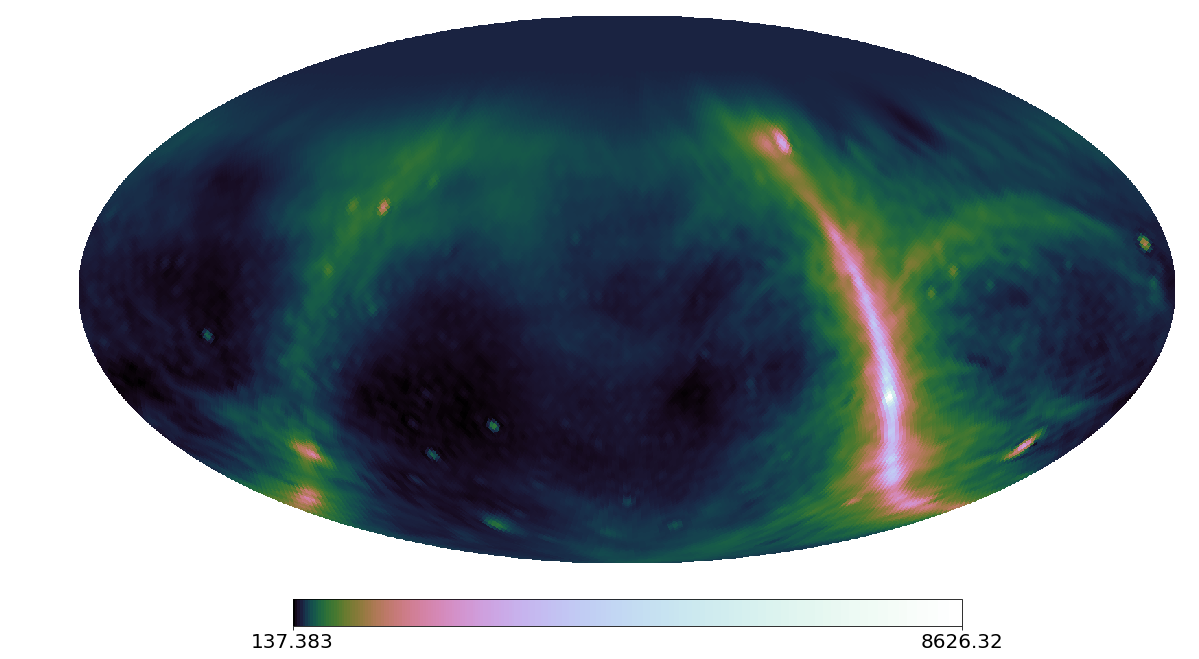}
    \caption{159\,MHz diffuse EDA2 map (equatorial view), log-scale. Generated without the use of a prior model, the global sky component is reinserted.}
    \label{fig:EDA2_nopriorCel}
\end{figure*}

\begin{figure*}
    \centering
    \includegraphics[width=\linewidth]{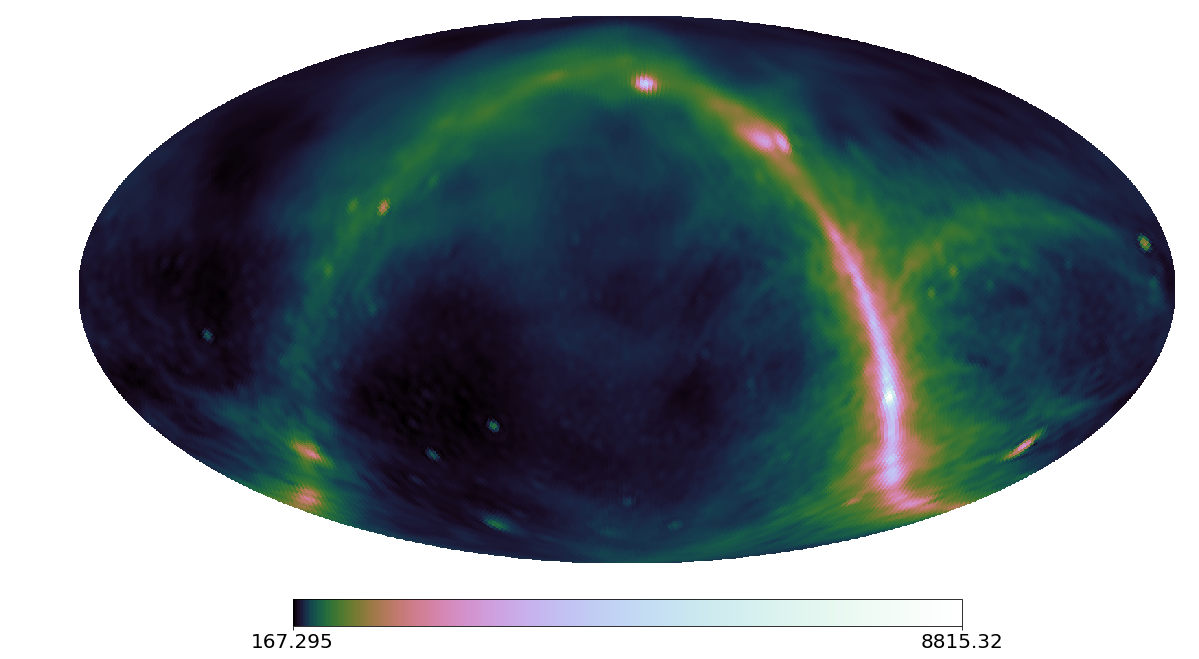}
    \caption{159\,MHz diffuse EDA2 map (equatorial view), log-scale. Generated with the use of a prior model to constrain the beam-inverse, the Northern hemisphere is therefore equivalent to the diffuse reprocessed Haslam map depicted in \autoref{fig:model_map}.}
    \label{fig:EDA2_priorCel}
\end{figure*}

The resulting \ac{EDA2} sky maps are presented in \autoref{fig:EDA2_nopriorCel} and \autoref{fig:EDA2_priorCel}. Two versions of sky maps are made, one without prior fitting (\autoref{fig:EDA2_nopriorCel}  and one with the updated desourced and destriped Haslam map \citep{Remazeilles2015} used as a prior (\autoref{fig:EDA2_priorCel}). Both sky maps have been presented in equatorial coordinates as that is the coordinate system we measured the sky in. Versions of the sky maps in their default \ac{HEALPix} representation, with coordinate grid, can be found in \hyperref[apdx:HEALPix-maps]{Appendix B}. These maps have been created by combining two 24-hour observations in different epochs (September and April) to be able to remove the Sun using \autoref{fig:epoch-weight}. Both maps have a resolution of approximately $3.1^{\circ}$, supersampled to a $0.916^{\circ}$ ($0.84$ square degrees) pixel scale. The \ac{FWHM} of the synthesized beam at $\delta=-40^{\circ}$, $\delta=0^{\circ}$, and $\delta=12^{\circ}$ are $3.10^{\circ}\times3.01^{\circ}$, $3.30^{\circ}\times3.06^{\circ}$, and $3.71^{\circ}\times3.05^{\circ}$ respectively (major $\times$ minor axis). These maps have been corrected for systematic and regression bias (\autoref{subsec:biasCorr}) and common-mode noise has been removed (\autoref{subsec:noiseCorr})

\subsection{Difference Visibility Dataset}
We create a difference visibility dataset by averaging the subtracted odd and even non-averaged samples from each other within each one minute averaged visibility sample following \autoref{eq:visdiff}.

\begin{align}
    \label{eq:visdiff}
	\Delta\sigma_{\mathrm{ij}} &= \sqrt{\frac{\sum\limits_{s=0}^{S_{\mathrm{max, \Delta V}}/2}\frac{\left(V_{s,\Delta V}^{\mathrm{ij}, \mathrm{even}}-V_{s,\Delta V}^{\mathrm{ij}, \mathrm{odd}}\right)}{\sqrt{2}}\frac{\left(V_{s,\Delta V}^{\mathrm{ij}, \mathrm{even}}-V_{s,\Delta V}^{\mathrm{ij}, \mathrm{odd}}\right)^*}{\sqrt{2}}}{S_{\mathrm{max, \Delta V}}/2}}.
\end{align}

\noindent Here, $S_{\rm max}$ is the maximum number of samples within a one minute averaged visibility sample (in our case eight), $V_{s,\Delta V}^{\mathrm{ij}, \mathrm{even}}$ and $V_{s,\Delta V}^{\mathrm{ij}, \mathrm{odd}}$ describe the odd and even time samples for each 1 minute average for sample point $s$ at baseline $(i,j)$ respectively, and $\Delta\sigma_{\mathrm{ij}}$ is the average visibility noise within average sample. Each difference visibility noise sample is then averaged together to within the same one-minute average bin. We use this dataset to estimate the noise in our sky maps as well as correct for any common-mode noise that occurs.

\subsubsection{Noise Correction}
\label{subsec:noiseCorr}
\citet{Eastwood2017} has shown that \ac{RFI} from sources that do not follow the rotation of the sky, or that common-mode pickup between interferometer elements can generate unwanted contribution to the visibilities. This will smear out across declination in image space, manifesting as concentric rings at various declinations across the sky.
To see these effects in the \ac{EDA2} sky maps we used the difference visibility dataset.

\begin{figure*}
    \centering
    \begin{subfigure}[b]{0.9\columnwidth}
        \includegraphics[width=0.9\linewidth]{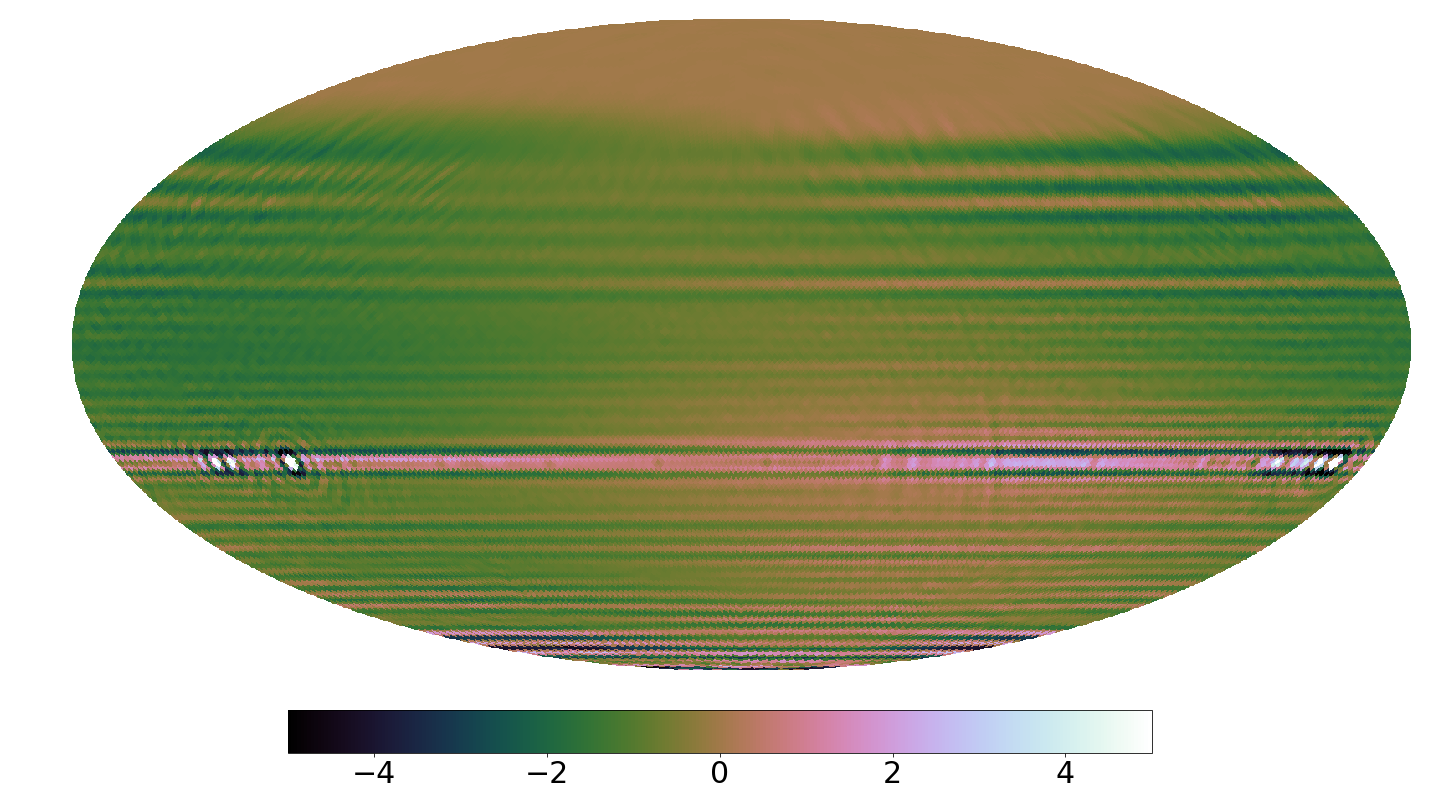}
        \caption{X-pol September noise in Kelvin}
        \label{subfig:xpol_sep_noise}
    \end{subfigure}
    \hfill
    \begin{subfigure}[b]{0.9\columnwidth}
        \includegraphics[width=0.9\linewidth]{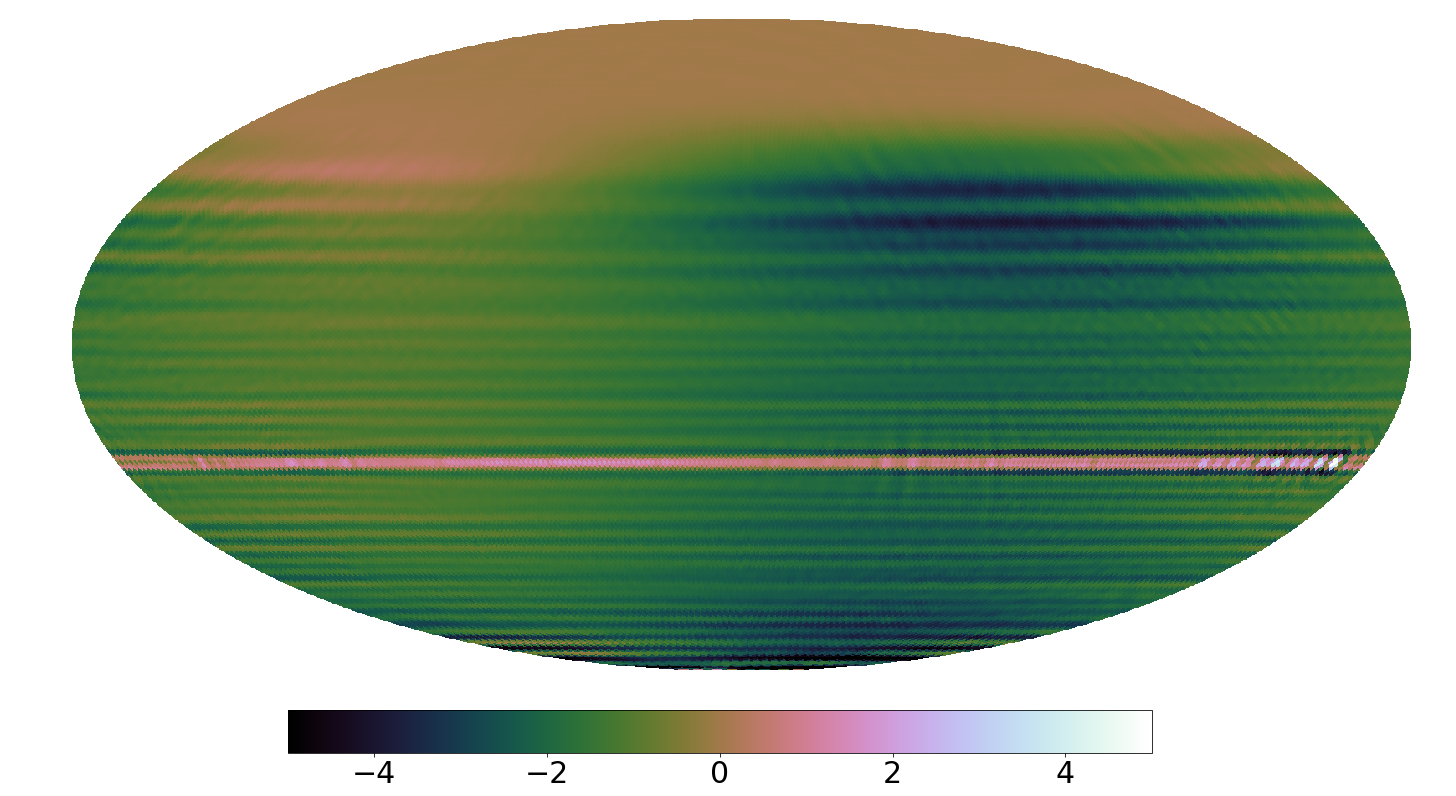}
        \caption{Y-pol September noise in Kelvin}
        \label{subfig:ypol_sep_noise}
    \end{subfigure}
    \\
    \begin{subfigure}[b]{0.9\columnwidth}
        \includegraphics[width=0.9\linewidth]{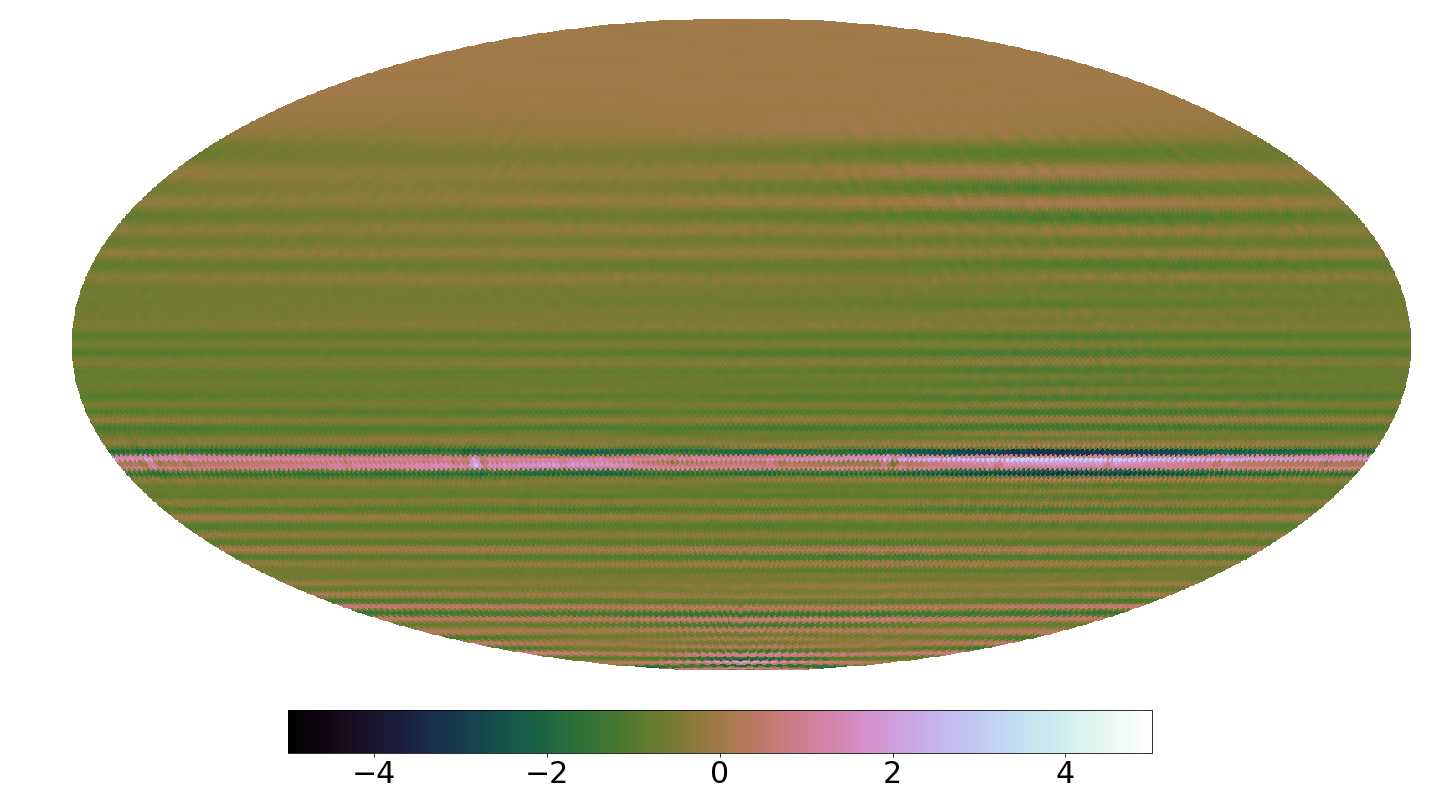}
        \caption{X-pol April noise in Kelvin}
        \label{subfig:xpol_apr_noise}
    \end{subfigure}
    \hfill
    \begin{subfigure}[b]{0.9\columnwidth}
        \includegraphics[width=0.9\linewidth]{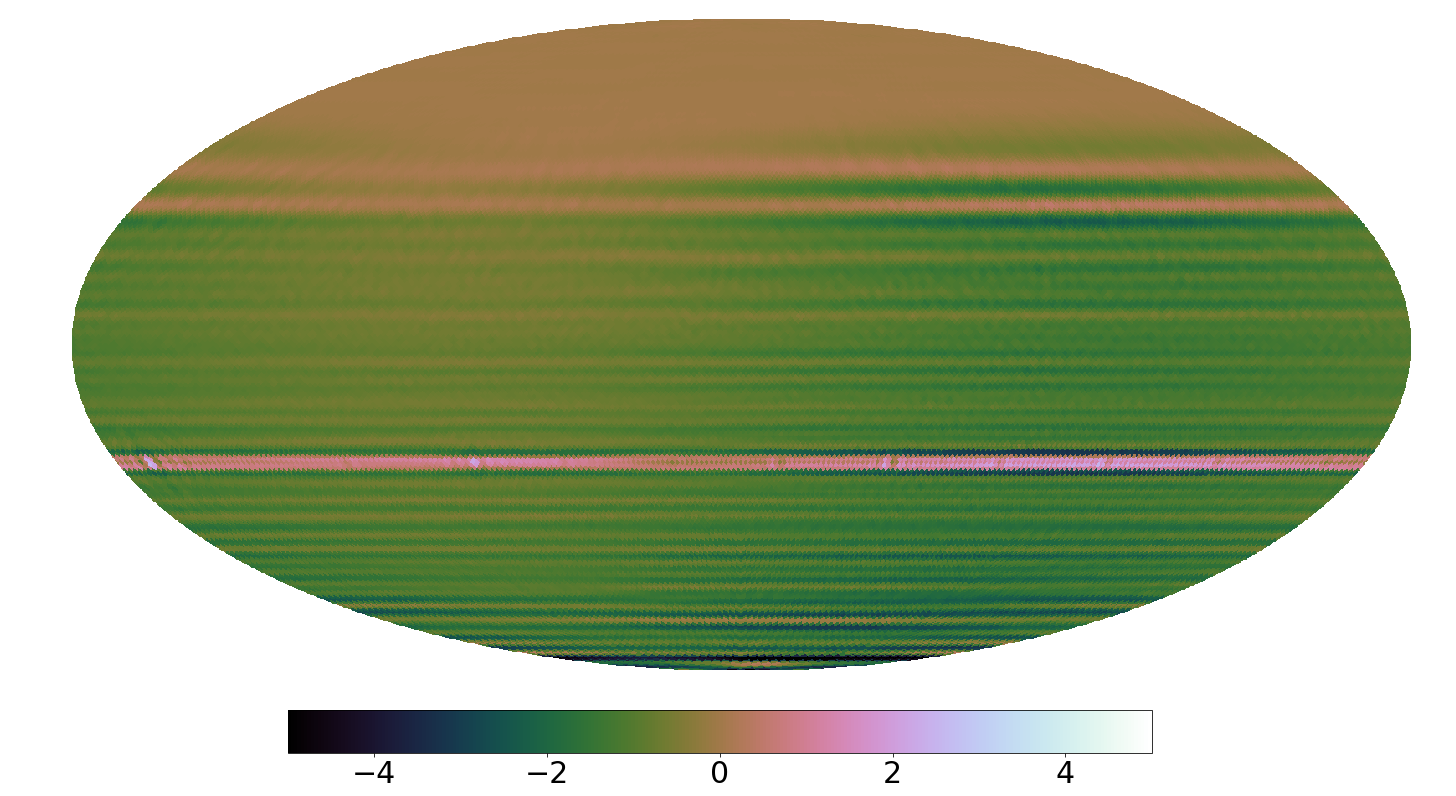}
        \caption{Y-pol April noise in Kelvin}
        \label{subfig:ypol_apr_noise}
    \end{subfigure}
    \caption{Equatorial projection of noise separated from measured visibilities after passing through the $m$-mode pipeline. A clear concentric ringing at multiple declinations is present clearly indicating a form of terrestial \ac{RFI} or stationary noise. These noise-modes manifest in spherical harmonic modes $m\leq1$.} 
    \label{fig:common_noise}
\end{figure*}

To get an image representation of the noise, the noise samples are inserted in our $m$-mode imaging pipeline with equal settings as to imaging the measured sky. This operation is performed for both X and Y polarisations for the April and September data. The resulting noise images are shown in \autoref{fig:common_noise}. It is evident that similar to \citep{Eastwood2017} we also have concentric rings at varying \ac{DEC}, that correspond to modes of $m\leq1$. For noise to smear out like this on the sky the source has to be common across all elements or stationary in nature relative to the array, likely this is therefore a product of terrestrial noise or common-mode pickup interfering with our 24 hour observation's signal path. These noise artifacts range from -7.7\,K to 2.6\,K and hence are small relative to our sky signal.

\begin{figure}
    \centering
    \includegraphics[width=\columnwidth]{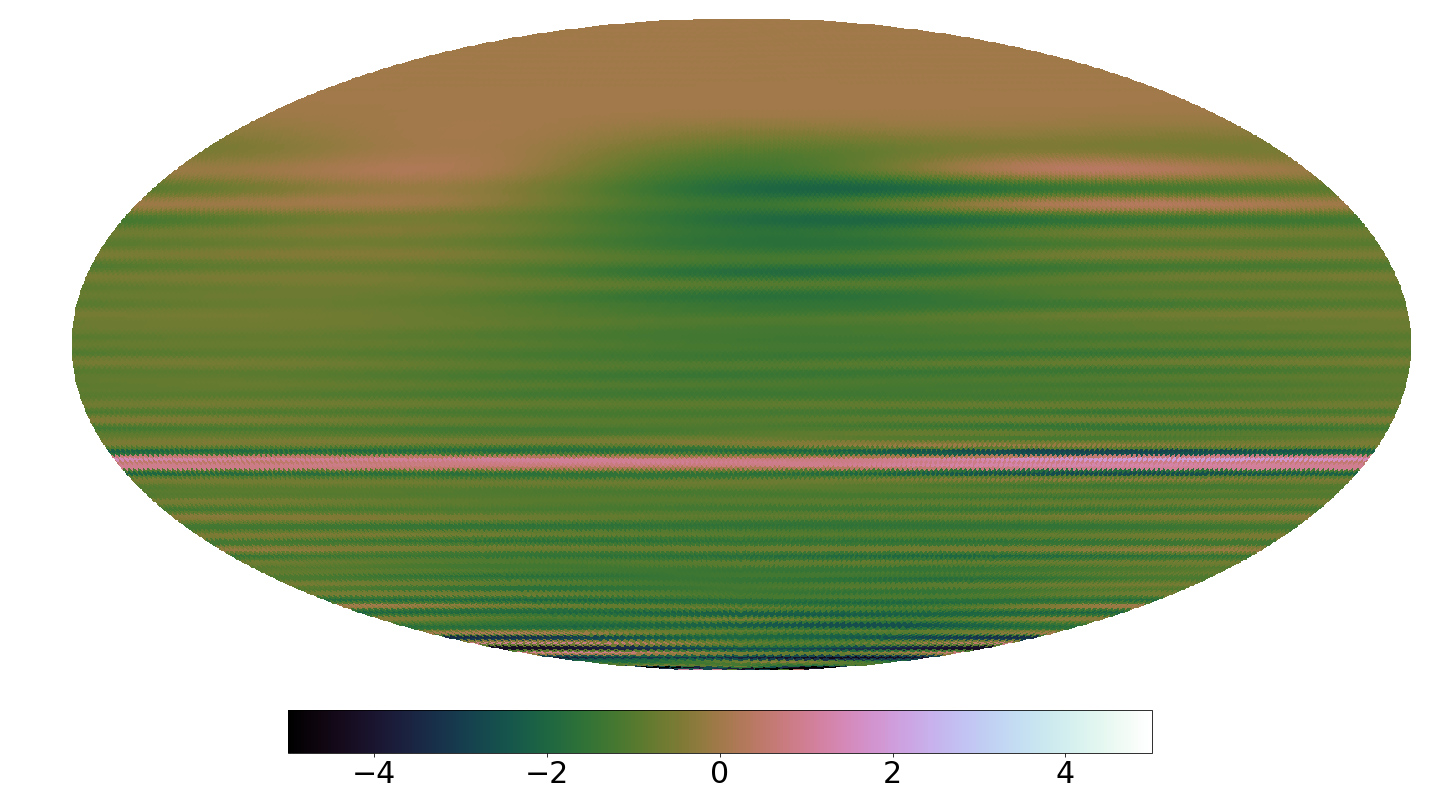}
    \caption{Map of the total systematic noise we removed from our final intensity maps in Kelvin. This map was generated by calculating the difference between a noise-corrected version of our final maps and an uncorrected version.}
    \label{fig:noisesub}
\end{figure}

To correct for these artifacts, \citet{Eastwood2017} removed all the $m\leq1$ modes and $l>100$ spherical harmonic coefficients from their sky maps. However, doing so has the risk of also throwing away actual sky information. Instead, we decomposed all noise maps and subtracted their actual contributions in $m=0$ and $m=1$ from our spherical harmonic sky coefficients in both the prior and non-prior fit September and April data; before generating the total intensity map. We do not lose actual sky information, but purely subtract what is present in modes of $m\leq1$ in the noise maps. After generating the total intensity map we subtract the noise-subtracted intensity maps from the same combined intensity maps where we've omitted the noise subtraction. The residual is shown in \autoref{fig:noisesub}, it can be seen no additional artifacts have been introduced.

\subsubsection{Thermal Noise}
We expect the maps to be confusion limited because many hours of data goes into each pixel on the sky. To verify this we have extracted the visibility noise of our two observations, using the $m>1$ modes of our imaged difference visibility data set of \autoref{subsec:noiseCorr}. We measured the \ac{RMS} of the noise in a $10^{\circ}$ region independently in the September and April data; at $49.8^{\circ}$\,\ac{RA}, $-13^{\circ}$\,\ac{DEC} and $204.7^{\circ}$\,\ac{RA}, $-13^{\circ}$\,\ac{DEC} respectively (which is well away from the galactic plane and other strong sources). This resulted in a noise estimate of 0.024\,K for September and 0.037\,K for April. The magnitude of the noise measurements for X and Y were virtually identical. We compared these values to the standard deviations we extracted from the final sky map in the same regions; which are 10.06\,K and 14.8\,K respectively.

\begin{figure}
    \centering
    \includegraphics[width=\columnwidth]{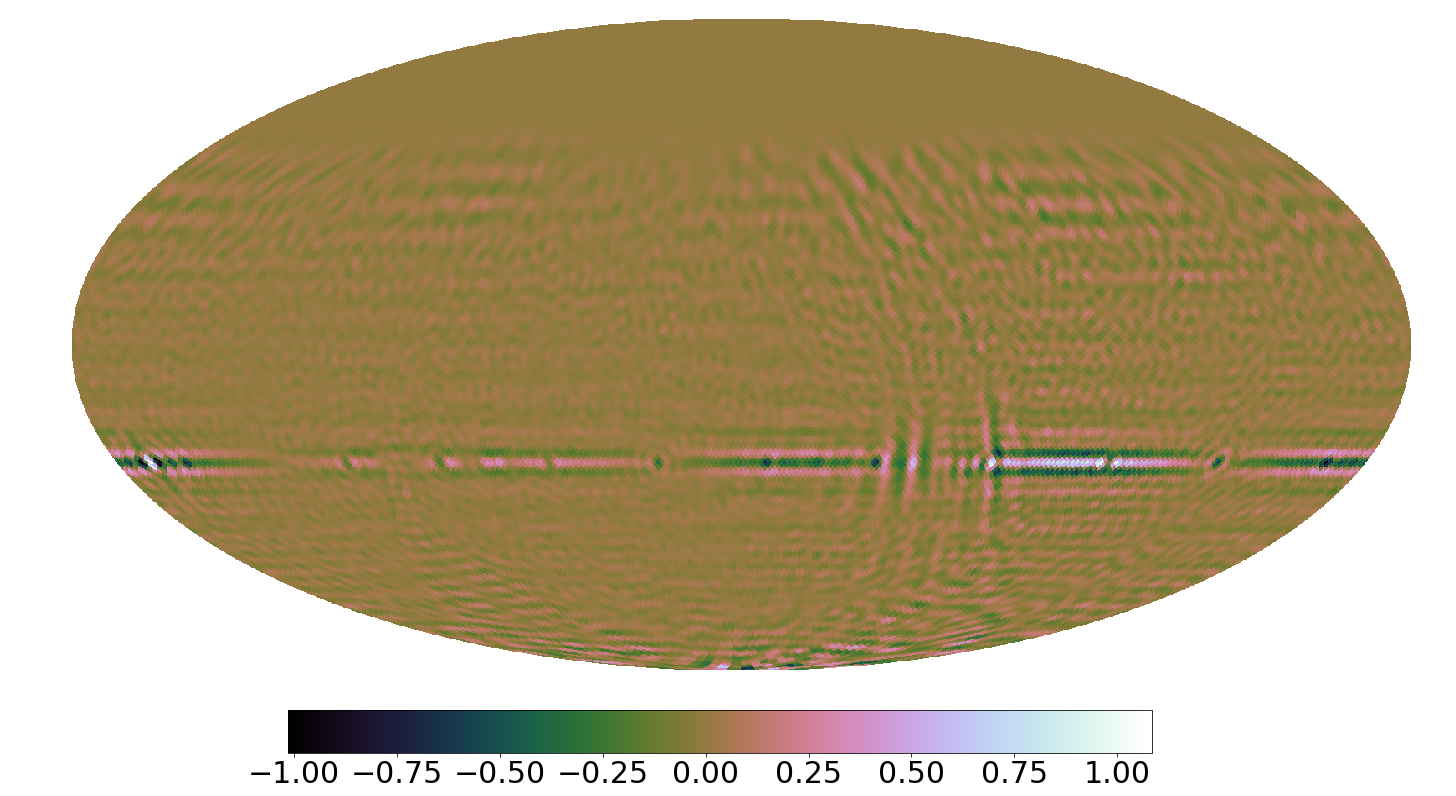}
    \caption{Noise intensity map in Kelvin, generated after removing the $m\leq1$ modes from the noise maps in \autoref{fig:common_noise}, then applying bias correction and weighted averaging as is performed on the sky maps. }
    \label{fig:resnoise}
\end{figure}

A total intensity noise map with $m\leq1$ subtracted is shown in \autoref{fig:resnoise}. This map is bias corrected on the X and Y polarisations for both September and April data and is then weighted averaged together, identical to as is performed on the final sky maps. The total map \ac{RMS} is measured to be 0.073\,K. However, a clear systematic residual, albeit small in overall scale, is present in our September noise data between and RA of 19 and 21 hours. We've measured the RMS of this systematic to be 0.103\,K. Furthermore, there is a bright streak around -26.7$^\circ$ in declination (zenith), which is likely a source of interference. The RMS of this streak is 0.24\,K.

\begin{figure}
    \centering
    \includegraphics[width=\linewidth]{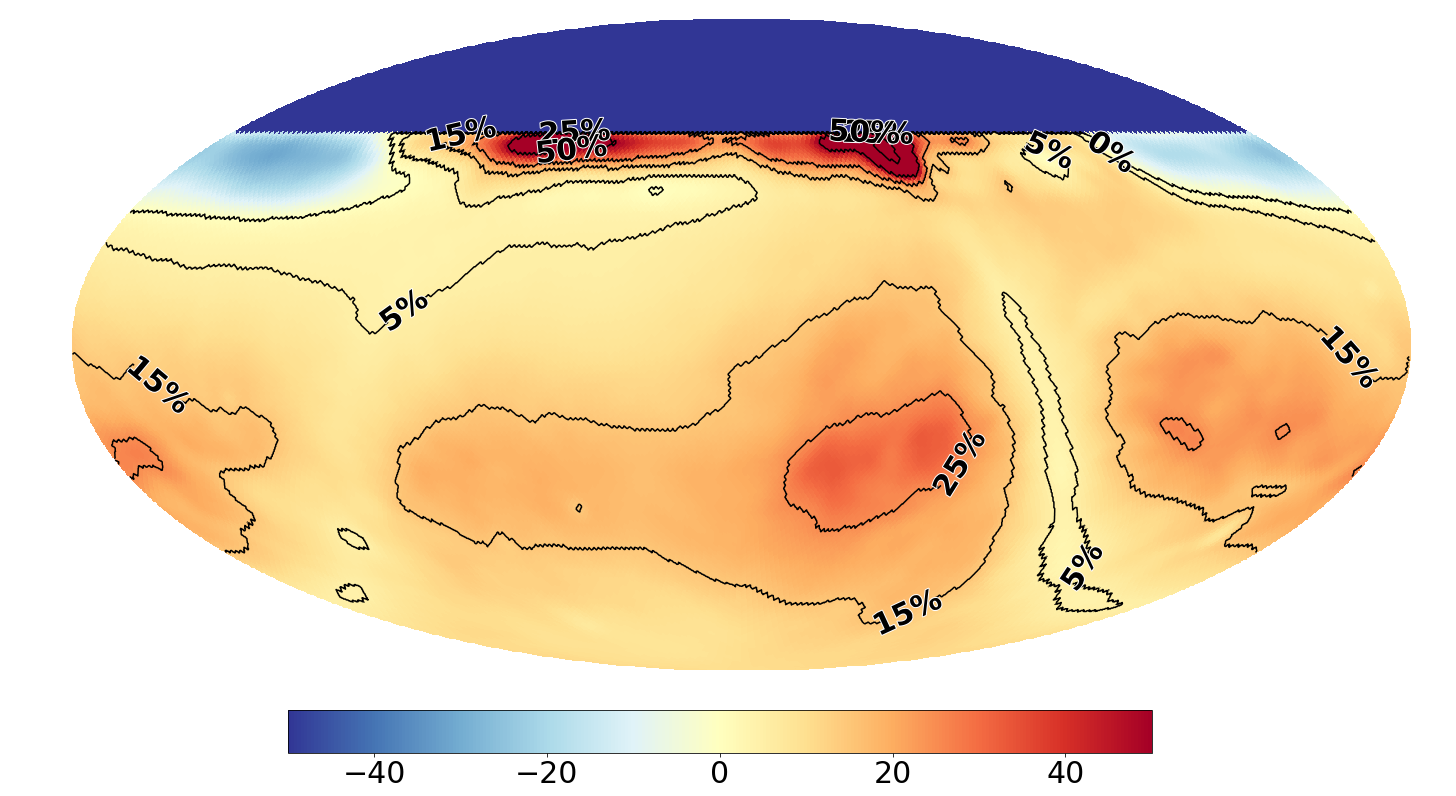}
    \caption{Relative difference between our prior-fit and unconstrained map (in \%, equatorial projection). Large scale diffuse emission matches between maps within 5\%, the galactic plane is in agreement too. Primarily the diffuse emission around the galactic plane is upscaled by the prior with 15\%-30\% more contribution.}
    \label{fig:celDiff}
\end{figure}

\subsection{Effects of Incorporating the Prior}
To show where the prior imposes constraints on the sky, a relative fraction difference map has been made between the unconstrained map and the map with the prior. We calculated this difference map by subtracting our non-prior map from our prior constrained map and then dividing by our non-prior map. It should be noted that in the unconstrained map a global sky component has been inserted to properly subtract the maps from each other, a sky temperature of 247\,K has been selected in accordance to global sky measurements in \citet{McKinley2020}. The resulting difference map is depicted in \autoref{fig:celDiff}, here it can be seen that both of our non-prior and prior fit sky maps seem to be in agreement on the galactic plane and most diffuse regions with up to 5\% difference. The prior does seem to impose constraints on the diffuse emissions around the galactic plane, increasing temperatures between 15\% -- 25\% relative to the non-prior fit map. The bright area at the top is likely to be an artifact of our weighting scheme shown in \autoref{fig:epoch-weight} where we down-weigh the sky towards the global sky component in the non-prior fit map, which has added diffuse galactic plane emissions in the prior fit map; resulting in a temperature discrepancy.

\subsection{Bias Correction}
\label{subsec:biasCorr}
The process of regularisation inevitably introduces bias when minimising the overall cost function. To calculate the bias introduced, we simulated visibilities with a known input map and compared the resulting output which we constrained with a desourced version of the input. For the known map, we used the Haslam map by \citet{Remazeilles2015}, which we reprojected to equatorial coordinates and rescaled to our \ac{HEALPix} pixel grid; similar to how we created the prior for the \ac{EDA2} map, but excluding the Gaussian smoothing. We subsequently generated two sets of images of the sky, separated by 1 minute of LST, whilst embedding the \ac{EDA2} X and Y polarisation beam patterns; this to cover a full sidereal day. Using these beam-weighted sky models, we employed Miriad \citep{Sault1995} using the \ac{EDA2}'s layout to generate 24 hours worth of visibilities which were provided to our $m$-mode pipeline as input to image the sky. Given the input model map is at the same frequency and angular scales as the \ac{EDA2} 159\,MHz sky map, we do not expect the Tikhonov factor to deviate and kept $\varepsilon=0.1$ for the imaging process.

\begin{figure*}
    \centering
    \begin{subfigure}[b]{0.9\columnwidth}
        \includegraphics[width=0.9\linewidth]{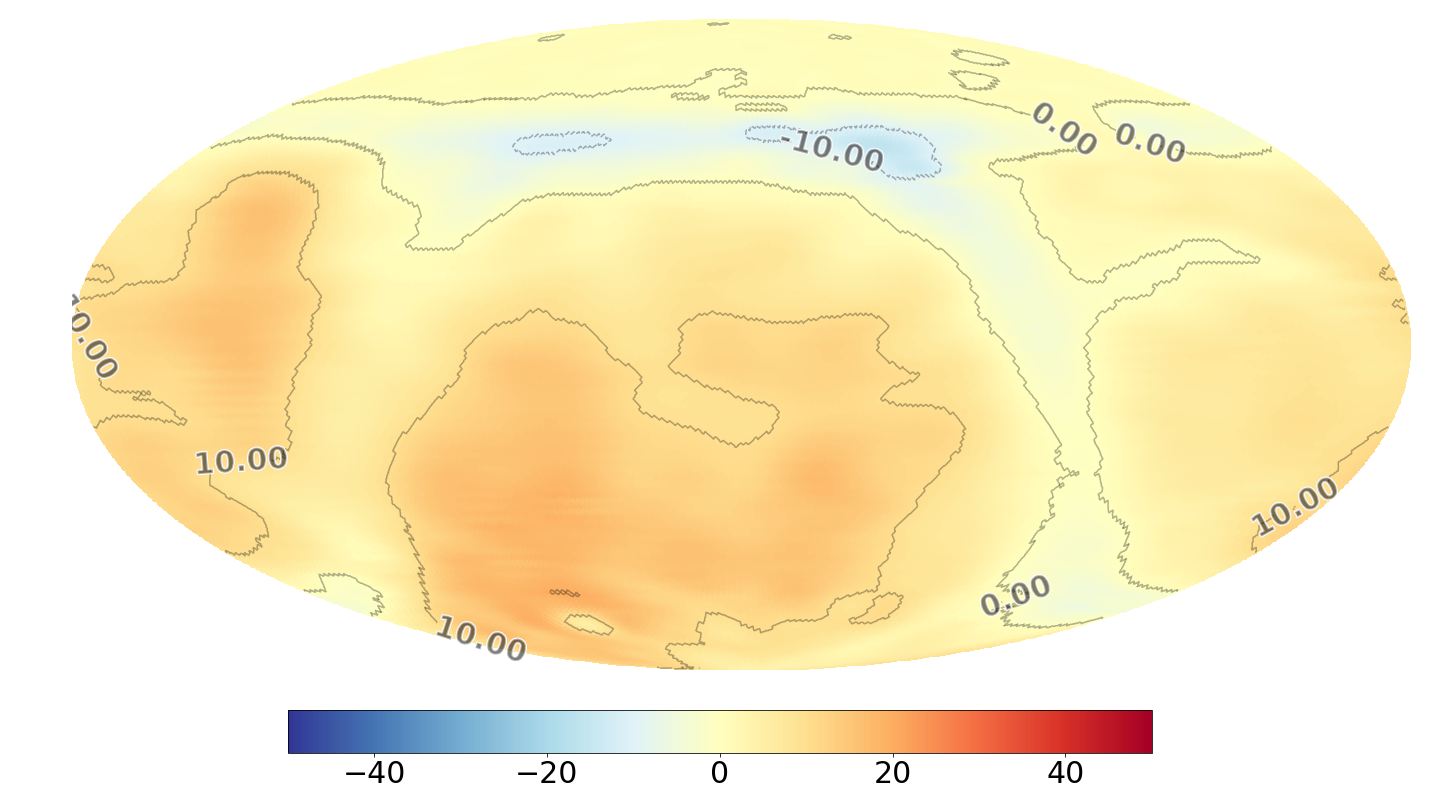}
        \caption{X-pol Bias as fractional difference in percentage \%. Calculated by subtracting the Haslam Map from the X-pol output map and dividing it by the Haslam map.}
    \end{subfigure}
    \hfill
    \begin{subfigure}[b]{0.9\columnwidth}
        \includegraphics[width=0.9\linewidth]{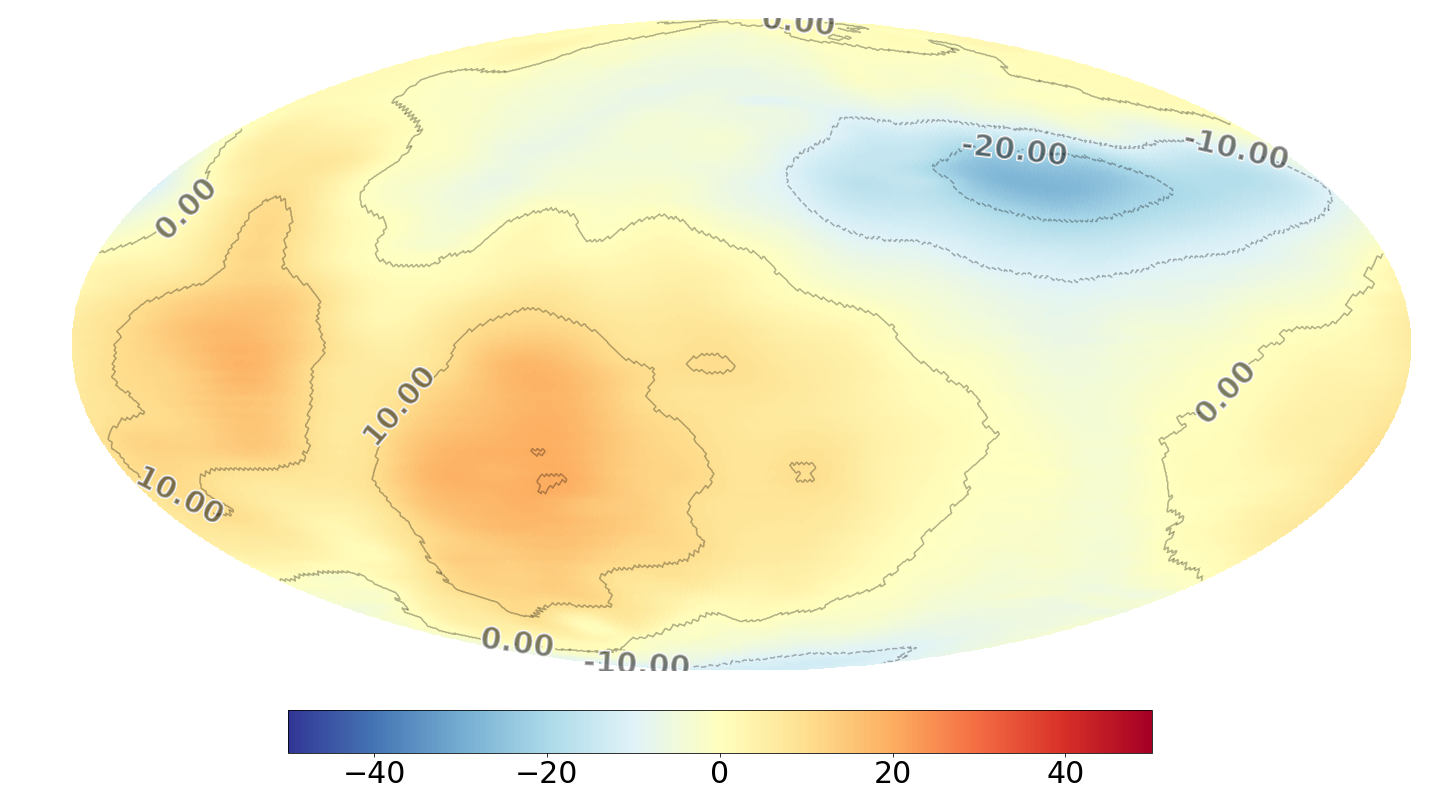}
        \caption{Y-pol Bias as fractional difference in percentage \%. Calculated by subtracting the Haslam Map from the Y-pol output map and dividing it by the Haslam map.}
    \end{subfigure}
    \caption{Average bias between our known Haslam input map and our prior fit output maps (in \%, equatorial projection). Left: X-polarisation bias, Right: Y-polarisation bias. A clear dipole effect is presents with 10\% deviation.}
    \label{fig:biasmap}
\end{figure*}

The resulting output X-polarisation and Y-polarisation maps we used to calculate the percentage fractional difference relative to the known input map. This sets all correct values on the sky to zero and any bias shows up as a percentage offset. These bias maps are shown in \autoref{fig:biasmap}. From \autoref{fig:biasmap} it is evident a dipole-like structure is present overestimating the one side of the image and underestimating the other. This can be clearly attributed to improper constraints on the $l=1$ spherical harmonic coefficients; smaller angular scales on the sky match the input on the sky and therefore do not appear in the bias. These offsets range from +10\% at its peak to -10\% at its lowest for the X-polarisation, similar for the Y-polarisation with a small negative region of -20\% offset.

Since we divided out the true sky, \textit{i.e.} our known input, we can correct for the bias as the bias map only represents relative over- and under estimations. We apply these correction through employing the bias maps as weighting schemes on our noise-corrected X-polarisation and Y-polarisation maps before generating our weighted intensity maps. The weighting scheme is defined by

\begin{align}
    I_{\rm corrected} &= I_{\rm sky} \times \left(1-\frac{I_{\rm bias}}{100}\right)\,.
\end{align}

\noindent This allows us to down weigh the overestimated regions on the sky and upscale the underestimated ones.

\section{Analysis}
In this section we compare how the non-prior map and prior constrained \ac{EDA2} 159\,MHz map compares to existing sky models. We compare the maps to the frequently used \ac{GSM} both the reprocessed version from 2016~\citep{Zheng2017} and the 2008 version~\citep{OLIVEIRA-COSTA2008}, the reprocessed desourced Haslam map~\citep{Remazeilles2015}, and the \ac{LWA}1 \ac{LFSS}~\citep{Dowell2017}. For all these sky models we use pyGDSM~\citep{Price2016} to rescale the maps to 159\,MHz to which we then applied NSIDE rescaling and smoothing to the \ac{EDA2} angular resolution; prior to comparing the data. We compare the \ac{EDA2} 159\,MHz maps to the sky models by calculating the relative fractional difference between the two maps. This is defined by

\begin{align}
    \label{eq:fracdiff}
    I_{\rm diff,\%} &= 100\% \times \left(\frac{I_{\rm EDA2}}{I_{\rm Model}}-1\right)\,,
\end{align}

\noindent Positive values indicate the \ac{EDA2} maps are brighter, negative values indicate the sky model maps are brighter. 

\subsection{2008 Global Sky Model}
\begin{figure}
    \centering
    \includegraphics[width=\columnwidth]{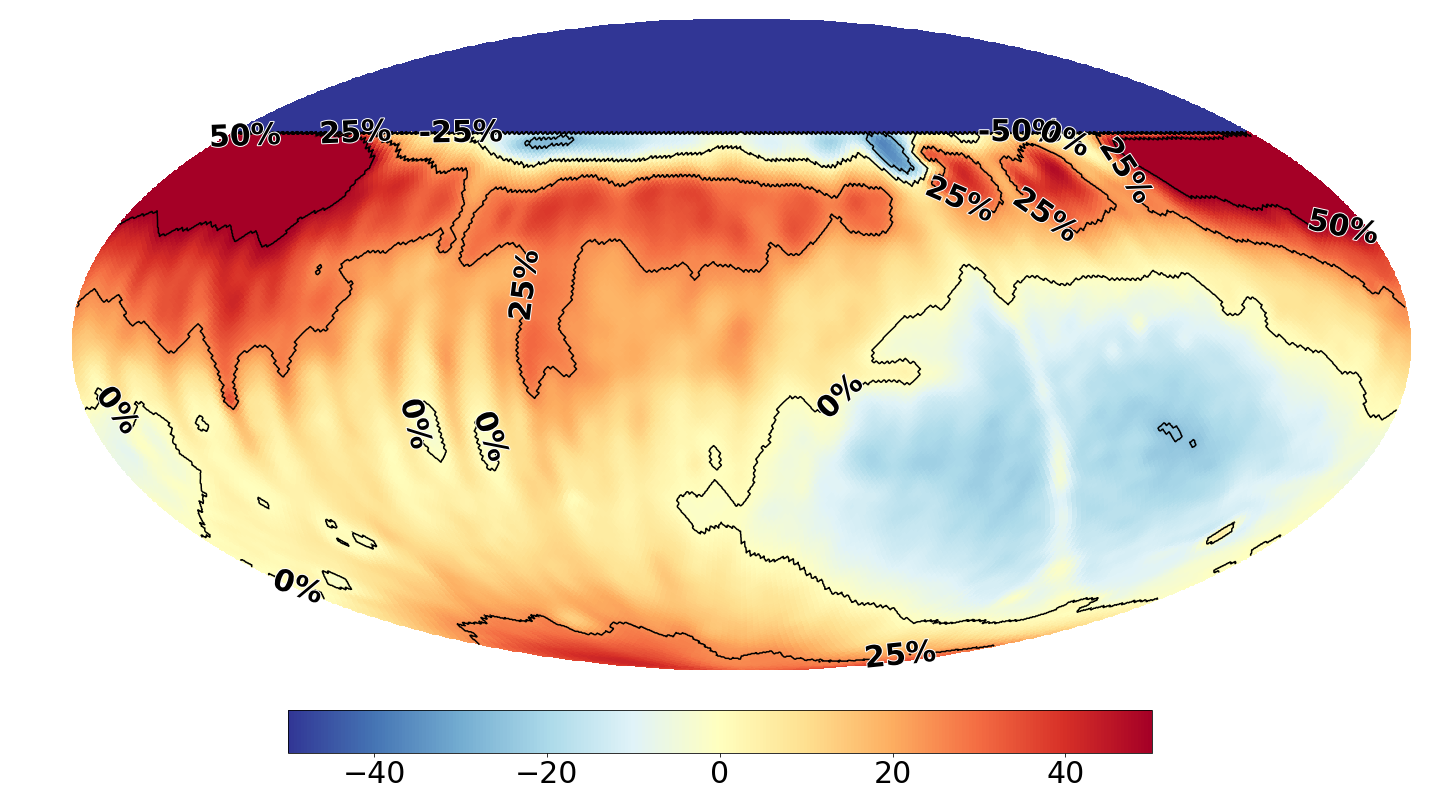}
    \caption{Comparison between our non-prior \ac{EDA2} 159\,MHz sky map and the 2008 \ac{GSM} of \citep{OLIVEIRA-COSTA2008} rescaled to 159\,MHz; in percentage and equatorial coordinates. The comparison is made by dividing our sky map by the \ac{GSM} at 159\,MHz and is then offset by 1 to put zero difference on regions that agree.  Contours have been overlaid to show a difference in scales across the map. Our map shows 18\% less contribution in the diffuse emission around the Galactic Centre, but is generally approximately 12\% brighter.}
    \label{fig:noprior_GSM2008}
\end{figure}

\begin{figure}
    \centering
    \includegraphics[width=\columnwidth]{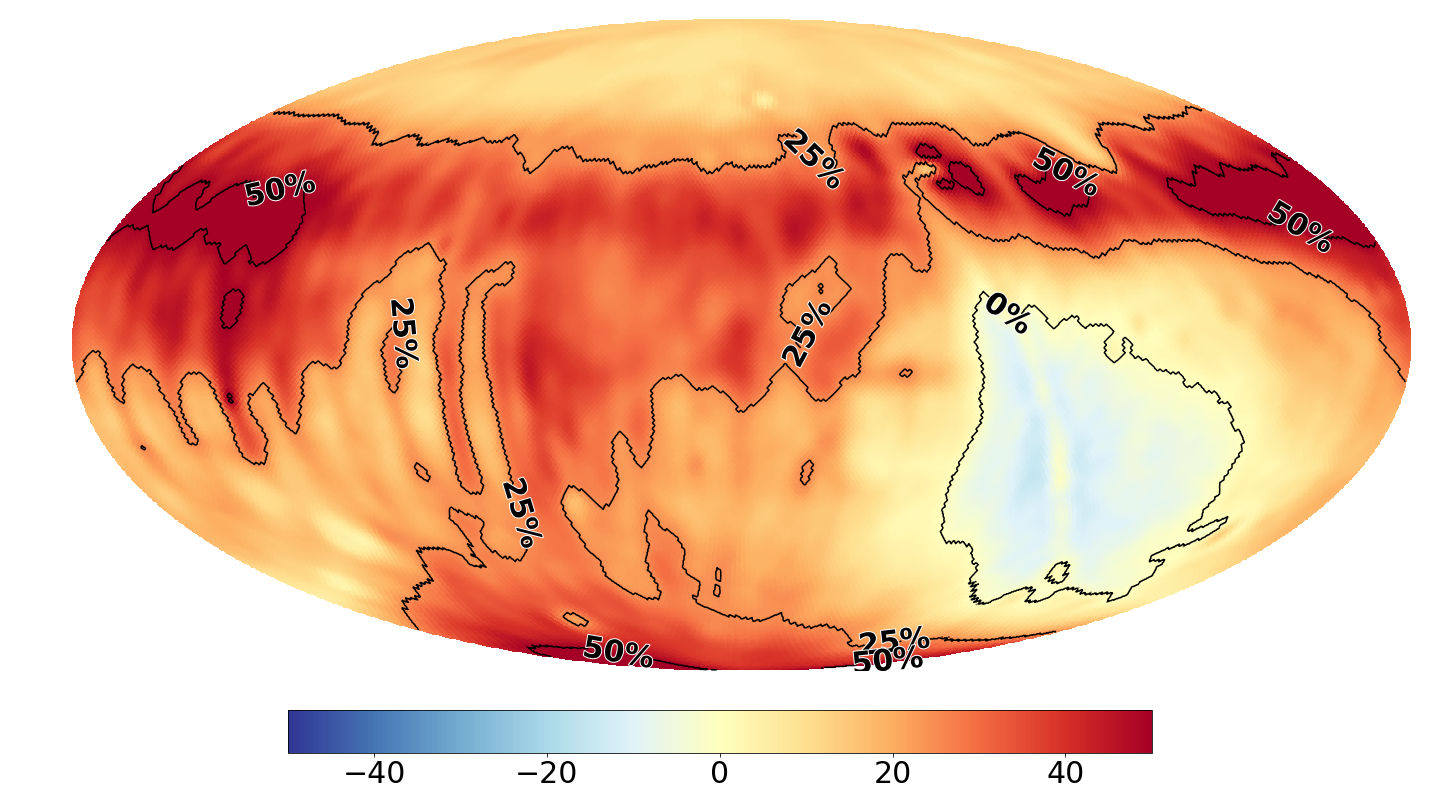}
    \caption{Comparison between our Haslam prior constrained \ac{EDA2} 159\,MHz sky map and the 2008 \ac{GSM} of \citep{OLIVEIRA-COSTA2008} rescaled to 159\,MHz; in percentage and equatorial coordinates. The comparison is done using the same method as in to \autoref{fig:noprior_GSM2008}. With the prior-fit map, the diffuse emission around the galactic plane matches better; ranging from 0\% to 10\% difference. However, in general, our prior-fit sky map is approximately 25\% brighter.}
    \label{fig:prior_GSM2008}
\end{figure}

The difference maps between our \ac{EDA2} maps and the 2008 \ac{GSM}~\citep{OLIVEIRA-COSTA2008} are shown in \autoref{fig:noprior_GSM2008} and \autoref{fig:prior_GSM2008}, we compare the non-prior vs the sky model and the prior vs the sky model respectively. Since the pyGDSM rescaled 2008 \ac{GSM} map is desourced we mask off any bright sources in our \ac{EDA2} map, as well as the region in the sky we do not observe in our map that has not been constrained by a prior. \autoref{fig:noprior_GSM2008} shows that generally we are 12\% higher in temperature compared to the \ac{GSM} off the Galactic Centre. Near the Galactic Centre, we notice we are on average 18\% lower in diffuse emission. \citet{Dowell2017} noticed a similar effect in their comparison to the \ac{GSM}. The difference is likely a cause of improper constraints on the free-free emission in the higher frequency maps that primarily make up the sky model. There is also a clear striping artifact present in the fractional difference maps between an \ac{RA} of approximately 5 and 12 hours, which is an imaging artifact in the \ac{GSM}, likely caused when combining the individual maps to make up the model.

Comparing our Haslam prior constrained \ac{EDA2} sky map to the 2008 \ac{GSM} at 159\,MHz shows that the prior constrains the diffuse emission around the Galactic Centre and more closely matches the diffuse emissions from the \ac{GSM} with between 0\% to 10\% difference. However, anywhere else our prior constraint increased our relative brightness compared to the \ac{GSM} to 25\% difference on average. It should be noted that for both the non-prior and prior constrained \ac{EDA2} sky maps we have 25\% more contribution around the southern celestial pole.

\subsection{2016 Global Sky Model}
\begin{figure}
    \centering
    \includegraphics[width=\columnwidth]{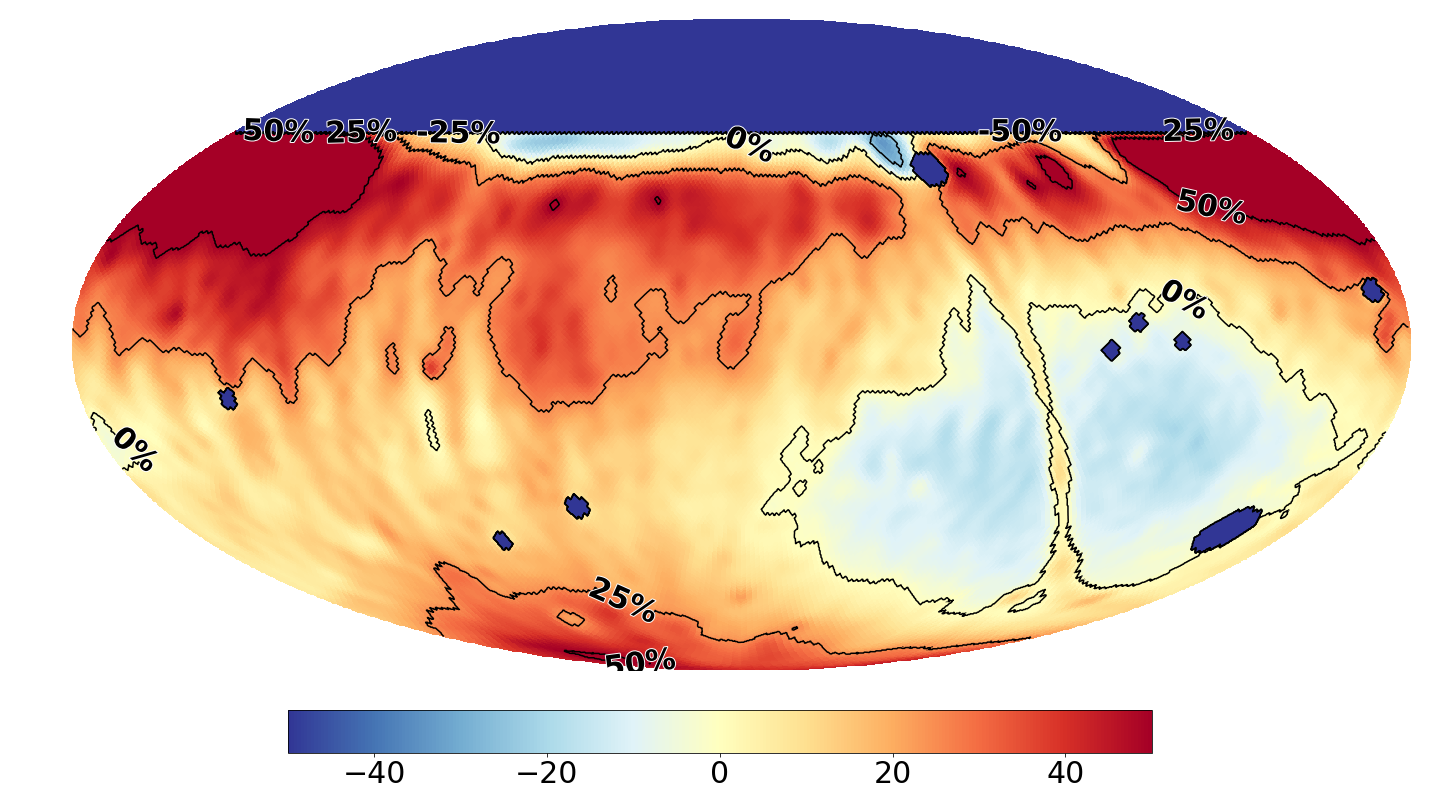}
    \caption{Comparison between non-prior \ac{EDA2} 159\,MHz sky map and the 2016 \ac{GSM} of \citep{Zheng2017} rescaled to 159\,MHz; in percentage and equatorial coordinates. In general, our map is on average 17\%-25\% brighter, but our maps closer resembles the diffuse emission around the Galactic Centre, with an average of 12\% difference, compared to the 2008 \ac{GSM}}
    \label{fig:noprior_GSM2016}
\end{figure}

\begin{figure}
    \centering
    \includegraphics[width=\columnwidth]{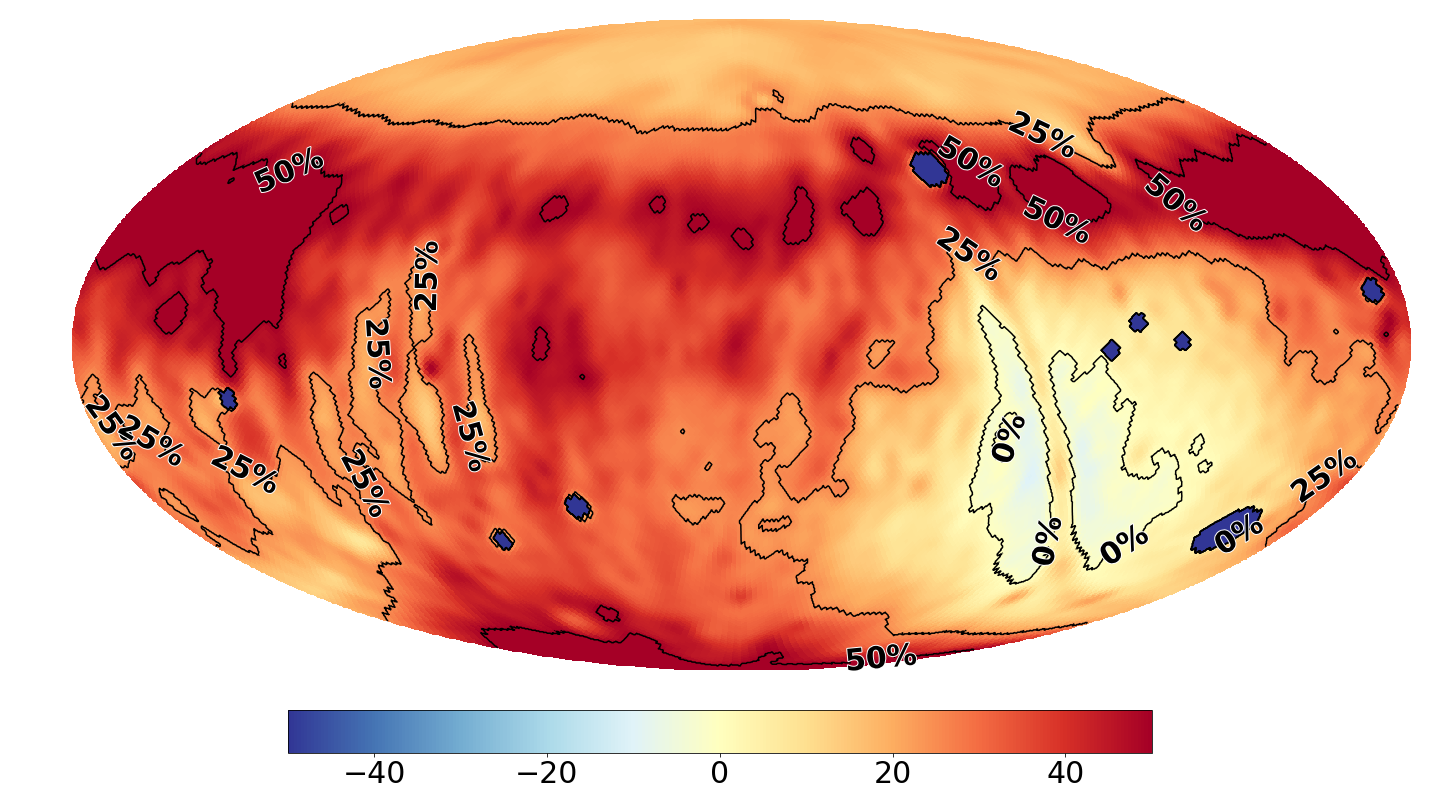}
    \caption{Comparison between our Haslam prior constrained \ac{EDA2} 159\,MHz sky map and the 2016 \ac{GSM} of \citep{Zheng2017} rescaled to 159\,MHz; in percentage and equatorial coordinates. We have a consistent offset of 25\%, however are more in agreement with the galactic plane with 3\% difference on average.}
    \label{fig:prior_GSM2016}
\end{figure}

We also generated the fractional differences between the \ac{EDA2} non-prior and prior constrained map and the 159\,MHz 2016 \ac{GSM} of \citet{Zheng2017}. The difference maps are shown in \autoref{fig:noprior_GSM2016} and \autoref{fig:prior_GSM2016}. \autoref{fig:noprior_GSM2008} shows that in the no-prior map we are 17\% brighter on average, with regions of 25\% difference under the galactic plane. The reprocessed \ac{GSM}, however, is more in agreement with our maps in the diffuse emission around the Galactic Centre, with an average of 12\% difference. Imaging artifacts from the 2008 \ac{GSM}, although less prevalent, are still present.

The prior constrained \ac{EDA2} sky map closer matches the diffuse emission around the Galactic Centre, up to a few percent difference. However, the constrained map is on average 25\% brighter than the 2016 \ac{GSM}. The south celestial poles for both \ac{EDA2} sky maps show up to 25\%-50\% difference compared to the 2016 \ac{GSM}. Comparing this to \ac{GSM} comparisons with the \ac{LWA} sky maps by \citet{Dowell2017}, we see differences of similar magnitudes in the diffuse emission regions.

\subsection{2014 Reprocessed Haslam map}
\begin{figure}
    \centering
    \includegraphics[width=\columnwidth]{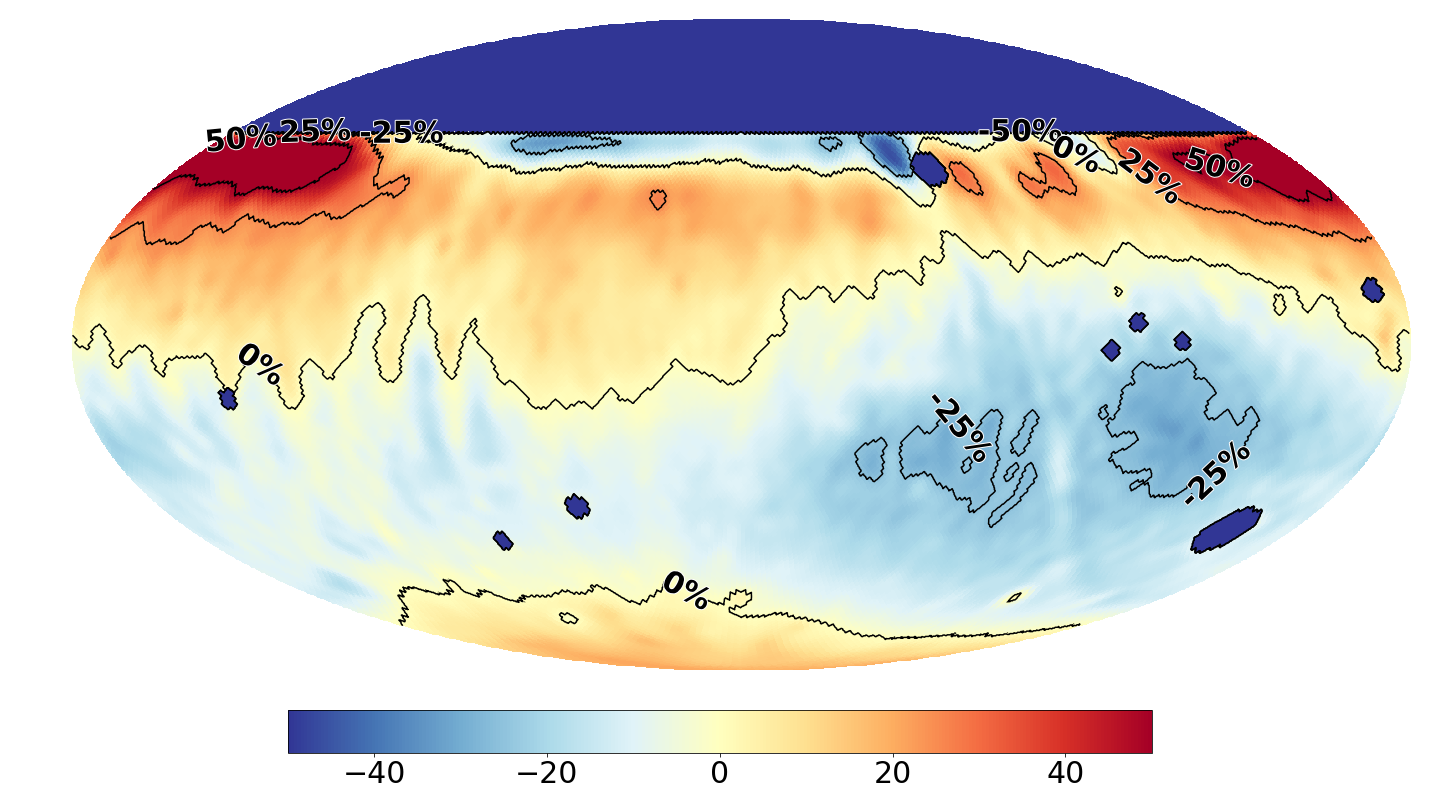}
    \caption{Comparison between non-prior \ac{EDA2} 159\,MHz sky map and the desourced 2014 reprocessed Haslam map~\citep{Remazeilles2015} rescaled to 159\,MHz; in percentage and equatorial coordinates. In general we are better in agreement compared to both \acp{GSM} and are on average -3\% -- 6\% different. However, the Haslam also shows excess (25\%) in galactic diffuse emission near the Galactic Centre compared to the \ac{EDA2}}
    \label{fig:noprior_Haslam}
\end{figure}

Since we constrain our map with the desourced reprocessed 2014 Haslam map by \citet{Remazeilles2015}, we compare our \ac{EDA2} sky maps to the Haslam map to make sure we do not force our map to be identical to Haslam. In \autoref{fig:noprior_Haslam} we compare our non-prior \ac{EDA2} map with the Haslam map. We are 3\% -- 6\% less bright on average and 6\% brighter at declinations higher than $30^{\circ}$. However, the Haslam map is up to 25\% brighter in the diffuse emission around the Galactic Centre.

\begin{figure}
    \centering
    \includegraphics[width=\columnwidth]{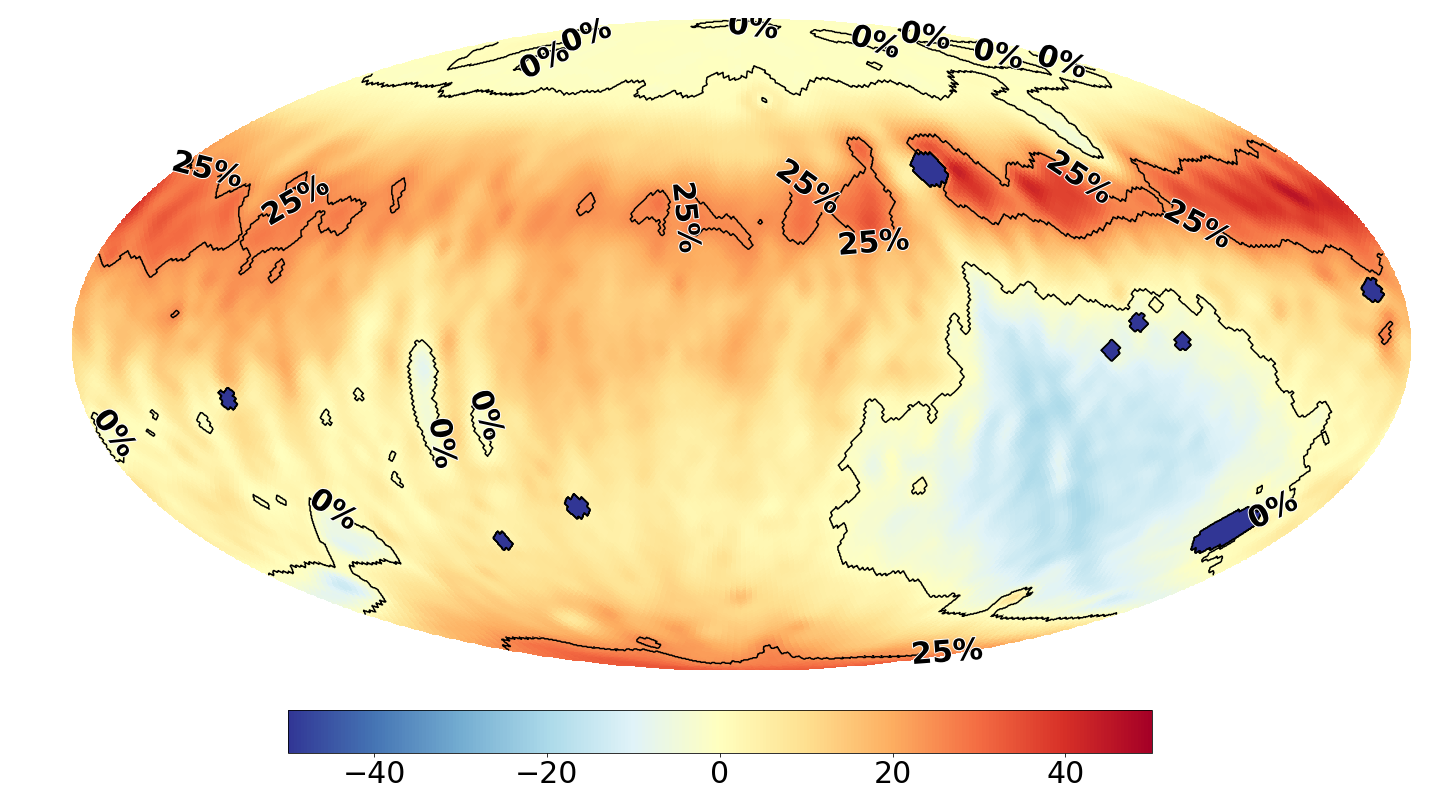}
    \caption{Comparison between our Haslam prior constrained \ac{EDA2} 159\,MHz sky map and the desourced 2014 reprocessed Haslam map~\citep{Remazeilles2015} rescaled to 159\,MHz; in percentage and equatorial coordinates. We have better overall agreement with 3\% -- 6\% difference, which is expected as we use the same map to fit the prior. We also closely match the galactic plane with an average of 3\% in excess. However, we see excess emissions in our map up to 12\% -- 25\% at declinations $\ge30^{\circ}$.}
    \label{fig:prior_Haslam}
\end{figure}

For our prior constrained map (\autoref{fig:prior_Haslam}) we are in overall better agreement. This is expected since we use the Haslam map as a prior to constrain our coefficients. However, there are still differences. In most regions above the southern celestial pole and below 30$^{\circ}$ declination we are on average 3\% -- 6\% brighter than Haslam. The overall contribution of \ac{EDA2} increases up to 12\% -- 25\% above 30$^{\circ}$ declination. We also match the galactic plane closely with 3\% difference. The diffuse emissions around the Galactic Centre more closely matches the Haslam map, where we are 7\% less bright on average. The Southern celestial pole still shows similar contribution of 10\% -- 25\% brighter compared to Haslam.

Similar to the 2008 and 2016 \ac{GSM} we see similar striping between PicA and VirA in the Haslam map; albeit more subtle.

\subsection{LWA1 Low-frequency Sky Survey}
\begin{figure}
    \centering
    \includegraphics[width=\columnwidth]{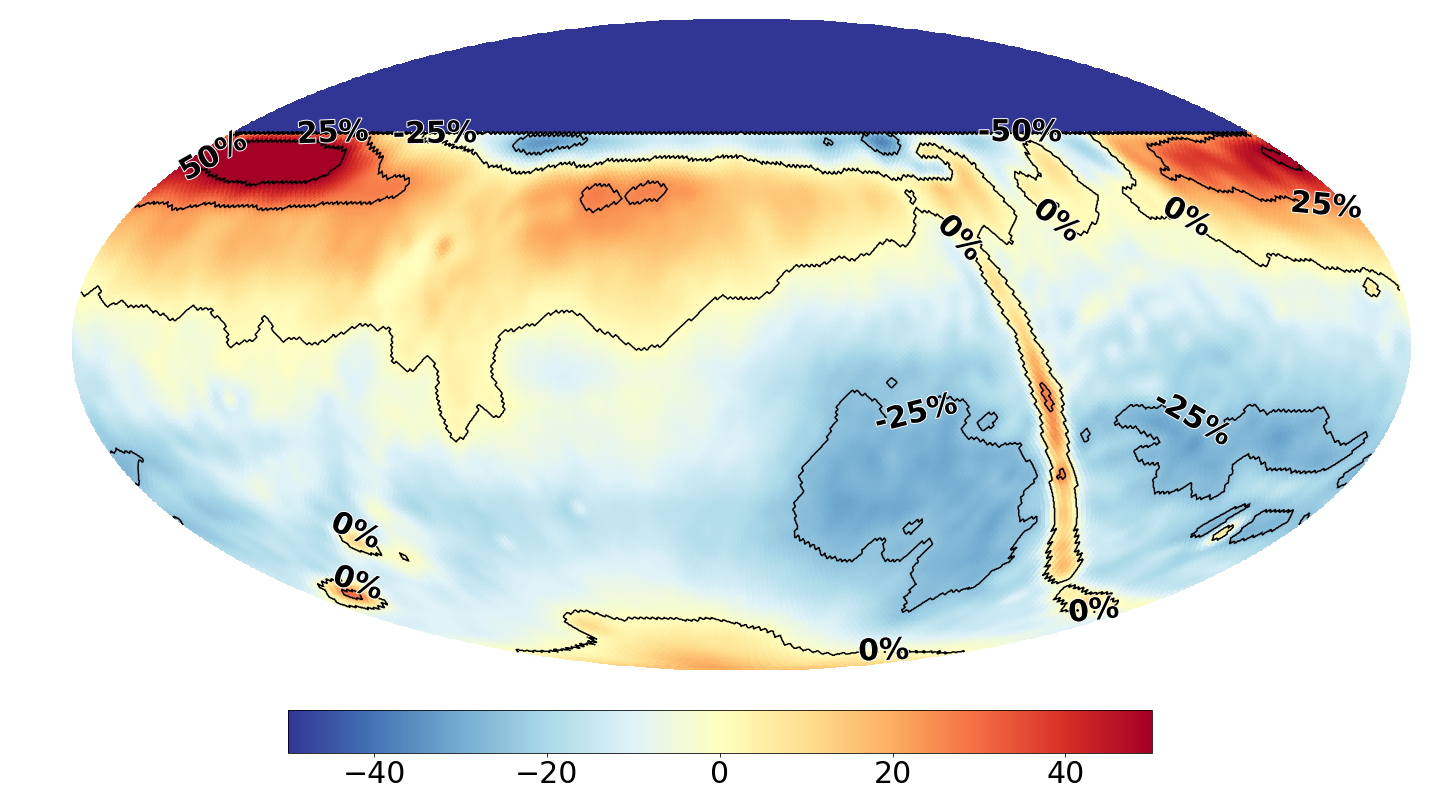}
    \caption{Comparison between our non-prior \ac{EDA2} 159\,MHz sky map and the \ac{LFSS} \citep{Dowell2017} rescaled to 159\,MHz; in percentage and equatorial coordinates. We are lesser in agreement around the galactic plane, where we are 25\% less in contribution. This is likely caused do to the fact the \ac{LFSS} is more diffuse compared to the \ac{EDA2} map. Furthermore, the \ac{LFSS} has on average 25\% more contribution in the diffuse emissions around the galactic plane compared to the non-prior fit map. However, in all other regions on the sky we seem in overall better agreement than compared to any other sky model where we have between 0\% -- 10\% difference  on average.}
    \label{fig:noprior_LFSS}
\end{figure}

To see how the \ac{EDA2} sky maps compares to sky maps made at lower frequencies, we compare the \ac{EDA2} maps to the \ac{LWA}1 \ac{LFSS} by \citet{Dowell2017}. For our non-prior map comparison (\autoref{fig:noprior_LFSS}) we are in good agreement with the \ac{LFSS} with differences between 0\% to 10\%; except for near the galactic plane. In the galactic plane we have 25\% more contribution compared to the \ac{LFSS}, we expect this due to H$_{\rm II}$ regions in the galactic plane. We do not see any of the imaging artifacts we identified in the Haslam and \ac{GSM} maps.

\begin{figure}
    \centering
    \includegraphics[width=\columnwidth]{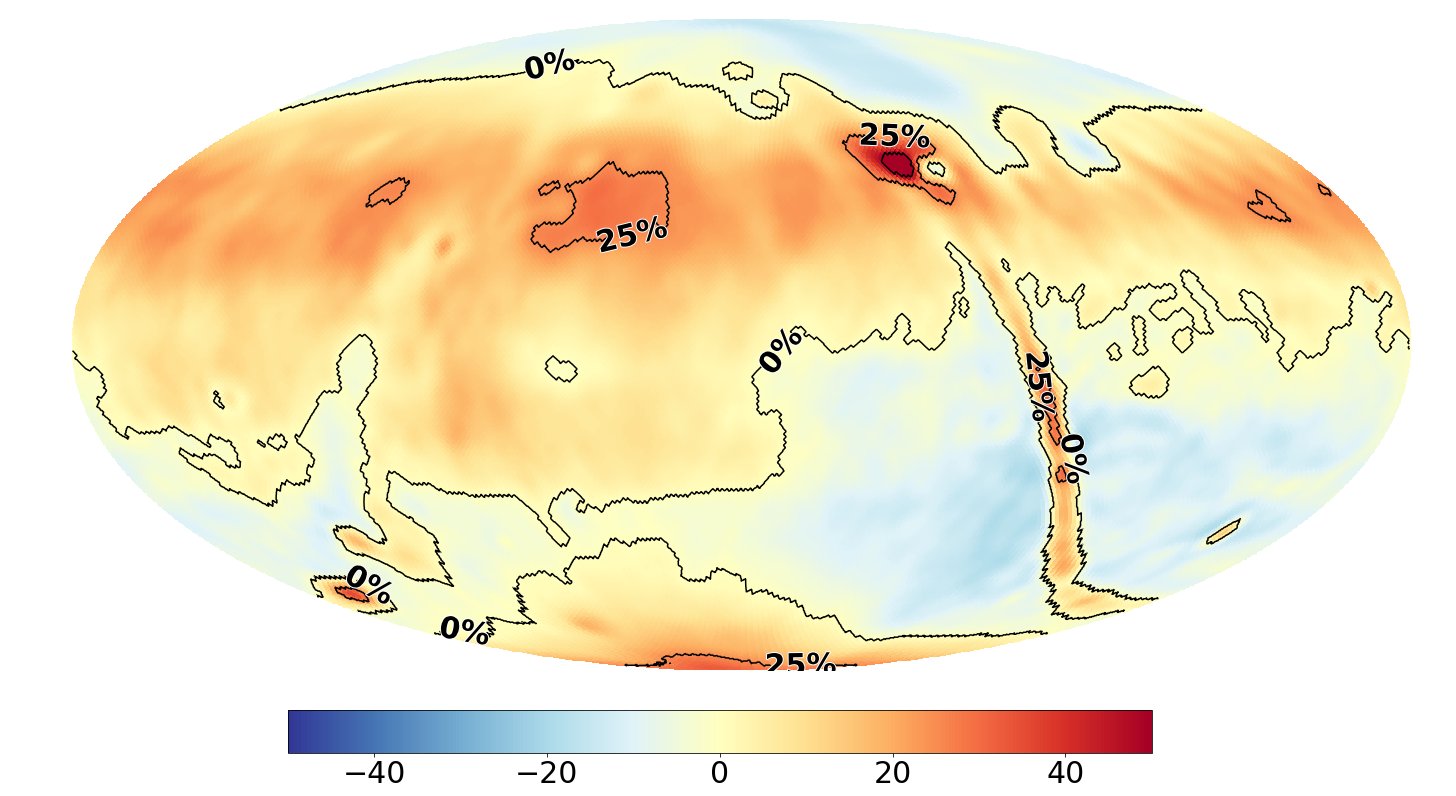}
    \caption{Comparison between the prior fit \ac{EDA2} 159\,MHz sky map and the \ac{LFSS} \citep{Dowell2017} rescaled to 159\,MHz; in percentage and equatorial coordinates. We have much better agreement compared to all other sky models, bar the galactic plane. In general we seem to closely match the diffuse emissions up to 40$^{\circ}$ in declination, where we have -4\% -- 4\% difference.}
    \label{fig:prior_LFSS}
\end{figure}

The prior constrained \ac{EDA2} sky map is in much better agreement with the \ac{LFSS} in general. In \autoref{fig:prior_LFSS} we closely follow the \ac{LFSS} up to declinations of $45^{\circ}$ and have on average -4\% -- 4\% difference. we have on average 7\% less contribution in diffuse emission around the Galactic Centre, however still see up to 25\% in excess on the galactic plane. The 50\% difference at Vela is likely a product of \ac{SI} between the Haslam prior and the \ac{LFSS}; since we do not see it in our map that is not prior constrained. It should be noted that compared to the \ac{LFSS} the \ac{EDA2} sky maps also show more contributions at the south celestial pole, which are again 10-25\% brighter.

\subsection{A Comment on Systematics}
\label{subsec:systematics}
To determine whether the relative fractional differences are in the same order of magnitude compared to the differences between the sky models we compare to, we calculated the mean temperatures in a 10 degree radius in the diffuse region at 1 hour in \ac{RA} and $-26.7^{\circ}$ in \ac{DEC} (zenith). These temperatures range from 202\,K for the 2016 \ac{GSM}~\citep{Zheng2017} to 247\,K for the \ac{LFSS}~\citep{Dowell2017}, resulting in an approximate 18\% fractional difference. For our maps, calculate the mean temperatures in the same region. In this region, our no-prior map has a mean temperature of 215\,K and our prior constrained map has a mean temperature of 253\,K. These values fall into the same approximate range of temperatures found when comparing the individual sky models, explaining the difference in fractional difference we see between our maps and the sky models.

When comparing \ac{EDA2} with all aforementioned sky models, we note that the \ac{EDA2} non-prior and prior-constrained sky maps both have a consistent relative offset in temperature most prominent at declinations between $50^{\circ}$ and $60^{\circ}$; which is an apparent systematic bias in our maps. We reiterate that the sky above 40$^{\circ}$ in \ac{DEC} is poorly measured by our system, as shown in \autoref{fig:xpol_beam} and \autoref{fig:ypol_beam} the y-polarisation dipole gain falls more quickly with increasing declination; it reaches 10\% at +32$^{\circ}$ in \ac{DEC} and 5\% at +40$^{\circ}$. We do not expect this to be due to other causes as we did not identify this systematic in our bias maps, as is evident in \autoref{subsec:biasCorr}.

We also note that the morphology of the fractional differences in all of the comparison maps, \textit{e.g.} where our map is less bright around the galactic plane, is similar. This morphology does not match the morphology of the biases seen in \autoref{fig:biasmap}. We therefore interpret these to be genuine differences in the local \acp{SI} of these maps, rather than being due to a systematic effect of the $m$-mode imaging.

\subsection{Spectral Index Map}
\begin{figure}
    \centering
    \includegraphics[width=\columnwidth]{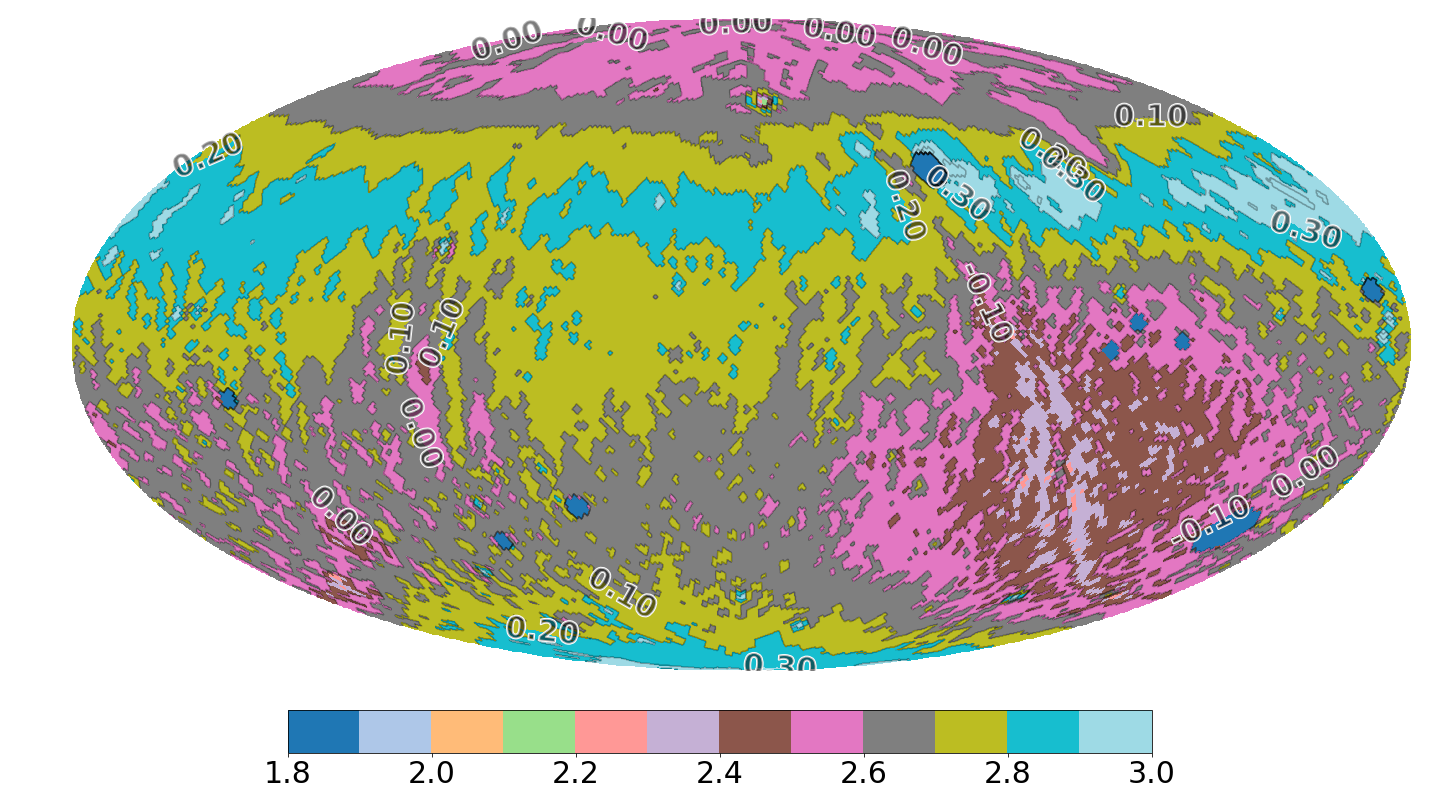}
    \caption{Spectral index map calculated between our prior-constrained \ac{EDA2} map and the desourced Haslam map of~\citet{Remazeilles2015}. We overlaid the absolute difference in \ac{SI} between our map and the Haslam map as labeled contours. Bright sources have been masked off.}
    \label{fig:EDA2vHaslam_SI}
\end{figure}

\autoref{fig:EDA2vHaslam_SI} shows the spectral index derived from our \ac{EDA2} map when rescaled to the 408\,MHz Haslam map which has been reprojected to equatorial coordinates and smoothed to the same angular resolution of 3.1 degrees. The Haslam map has been extracted from pyGDSM~\citep{Price2016} which is assumed to have a flat \ac{SI} of 2.6 across the sky. We see the largest difference in our maps at the galactic plane, where we measure an \ac{SI} of 2.4--2.45 more in accordance to the \acp{SI} derived by~\citet{Dowell2017}. The diffuse emission around the Galactic Centre we calculated to have an SI of 2.5 on average. The remaining diffuse emission ranges from 2.6--2.7 in \ac{SI}. \acp{SI} calculated above $40^{\circ}$ in declination we do not deem trustworthy due to the systematic bias we discussed in \autoref{subsec:systematics}. The reason we see the major difference around the galactic plane is likely due to the fact a single-power law assumption start breaking down when having significant discrepancy in frequency, as $\mathrm{H}_{\rm II}$ regions become more prevalent due to thermal absorption~\citep{Dowell2017, Kassim1989}. To properly define \acp{SI} for the \ac{EDA2} observed sky, more comparisons and observations across multiple frequencies have to be made. However, this is out of scope for this paper and we will address this in future work.

\section{Conclusions}
\label{sec:conclusions}
We present two \ac{EDA2} 159\,MHz all-sky maps generated using the novel imaging method known as the $m$-mode formalism. We have shown how prior-fitting the Tikhonov regularisation, first suggested by \citet{Eastwood2017}, puts constraints on the diffuse emissions on the sky for modes we are not, or less, sensitive to. Furthermore, we have shown how we can remove systematic bias from our sky maps implemented as a weighting scheme, and how we can correct for terrestrial noise by down weighting those specific modes in coefficient space; without compromising actual information on the sky. To generate these maps, two observations have been performed, separated by a 7 month interval, in order to remove the sun. For both observations 24 hours of data have been used. The maps are created with a maximum angular resolution of $3.1^\circ$, within the time-frame of a single day. These maps have $<0.5$\,K measured thermal noise and are super sampled on a 0.91 degree pixel grid.

Between declination range of $-70^{\circ}$ and $+40^{\circ}$ our maps on average have approximately 10\% relative difference compared to the reprocessed Haslam map~\citep{Remazeilles2015} and the \ac{LFSS}~\citep{Dowell2017}; and 25\% difference of the \acp{GSM}~\citep{OLIVEIRA-COSTA2008, Zheng2017}. At higher declinations above 40$^{\circ}$ our maps are not well measured. We also show a consistent 10\%-25\% higher temperature around the southern celestial pole.

We also generated an all-sky thermal noise intensity map with an average \ac{RMS} of 0.073\,K across the observed sky. In this map, there is some systematic/interference between RA of 19 and 21 hours and at $-26.7^{\circ}$ in declination (zenith), with a maximum \ac{RMS} of 0.24\,K. However, compared to the temperature variations on quiet regions on the sky (approximately 10\,K -- 14\,K), the thermal noise (including the systematics) is roughly 2 orders of magnitude lower. Because our maps are confusion limited they would form one part of a complete foreground model for \ac{EoR} sciences, but can be used as-is for single-antenna total power measurements. \citet{Byrne2021} have shown that the angular scale at which power of diffuse emissions and point sources are equal is in the order of several degrees, therefore diffuse sky maps such as these will be required for complete calibration models in the future.

Furthermore, we introduced a variation to the image deconvolution algorithm of \citet{Eastwood2017}. This algorithm operates in image space by taking advantage of the fact the \ac{PSF} is shift-invariant in right ascension.

Finally, we introduced a formalism to inspect the interferometer's sensitivity in spherical harmonic coefficient space, analogous to the $u,v$-plane coverage in traditional interferometry; showing the $m$-mode formalism allows for sensitivity to diffuse emission much larger in angular scales than traditional snapshot imaging provides. We also show how the spherical harmonic beam coverage can aid in better constraining a prior.

\begin{acknowledgements}
We thank Daniel Ung for aiding in generating and providing the FEKO-generated \ac{EDA2} beam models. Additionally, we would like to thank Marcin Sokolowski and Ravi Subrahmanyan for their assistance and feedback on the \ac{EDA2} gain calibration methods. Furthermore, we thank Jaiden Cook for providing assistance and insights on Gaussian fitting algorithms. We would also like to thank the anonymous reviewer for their valuable comments, resulting in an overall improvement of this manuscript. This research was supported by the \ac{ASTRO3D}, through project number CE170100013. This scientific work makes use of the \acf{MRO}, operated by CSIRO. We acknowledge the Wajarri Yamatji People as the traditional owners of the observatory site. We acknowledge the Pawsey Supercomputing Centre which is supported by the Western Australian and Australian Governments. We acknowledge the use of the \ac{LAMDA}, part of the \ac{HEASARC}. \ac{HEASARC}/\ac{LAMDA} is a service of the Astrophysics Science Division at the NASA Goddard Space Flight Center. We acknowledge the work and the support of the developers of the following Python packages: pyGDSM \citep{Price2016}, Numpy \citep{harris2020}, Astropy \citep{astropy:2013,astropy:2018}, Healpy \citep{Zonca2019}, Scipy \citep{2020SciPy}, and Matplotlib \citep{Hunter:2007}. We acknowledge the work and support of the developers of the HEALPix software \citep{Gorski2005} and Miriad software \citep{Sault1995}.
\end{acknowledgements}

\bibliographystyle{pasa-mnras}
\bibliography{bibliography}

\begin{appendix}
\clearpage
\section{Examples of Beam-Coefficient Space}
\label{apdx:beam-coverage}

\begin{figure*}
    \centering
    \includegraphics[width=.95\linewidth]{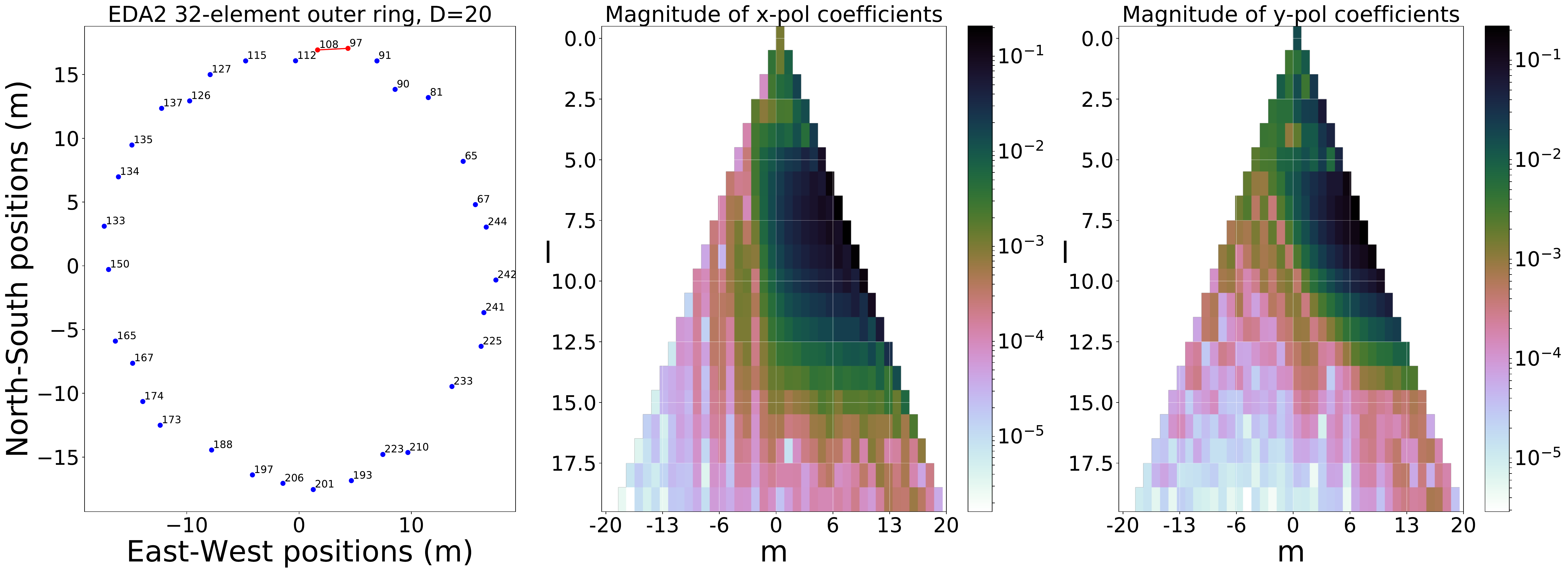}
    \caption{Example of spherical harmonic beam coefficients for a short East-West baseline (contribution is low on spatial coefficients $l$ and is primarily $m$ dependant).}
    \label{fig:baseD20}
\end{figure*}

\begin{figure*}
    \centering
    \includegraphics[width=.95\linewidth]{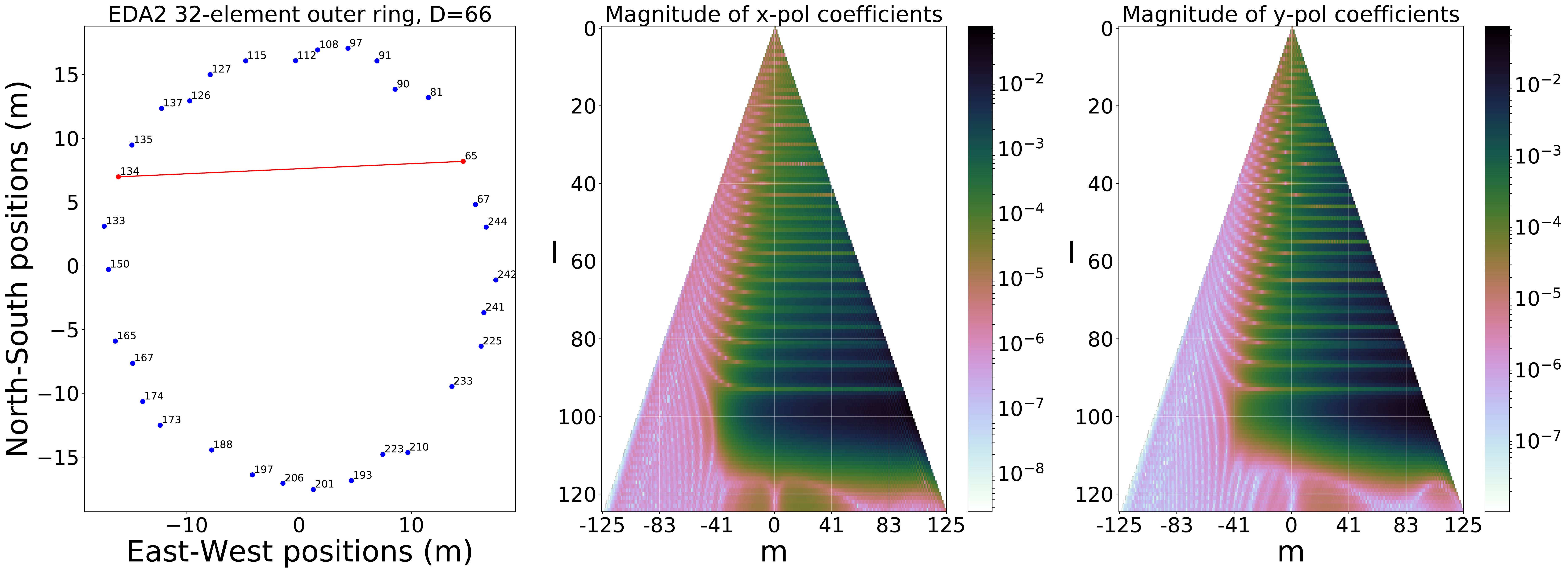}
    \caption{Example of spherical harmonic beam coefficients for a long East-West baseline (contribution is high on spatial coefficients $l$ and is primarily $m$ dependant).}
    \label{fig:baseD66}
\end{figure*}

\begin{figure*}
    \centering
    \includegraphics[width=.95\linewidth]{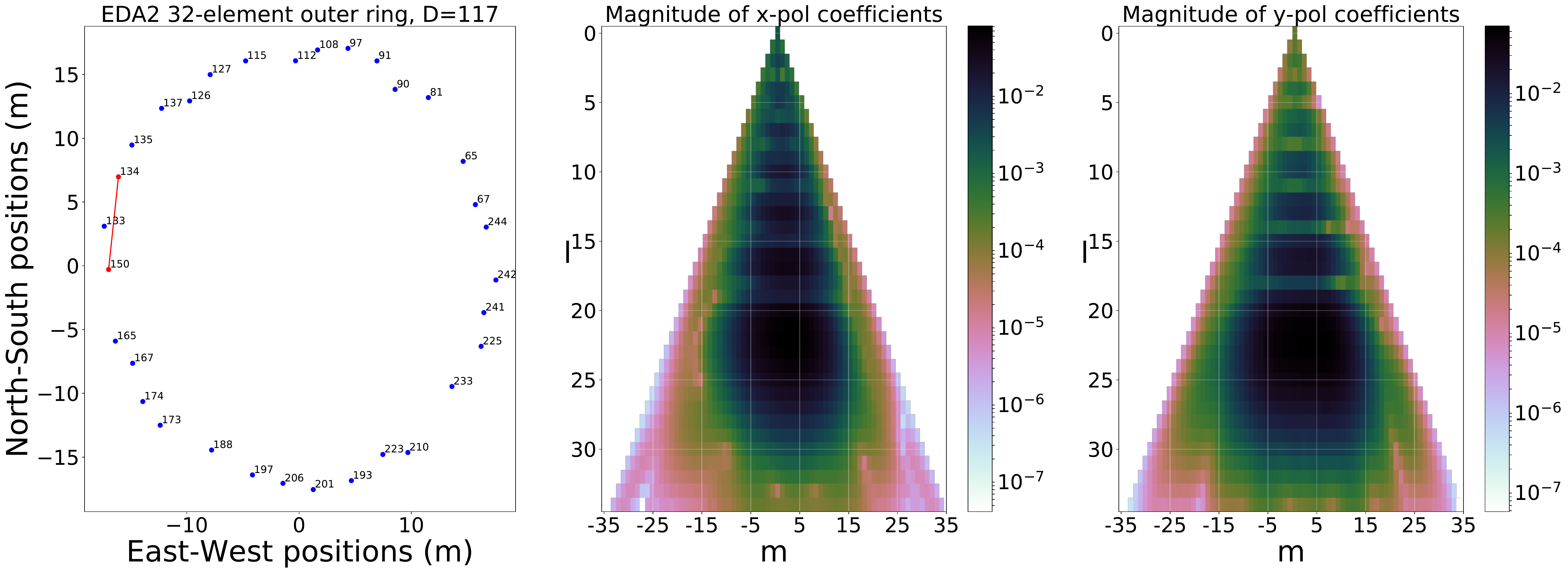}
    \caption{Example of spherical harmonic beam coefficients for a  North-South baseline (contribution is symmetric around $m=0$).}
    \label{fig:baseD117}
\end{figure*}

\begin{figure*}
    \centering
    \includegraphics[width=.95\linewidth]{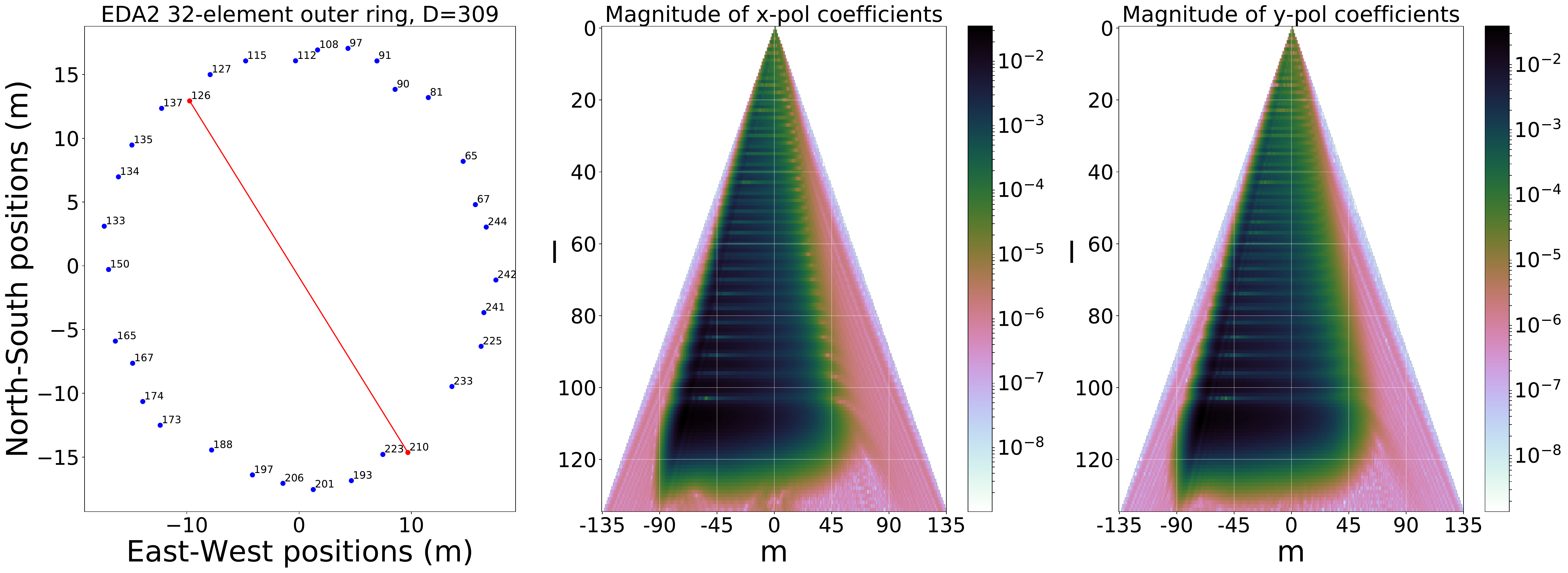}
    \caption{Example of spherical harmonic beam coefficients for a  diagonal baseline pointing North-West (contribution moves to the negative $m$-modes).}
    \label{fig:baseD309}
\end{figure*}

\begin{figure*}
    \centering
    \includegraphics[width=.95\linewidth]{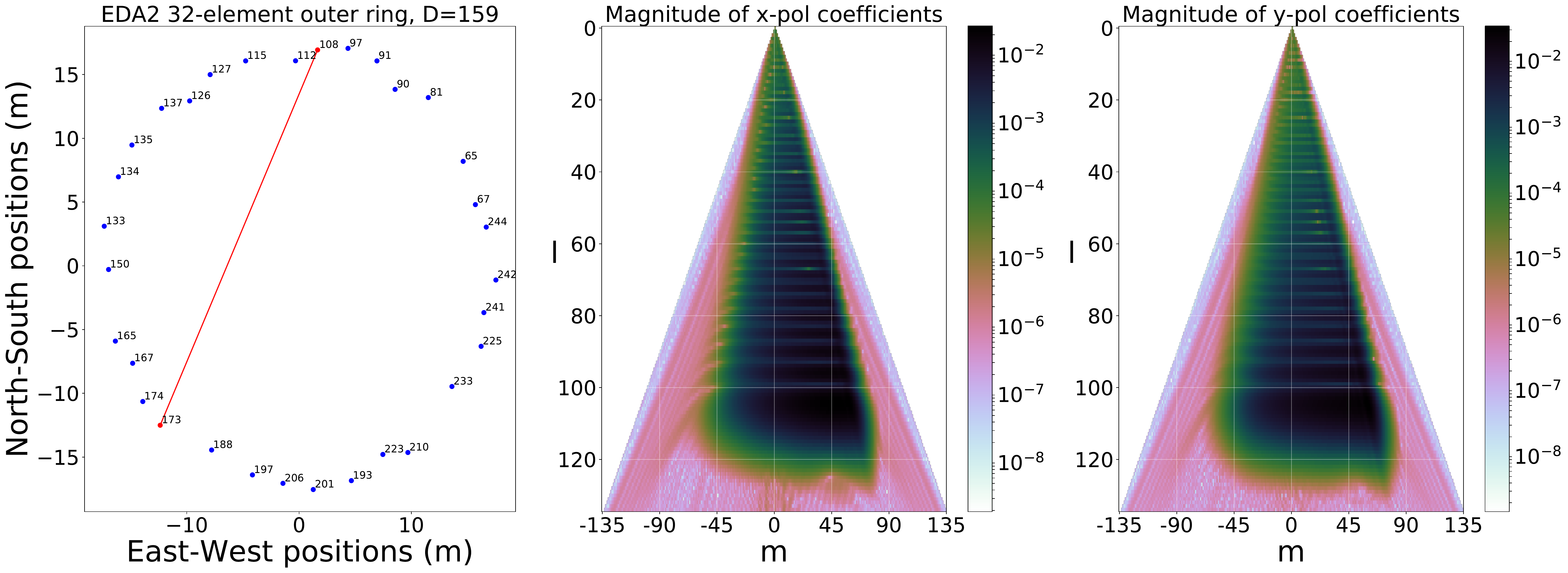}
    \caption{Example of spherical harmonic beam coefficients for a  diagonal baseline pointing North-East (contribution moves to the positive $m$-modes).}
    \label{fig:baseD159}
\end{figure*}

\twocolumn[\section{Sky Maps in HEALPix representation}]
\label{apdx:HEALPix-maps}
\begin{figure*}
    \centering
    \includegraphics[width=\linewidth]{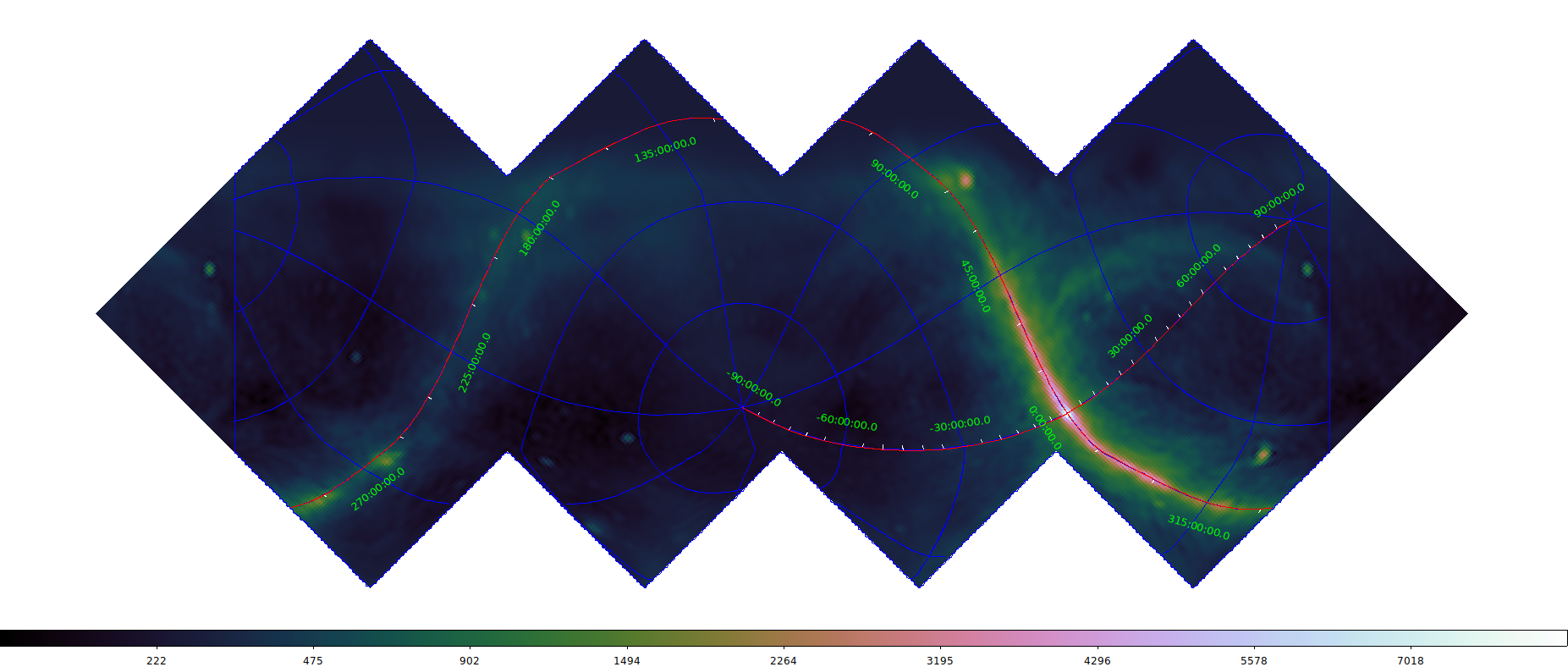}
    \caption{159\,MHz diffuse EDA2 map (equatorial view, HEALPix RING ordering scheme), log-scale. Generated without the use of a prior model, the global sky component is reinserted.}
    \label{fig:EDA2_nopriorCelHPX}
\end{figure*}

\begin{figure*}
    \centering
    \includegraphics[width=\linewidth]{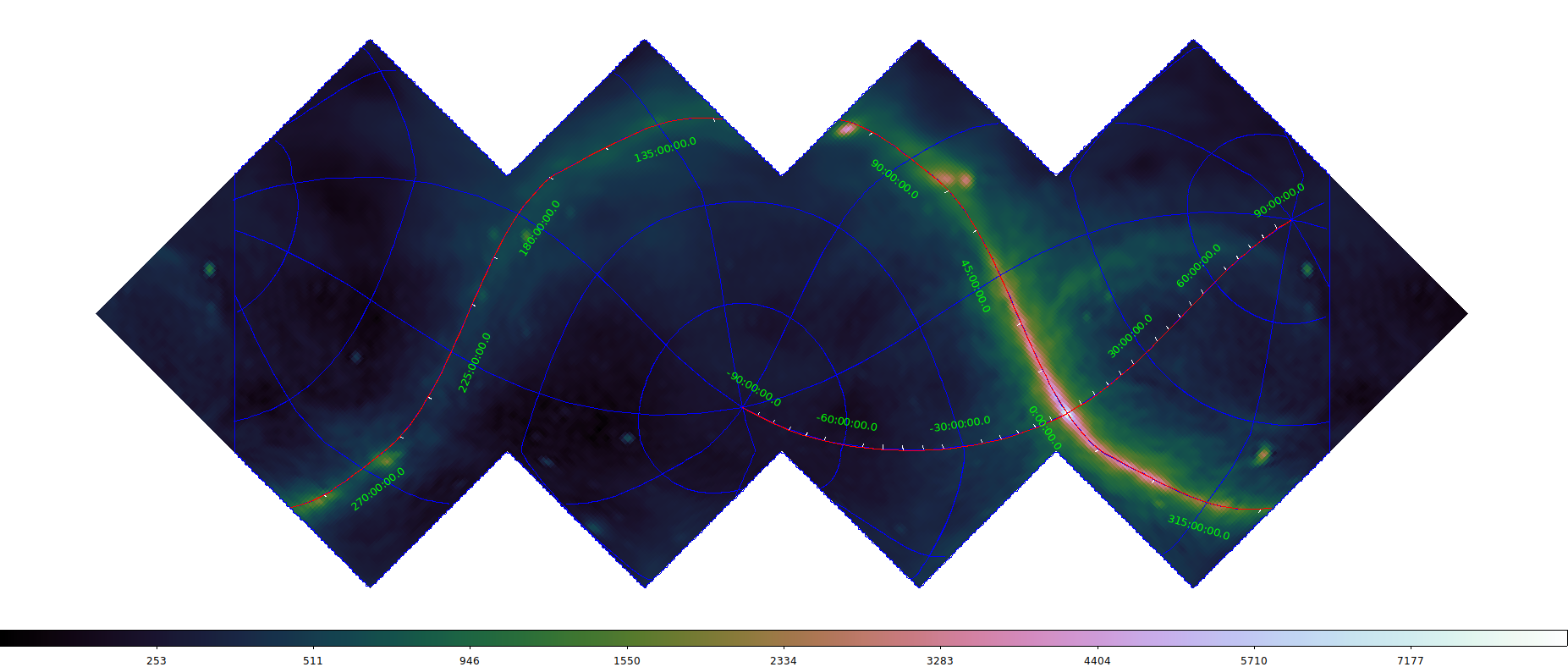}
    \caption{159\,MHz diffuse EDA2 map (equatorial view, HEALPix RING ordering scheme), log-scale. Generated with the use of the reprocessed desourced Haslam map as a prior model to constrain the largest scales. Since we cannot observe at declinations $>60^{\circ}$, the northern hemisphere is equivalent to the diffuse Haslam map depicted in \autoref{fig:model_map}.}
    \label{fig:EDA2_priorCelHPX}
\end{figure*}

\end{appendix}

\end{document}